\documentclass{article}
\usepackage{epsfig}
\usepackage{amsmath}
\usepackage{amsfonts}
\usepackage{amssymb}
\usepackage{euscript}
\usepackage{graphicx}
\usepackage{psfrag}

\begin{document}

\sloppy

\begin{center}
{{\Huge \bfseries Astrophysics from the physical point of view
}}
\end{center}
\vspace{4cm}

\begin{center}
{\Huge
\itshape{ B.V.Vasiliev}}
\end{center}

\vspace{4cm}
\clearpage

\clearpage
\hspace{2cm} {\Large{The Annotation}} \vspace{1cm}

    The physics of stars requires from its models and theories to be in accordance with totality of existing astrophysical knowledge. It is basis for the star physics. This is the main difference of astrophysics from all other branches of physics, for which foundations are  measurement data.

However at an using the progress of a measuring technics, the important measurement data were obtained by astronomers for few last decades. The existence of these data gives a possibility to establish the star physics on the solid base of an agreement of the theory and data of measurements.

To reach the consent of the theory of stars with data of astronomical measurements, it is necessary to abandon to some astrophysical constructions, which are commonly accepted today. At first, necessary to take into account that the interior of the star is formed by electron-nuclear plasma - the electrically polarized medium. So the equation of the balance of plasma in gravitational field must take into account the role of the gravity-induced electric polarization. The taking into account of the gravity-induced electric polarization of plasmas allows to formulate the theory of a star, in which all main parameters - a mass of the star, its temperature, radius and luminosity - are expressed by determined combinations of the world constants, but star individuality is depended of only two parameters - mass and charge numbers of atomic nuclei from which stellar plasma is composed. This theory gives quantitative explanations with satisfactory accuracy for all dependencies, which are measured by astronomers now.

The special attention attracts the stellar mass spectrum. Theoretically the mass of the star can be calculated on a base of balance equations of the intra-stellar plasma.  It turns out that, the majority of stars, except most heavy, is composed by plasma with  neutron-excess atomic nuclei. The stability of these nuclei is attached by the specifical mechanism of neutronization acting into dense plasma.

Within the framework of this approach, the theories of the periastron rotation of close binary stars and the magnetic field of stars was considered.

It is important to note, that the taking into account of the gravity-induced electric polarization leads to the other conceptual changes. For instance, it rejects the collapse mechanism of stars at the last stage of their evolutions and hereunder puts the question about a mechanism of the "black holes" formation.

    Thereby, at taking into account the gravity-induced effects in plasma, we must refuse some models, which are commonly accepted today and seem obvious. The process of a refusing can be painful for a conservative part of astrophysical community. But this refusing is completely justified and absolutely necessary. Only herewith the quantitative explanation can be obtained for all corresponding measurement data without using of a any fitting parameter. The star physics obtains herewith the reliable foundation in the manner of result of these measurements, as a physical science must have.

\clearpage

\begin
{abstract}
Astrophysics = the star physics was beginning its development without a supporting of measurement data, which could not be obtained then. Still astrophysics exists without this support, although now astronomers collected a lot of valuable information. This is the main difference of astrophysics from all other branches of physics, for which foundations are measurement data. The creation of the theory of stars, which is based on the astronomical measurements data, is one of the main goals of modern astrophysics. Below the principal elements of star physics based on data of astronomical measurements are described.
The theoretical description of a hot star interior is obtained. It explains
the distribution of stars over their masses, mass-radius-temperature and
mass-luminosity dependencies. The theory of the apsidal rotation of binary
stars and the spectrum of solar oscillation is considered.
All
theoretical predictions are in a good agreement with the known measurement
data, which confirms the validity of this consideration.
\end{abstract}



\tableofcontents \clearpage

\section{Introduction} \label{Ch1A}

\subsection{That is the basis of astrophysics?}

Astrophysics\footnote{The modern
astrophysics has a whole series of different branches.
It is important to stress that all of them except the physics of hot stars beyond the scope of  this consideration; we shall use the term
"astrophysics" here and below in its initial meaning - as the
physics of stars or more tightly as the physics of hot stars.} is linked with other physical sciences using of standard physical laws - laws of mechanics, statistical physics and thermodynamics, electrodynamics etc. However at an inevitable similarity of all physical sciences,
astrophysics has an unique peculiar "architecture". Its difference has historical roots.

The modern physics begins its formation at last 16 c. - middle 17 c. mainly with works of W.Gilbert and G.Galileo. They introduce into practice the main instrument of the present exact science -
the empirical testing of a scientific hypothesis.
Until that time false scientific statements weren't afraid
of an empirical testing. A flight of fancy was dainty and
refined  than an ordinary and crude material world. The exact
correspondence to a check experiment was not
necessary for a philosophical theory, it almost discredited the theory in the experts opinion.
The discrepancy of a theory and observations was not confusing at
that time. Now the empirical testing of all theoretical hypotheses
gets by a generally accepted obligatory method of the exact science. As a
result all basic statements of physics are sure established and based on the solid foundation of  an agreement with measurement data.

Astrophysicists were applying  standard physical laws for description of the interior of  distant  and mysterious stars, when there was nothing   known about them except that they exist. As a result, up to these days, astrophysical models of a star interior are  elaborated without some hope that they can be checked by comparison with measurement data.

The astrophysical community is thinking that a
foundation of their science is not a measuring data, but it
is an internal consensus of a totality of astrophysical knowledge, models of stars and their
evolution, which gives a sure in an objective character of their
adequacy. \footnote{According to words of one of known
European astrophysicist (on memories).}

At this basis, the first ideas of "apostles"{} - A.Eddington, S.Chandrasekhar, G.Bete, K.Schwazshchild and others - play a special "canonical" role. It makes the star physics very conservative.
In principle, any new results and ideas must be rejected if they were created  by physical science after  a "canon" formulating. Such happened with plasma physics. Its main results were obtained  in the second half of 20 c. only and
they are not included in the foundation of astrophysics despite the fact that stars are composed by plasma.

It is important that at present, the basis of astrophysics can be revised. Using the progress of a measuring technics, astronomers  were obtaining during last decades of 20 c. the important measuring data.
Now one can find approximately ten such measured dependencies and star parameters.
The existence of these data opens a new stage of  star  physics.  Now it can abandon to the star description in  pre-Galilean manner  and to get up on the solid base of an accord of the theory and  data measurement, as it accepted in all other branches of physics.

    The comparison of these measuring data and appropriate theoretical models  will be considered below in detail.

\subsection{The basic postulate of astrophysics}

The basic postulate of astrophysics - the Euler equation - was
formulated in a mathematical form by L.Euler in a middle of 18th
century for the  "terrestrial" effects description. This equation
determines the equilibrium condition of  liquids or gases in a
gravitational field:
\begin{equation}
\gamma {\mathbf g} = -{\mathbf \nabla} P.\label{Eu}
\end{equation}
According to it the action of a gravity forth  $\gamma {\mathbf g}$
($\gamma$ is density of substance,  ${\mathbf g}$ is the  gravity
acceleration) in equilibrium is balanced by a forth which is induced
by the pressure gradient in the substance. This postulate is surely
established and experimentally checked be "terrestrial" physics. It
is the base of an operating of  series of technical devices -
balloons, bathyscaphes and other.

All modern models of  stellar interior are obtained on the  base of
the Euler equation. These models assume that pressure inside a star
monotone increases depthward from the star surface. As a star
interior substance can be considered as an ideal gas which pressure
is proportional to its temperature and density, all astrophysical
models predict  more or less monotonous increasing of temperature
and density of the star substance in the direction of the center of
a star.

\subsection{The Galileo's method}
The correctness of fundamental scientific postulate has to be
confirmed by empirical testing. The basic postulate of astrophysics
- the Euler equation (Eq.(\ref{Eu})) - was formulated long before
the  plasma theory has appeared, even long before the discovering of
electrons and nuclei. Now we are sure that the star interior must
consist of electron-nuclear plasma. The Euler equation describes
behavior of liquids or gases - substances, consisting of neutral
particles. One can ask oneself: must the electric interaction
between plasmas particles  be taken into account at a reformulation
of the Euler equation, or does it give a small correction only,
which can be neglected in most cases, and how is it  accepted by
modern stellar theories?

To solve the problem of the correct choice  of the postulate, one
has  the Galileo's method. It consists of 3 steps:

{\it (1) to postulate a hypothesis about the nature of the
phenomenon, which is free from  logical contradictions;

(2) on the base of this postulate, using the standard
mathematical procedures, to conclude laws of the phenomenon;

(3) by means of empirical method to ensure, that the nature obeys
these laws (or not) in reality, and to confirm (or not) the basic
hypothesis.}
\bigskip

The use of this method gives a possibility to reject false
postulates and theories, provided there exist a necessary
observation data, of course.

The modern theory of the star interior is based on the Euler
equation Eq.(\ref{Eu}). Another possible approach is to take into
account all interactions acting in the considered system. Following
this approach one  takes into account in the equilibrium equation
the electric interactions too, because plasma particles have
electric charges. An energetic reason speaks that plasma in gravity
field must stay electroneutral and surplus free charges must be
absent inside it.The gravity action on a neutral plasma may induce a
displacements of its particles, i.e. its electric polarization,
because electric dipole moments are lowest moments of the analysis
of multipoles. If we take into account the electric polarization
plasma $\mathfrak{P}$, the equilibrium equation obtains the form:
\begin{equation}
\gamma \mathbf {g}+\frac{4\pi}{3}\mathfrak{P}\nabla\mathfrak{P}+
\mathbf {\nabla} P =0,\label{Eu-1}
\end{equation}
On the basis of this postulate and using standard procedures, the
star interior structure can be considered (it is the main substance
of this paper).

Below it will be shown that the electric force in a very hot dense
plasma can fully compensate the gravity force and the equilibrium
equation then obtains the form:
\begin{equation}
\gamma
\mathbf{g}+\frac{4\pi}{3}\mathfrak{P}\nabla\mathfrak{P}=0\label{Eu-2}
\end{equation}
and
\begin{equation}
\mathbf {\nabla} P=0\label{P-0},
\end{equation}
which points to  a radical distinction of this approach in
comparison to the standard scheme based on Eq.(\ref{Eu}).

\subsection[What does the  measurement data express?]{What does the the astronomic measurement data express?}
Are there actually  astronomic measurement data, which can give
possibility to  distinguish "correct" and "incorrect" postulates?
What must one do, if the direct measurement of the star interior
construction is impossible?

Are there astronomical measurement data which can clarity this
problem? Yes. They are:

1. There is the only method which gives information about the
distribution of the matter inside stars. It is the measurement of
the  periastron rotation of closed double stars. The results of
these measurements give a qualitative confirmation that the  matter
of a star is more dense near to its center.

2. There is the measured distribution of star masses. Usually it
does not have any interpretation.

3. The astronomers detect the dependencies of radii of stars and
their surface temperatures from their masses.  Clear quantitative
explanations of these dependencies are unknown.

4. The existence of these dependencies leads to a mass-luminosity
dependence. It is indicative that the astrophysical community does
not find any  quantitative explanation to  this dependence almost
100 years after as it was discovered.

5. The discovering of the solar surface oscillations   poses a new
problem that needs a quantitative explanation.

6. There are measured data showing that a some stars are possessing
by the  magnetic fields. This date can be analyzed at taking into
account of the electric polarization inside these stars.

It seems that all measuring  data are listed which can determine the
correct choice of the starting postulate. \footnote{ From this point
of view one can consider also the measurement of solar neutrino flux
as one of similar phenomenon. But its result can be interpreted
ambiguously because there are a bad studied mechanism of their
mutual conversation and it seems prematurely to use this measurement
for a stellar models checking.}

It is important to underline that all above-enumerated dependencies
are known but they don't have a quantitative  (often even
qualitative) explanation in frame of the standard theory based on
Eq.(\ref{Eu}). Usually one does not consider the existence of these
data as a possibility to check the theoretical constructions. Often
they are ignored.

It will be shown below that all these dependencies obtain a
quantitative explanation in the theory based on the postulate
Eq.(\ref{Eu-2}), which takes in to account the  electric interaction
between particles in the stellar plasma. At that all basic measuring parameters of stars - masses, radii, temperatures -  can be described by definite rations of world constants, and it gives a  good agreement with measurement data.

\subsection[About a star theory development]{About a star theory development}
The correct choice of the substance equilibrium equation is absolute requirement  of   an development  of the star theory which can be in agreement with measuring data.

To simplify a task of formulation of the such theory , we can accept two
additional postulates.

  A hot star generates an energy into its central region continuously. At the same time this energy radiates from the star surface.
This radiation is not in equilibrium relatively stellar substance.
It is convenient to consider that the star is existing in its stationary state.
It means that the star radiation has not  any changing in the time,
and at that the star must radiate from its surface as much energy as many it generates into its core. At this condition, the stellar substance is existing in stationary state and time derivatives from any thermodynamical functions which is characterizing for stellar substance are equal to zero:
\begin{equation}
\frac{dX}{dt}=0.
\end{equation}
Particularly, the time derivative of the entropy must be equal to zero in  this case. I.e. conditions of an existing of each small volume of stellar substance can be considered as adiabatic one in spite  of the presence of the non-equilibrium radiation. We shall use this simplification in Chapter VI.

The second simplification can be obtained if to suppose that a stationary star  reaches the minimum of its energy after milliards years of development.
(Herewith we exclude from our consideration stars with "active lifestyle". The interesting problem of  the transformation of a star falls out of examination too).

The minimum condition of the star energy gives possibility to determine main parameters of equilibrium stellar substance - its density and temperature.

It is reasonable to start the development of the star theory  from this point. So the problem of existing of the energy-favorable  density of the stellar substance and its temperature will be considered in the first place in the next Chapter.

\clearpage
\section{The energy-favorable state of hot dense plasma} \label{Ch2}

\subsection{The properties of a hot dense plasma}

\subsubsection{A hot plasma and Boltzman's distribution}

Free electrons being fermions obey the Fermi-Dirac statistic at
low temperatures. At high temperatures, quantum distinctions in
behavior of electron gas disappear and it is possible to
consider electron gas as the ideal gas which obeys the Boltzmann's
statistics. At high temperatures and high densities, all substances
transform into electron-nuclear plasma. There are two tendencies in
this case. At temperature much higher than the Fermi temperature
$T_F=\frac{\mathcal{E}_F}{k}$ (where $\mathcal{E}_F$ is Fermi
energy), the role of quantum effects is small. But their role grows
with increasing of the pressure and density of an electron gas. When
quantum distinctions are small, it is possible to describe the
plasma electron gas as a the ideal one. The criterium of
Boltzman's statistics applicability
\begin{equation}
T\gg\frac{\mathcal{E}_F}{k}.\label{a-6}
\end{equation}
hold true for a non-relativistic electron gas with density
$10^{25}$ particles in $cm^{3}$ at $T\gg10^6 K$.

At this temperatures, a plasma has energy
\begin{equation}
\mathcal{E}=\frac{3}{2}kTN
\end{equation}
and its EOS is the ideal gas EOS:
\begin{equation}
P=\frac{Nk T}{V}\label{a-7}
\end{equation}

But even at so high temperatures, an electron-nuclear plasma can be
considered as ideal gas in the first approximation only. For more
accurate description its properties, the specificity of the plasma
particle interaction must be taken into account and two main
corrections to ideal gas law must be introduced.

The first correction takes into account the quantum character of
electrons, which obey the Pauli principle, and cannot occupy levels
of energetic distribution  which are already occupied by other
electrons. This correction must be positive because it leads to an
increased gas incompressibility.

Other correction takes into account the  correlation of the screening
action of charged particles inside dense plasma. It is the so-called
correlational correction. Inside a dense plasma, the charged particles
screen the fields of other charged particles. It leads to a decreasing
of the pressure of charged particles. Accordingly, the correction for the
correlation of charged particles must be negative,because it
increases the compressibility of electron gas.

\subsubsection[The correction for the  Fermi-statistic]{The hot plasma energy with taking into account the correction for the  Fermi-statistic}
The energy of the electron gas  in the Boltzmann's case $(kT\gg
\mathcal{E}_F)$ can be calculated using the expression of the
full energy of a non-relativistic Fermi-particle system \cite{LL}:
\begin{equation}
\mathcal{E}=\frac{2^{1/2}V m_e^{3/2}}{\pi^2 \hbar^3} \int_0^\infty
\frac{\varepsilon^{3/2}d\varepsilon} {e^{(\varepsilon-\mu_e)/kT}+1},
\label{eg}
\end{equation}
expanding it in a series. ($m_e$ is electron mass,
$\varepsilon$ is the energy of electron and $\mu_e$ is its chemical
potential).

In the Boltzmann's case, $\mu_e<0$ and $|\mu_e/kT|\gg 1$ and the
integrand at $e^{\mu_e/kT}\ll 1$ can be expanded into a series
according to powers $e^{\mu_e/kT-\varepsilon/kT}$. If we
introduce the notation $z=\frac{\varepsilon}{kT}$ and conserve the
two first terms of the series,  we obtain
\begin{eqnarray}
I\equiv (kT)^{5/2}\int_0^\infty\frac{z^{3/2}dz} {e^{z-\mu_e/kT}+1}
\approx \nonumber \\ \approx (kT)^{5/2} \int_0^\infty z^{3/2}
\biggl(e^{\frac{\mu_e}{kT}-z}-e^{2(\frac{\mu_e}{kT}-z)}+...
\biggr)dz
\end{eqnarray}
or
\begin{eqnarray}
\frac{I}{(kT)^{5/2}}\approx
e^{\frac{\mu_e}{kT}}\Gamma\biggl(\frac{3}{2}+1\biggr)
-\frac{1}{2^{5/2}}e^{\frac{2\mu_e}{kT}}
\Gamma\biggl(\frac{3}{2}+1\biggr)\approx\nonumber \\ \approx
 \frac{3\sqrt{\pi}}{4}
e^{\mu_e/kT}\biggl(1-\frac{1}{2^{5/2}}e^{\mu_e/kT}\biggr).
\end{eqnarray}
Thus, the full energy of the hot electron gas is
\begin{equation}
\mathcal{E}\approx
\frac{3V}{2}\frac{(kT)^{5/2}}{\sqrt{2}}\biggl(\frac{m_e}{\pi
\hbar^2}\biggr)^{3/2} \biggl(e^{\mu_e/kT}-\frac{1}{2^{5/2}}
e^{2\mu_e/kT}\biggr)
\end{equation}
Using the definition of a chemical potential of ideal gas (of
particles with spin=1/2) \cite{LL}
\begin{equation}
\mu_e= kT log \biggl[\frac{N_e}{2V}\biggl(\frac{2\pi \hbar^2}{m_e
kT}\biggr)^{3/2}\biggr]\label{chem}
\end{equation}
we obtain the full energy of the hot electron gas
\begin{equation}
\mathcal{E}_e\approx\frac{3}{2}kTN_e\left[1+\frac{\pi^{3/2}}{4}\left(\frac{a_B
e^2}{kT}\right)^{3/2} n_e\right], \label{a-16}
\end{equation}
where $a_B=\frac{\hbar^2}{m_ee^2}$ is the Bohr radius.

\subsubsection[The correction for correlation  of particles]{The correction  for  correlation  of charged particles in a hot plasma}

At high temperatures, the plasma particles are uniformly distributed in space.
At this limit, the energy of ion-electron interaction
tends to zero. Some correlation in space distribution of particles
arises as the positively charged particle groups around itself
preferably particles with negative charges and vice versa. It is
accepted to estimate the energy of this correlation by the method
developed by Debye-H$\ddot u$kkel for strong electrolytes \cite{LL}.
The energy of a charged particle inside plasma is equal to
$e\varphi$, where $e$ is the charge of a particle, and $\varphi$ is
the electric potential induced by other particles  on the considered particle.

This potential inside plasma is determined by the Debye law
\cite{LL}:

\begin{equation}
\varphi(r)=\frac{e}{r} e^{-\frac{r}{r_D}}\label{vr}
\end{equation}
where the Debye radius is
\begin{equation}
r_D=\left({\frac{4\pi e^2 }{kT}~\sum_a n_{a}
Z_a^2}\right)^{-1/2}\label{a-18}
\end{equation}
For small values of ratio $\frac{r}{r_D}$, the potential can be
expanded into a series
\begin{equation}
\varphi(r)=\frac{e}{r}-\frac{e}{r_D}+...\label{rr}
\end{equation}
The following terms are converted into zero at $r\rightarrow 0$. The
first term of this series is the potential of the considered
particle. The second term
\begin{equation}
\mathcal{E}=-e^3 \sqrt{\frac{\pi}{kTV}}\left(\sum_a N_a
Z_a^2\right)^{3/2}
\end{equation}
is a potential induced by other particles of plasma on the charge
under consideration. And so the correlation energy of plasma
consisting of $N_e$ electrons and $(N_e/Z)$ nuclei with charge $Z$
in volume $V$ is \cite{LL}
\begin{equation}
\mathcal{E}_{corr}=-e^3 \sqrt{\frac{\pi n_e}{kT}}Z^{3/2}N_e \label{a-20}
\end{equation}

\subsection[The energy-preferable state]{The energy-preferable state of a hot plasma}
\subsubsection[The energy-preferable density]{The energy-preferable density  of a hot plasma}
Finally, under consideration of both main
corrections taking into account the inter-particle interaction, the full energy of plasma is given by
\begin{equation}
\mathcal{E}\approx\frac{3}{2}kTN_e\biggl[1+\frac{\pi^{3/2}}{4}\biggl(\frac{a_B
e^2}{kT}\biggr)^{3/2}n_e - \frac{2\pi
^{1/2}}{3}\biggl(\frac{Z}{kT}\biggr)^{3/2}e^3
n_e^{1/2}\biggr]\label{a-21}
\end{equation}
The plasma into a star has a feature. A star generates the energy into its inner region and radiates it from the surface. At the
steady state of a star, its substance must  be in the equilibrium state
with a minimum of its energy. The radiation is not in equilibrium of course
and can be considered as a star environment.
The equilibrium state of a body in an environment is
related to the minimum of the function (\cite{LL}§20):
\begin{equation}
\mathcal{E}-T_o S+P_oV, \label{a-22}
\end{equation}
where  $T_o$ and $P_o$ are the temperature and the pressure of an
environment. At taking in to account that the star radiation is
going away into vacuum, the two last items can be neglected and one can
obtain the equilibrium equation of a star substance  as the minimum of
its full energy:
\begin{equation}
\frac{d\mathcal{E}_{plasma}}{dn_e}=0.\label{a-24}
\end{equation}
Now taking into account Eq.({\ref{a-21}}), one obtains that an
equilibrium condition corresponds to the equilibrium density of the
electron gas of a hot plasma
\begin{equation}
n_e^{equilibrium}\equiv{n_\star}=\frac{16}{9\pi}\frac{Z^3}{a_B^3}\approx
1.2 \cdot 10^{24} Z^3 cm^{-3},\label{eta1}
\end{equation}
It gives the electron density $\approx3\cdot 10^{25}cm^{-3}$ for
the equilibrium state of the hot plasma of helium.

\subsubsection[The energy-preferable temperature]{The estimation of temperature of energy-preferable state of a hot stellar plasma}
As the steady-state value of the density of a hot non-relativistic
plasma is known, we can obtain an energy-preferable
temperature of a hot non-relativistic plasma.

The virial theorem \cite{LL,VL} claims that the full energy of
particles $E$, if they form a stable system with the Coulomb law
interaction, must be equal to their kinetic energy $T$ with a negative sign.
Neglecting small corrections at a high temperature, one can
write the full energy of a hot dense plasma as
\begin{equation}
\mathcal{E}_{plasma}= U + \frac{3}{2}kTN_e = - \frac{3}{2}kT
N_e.\label{Ts11}
\end{equation}
Where $U\approx-\frac{G\mathbb{M}^2}{\mathbb{R}_0}$ is the potential
energy of the system, $G$ is the gravitational constant, $\mathbb{M}$
and $\mathbb{R}_0$ are the mass and the radius of the star.

As the plasma temperature is high enough, the energy of the black
radiation cannot be neglected. The full energy of the  stellar plasma
depending on the particle energy and the black radiation energy
\begin{equation}
\mathcal{E}_{total}=-\frac{3}{2}kT N_e + \frac{\pi^2}{15}
\biggl(\frac{kT}{\hbar c}\biggr)^3 V kT
\end{equation}
at equilibrium state must be minimal, i.e.
\begin{equation}
\biggl(\frac{\partial \mathcal{E}_{total}}{\partial T}\biggr)_{N,V}
=0.\label{a-27}
\end{equation}
This condition at $\frac{N_e}{V}=n_\star$ gives a possibility to
estimate the temperature of the hot stellar plasma at the  steady state:
\begin{equation}
\mathbb{T}_\star\approx Z\frac{\hbar c}{ka_B}\approx
10^7Z~K.\label{Ts1}
\end{equation}
The last obtained estimation can raise doubts. At  "terrestrial"
conditions, the energy of any substance reduces to a minimum at
$T\rightarrow 0$. It is caused by a positivity of a heat capacity
of all of substances. But the
steady-state energy of star is negative and its absolute value
increases with increasing of temperature (Eq.({\ref{Ts11}})).
It is the main property of a star as a thermodynamical object. This
effect is a reflection of an influence of the gravitation on a
stellar substance and is characterized by a negative effective heat
capacity. The own heat capacity of a stellar substance (without
gravitation) stays positive. With the increasing of the temperature, the role
of the black radiation increases ($\mathcal{E}_{br}\sim T^4$). When
its role dominates, the star obtains a positive heat capacity. The
energy minimum corresponds to a point between these two branches.
\subsubsection{Are accepted assumptions correct?}
At expansion in series of the full energy of a Fermi-gas, it was
supposed that the condition of applicability of Boltzmann-statistics
({\ref{a-6}}) is valid. The substitution of obtained values of the
equilibrium density  $n_\star$ ({\ref{eta1}}) and equilibrium
temperature $\mathbb{T}_\star$ ({\ref{Ts1}}) shows that the ratio
\begin{equation}
\frac{\mathcal{E}_F(n_\star)}{k\mathbb{T}_\star}\approx
3.1Z\alpha\ll 1.\label{alf}
\end{equation}
Where $\alpha\approx\frac{1}{137}$ is fine structure constant.

At appropriate substitution, the condition of expansion in series of
the electric potential ({\ref{rr}}) gives
\begin{equation}
\frac{r}{r_D}\approx (n_\star^{1/3}r_D)^{-1}\approx
{\alpha}^{1/2}\ll 1.
\end{equation}
Thus, obtained values of steady-state parameters of plasma are in
full agreement with assumptions which was made above.
\clearpage
\section[The gravity induced electric polarization]{The gravity induced electric polarization in a dense hot plasma}\label{Ch3}

\subsection{Plasma cells}

The existence of plasma at energy-favorable state with
the constant density $n_\star$ and the constant temperature
$\mathbb{T}_\star$  puts a question about equilibrium of this plasma
in a gravity field. The Euler equation in commonly accepted form
Eq.({\ref{Eu}) disclaims a possibility to reach the equilibrium in a
gravity field at a constant pressure in the substance: the gravity
inevitably must induce a pressure gradient into gravitating matter.
To solve this problem, it is necessary to consider  the equilibrium
of a dense plasma in an gravity field in detail. At zero approximation, at
a very high temperature, plasma can be considered as a "jelly", where
electrons and nuclei are "smeared" over a volume. At a lower
temperature and a high density, when an interpartical interaction
cannot be neglected, it is accepted to consider a plasma dividing in
cells \cite{Le}. Nuclei are placed at centers of these cells, the rest of
their volume is filled by electron gas. Its density decreases from the
center of a cell to its periphery. Of course, this dividing is not
freezed. Under action of heat processes, nuclei move. But having a
small mass, electrons have time to trace this moving and to form a
permanent electron cloud around nucleus, i.e. to form a cell. So
plasma under action of a gravity must be characterized by two
equilibrium conditions:

- the condition of an equilibrium of the heavy nucleus inside a plasma
cell;

- the condition of an equilibrium of the electron gas, or equilibrium of
cells.

\subsection[The equilibrium of a nucleus]{The equilibrium of a nucleus inside plasma
cell filled by an electron gas}

At the absence of gravity, the negative charge of an electron cloud inside a
cell exactly balances the positive charge of the nucleus at its
center. Each cell is fully electroneutral. There is no  direct interaction
between nuclei.

The gravity acts on electrons and nuclei at the same time. Since the
mass of nuclei is large, the gravity force applied to them is  much larger
than the force applied to electrons. On the another hand, as
nuclei have no direct interaction, the elastic properties of plasma
are depending on the electron gas reaction. Thus er have a
situation, when the force applied to nuclei  must be balanced by the
force of the electron subsystem. The cell obtains an electric
dipole moment $d_s$, and  the plasma obtains polarization
$\mathfrak{P}=n_s~d_s$, where $n_s$ is the density of the cell.

It is known \cite{LL8}, that the polarization of neighboring cells
induces in the considered cell the electric field intensity
\begin{equation}
E_s=\frac{4\pi}{3}\mathfrak{P},
\end{equation}
and the cell obtains the energy
\begin{equation}
\mathcal{E}_s=\frac{d_s~E_s}{2}.
\end{equation}

The gravity force applied to the nucleus is proportional to its mass
$Am_p$ (where $A$ is a mass number of the nucleus, $m_p$ is the proton
mass). The cell consists of $Z$ electrons, the gravity force applied
to the cell electron gas is proportional to $Z m_e$ (where $m_e$ is
the electron mass). The difference of these forces tends to pull apart
centers of positive and negative charges and to increase the dipole
moment. The electric field  $E_s$ resists it. The process obtains
equilibrium at the balance of the arising electric force  $\nabla
\mathcal{E}_s$ and the difference of gravity forces applied to
the electron gas and the nucleus:
\begin{equation}
\mathbf{\nabla}\left(\frac{2\pi}{3}\frac{\mathfrak{P}^2}{n_s}\right)+(Am_p-Zm_e)\mathbf{g}=0\label{b2}
\end{equation}
Taking into account, that $\mathbf{g}=-\nabla \psi$, we obtain
\begin{equation}
\frac{2\pi}{3}\frac{\mathfrak{P}^2}{n_s}= (Am_p-Zm_e)\psi.\label{b3}
\end{equation}
Hence,
\begin{equation}
\mathfrak{P}^2 = \frac{3GM_r}{2\pi
r}n_e\left(\frac{A}{Z}m_p-m_e\right),\label{b5}
\end{equation}
where $\psi$ is the potential of the gravitational field,
$n_s=\frac{n_e}{Z}$ is the density of cell (nuclei), $n_e$ is the density of
the electron gas, $M_r$ is the mass of a star containing inside a sphere
with radius $r$.

\subsection[The equilibrium in electron gas subsystem]{The equilibrium in plasma electron gas subsystem}

Non-uniformly polarized matter can be represented by an electric
charge distribution with density  \cite{LL8}
\begin{equation}
\widetilde{\varrho}= \frac{div E_s}{4\pi}=\frac{div\mathfrak{P}}{3}.
\end{equation}
The full electric charge of cells placed inside the sphere with
radius $r$
\begin{equation}
Q_r= 4\pi \int_0^r \widetilde{\varrho}r^2 dr
\end{equation}
determinants the electric field intensity applied to a cell placed on
a distance  $r$  from center of a star
\begin{equation}
\widetilde{\mathbf{E}}=\frac{Q_r}{r^2}
\end{equation}
As a result, the action of a non-uniformly polarized environment can be
described by the force $\widetilde{\varrho}\widetilde{E}$. This
force must be taken into account in the formulating of equilibrium equation.
It leads to the following form  of the Euler equation:
\begin{equation}
\gamma\mathbf{g}+\widetilde{\varrho}\widetilde{\mathbf{E}}+\nabla
P=0\label{Eu1}
\end{equation}

\clearpage
\section{The internal structure of a star} \label{Ch4}

It was shown above that the state with the constant density is
energy-favorable for a plasma at a very high temperature. The
plasma in the central region of a star can possess by this property .
The  calculation  made below shows that  the mass of central region
of a star with the constant density  - the star core - is equal
to 1/2 of the full star mass. Its radius is approximately equal to
1/10 of radius of a star, i.e. the  core with high density take
approximately 1/1000 part of the full volume of a star. The other
half of a stellar matter is distributed over the region placed above the
core. It has a relatively small density and it could be called as a
star atmosphere.

\subsection{The plasma equilibrium in the star core}
In this case,  the equilibrium condition (Eq.(\ref{b2})) for steady density plasma is achieved at
\begin{equation}
\mathfrak{P}=\sqrt{G}\gamma_\star r,
\end{equation}
Here the mass density is $\gamma_\star\approx\frac{A}{Z}m_p n_\star$.
The polarized state of the plasma can be described by a state with  an
electric charge at the density
\begin{equation}
\widetilde{\varrho}=\frac{1}{3}div
\mathfrak{P}=\sqrt{G}\gamma_\star,\label{roe}
\end{equation}
and the electric field applied to a cell is
\begin{equation}
\widetilde{\mathbf{E}}=\frac{\mathbf{g}}{\sqrt{G}}.
\end{equation}
As a result, the electric force applied to the cell will fully balance
the gravity action
\begin{equation}
\gamma\mathbf{g}+\widetilde{\varrho}\widetilde{\mathbf{E}}=0\label{Eu2}
\end{equation}
at the zero pressure gradient
\begin{equation}
\nabla P=0\label{EuP}.
\end{equation}

\subsection[The main parameters of a star core]{The main parameters of a star core (in order of values)}

At known density $n_\star$ of plasma into a core  and its
equilibrium temperature $\mathbb{T}_\star$, it is possible to
estimate the mass $\mathbb{M}_\star$ of a star core  and its radius
$\mathbb{R}_\star$. In accordance with the virial
theorem\footnote{Below we shell use this theorem in its more exact
formulation.}, the kinetic energy of particles composing the steady
system  must be approximately equal to its potential energy with
opposite sign:
\begin{equation}
\frac{G\mathbb{M}_\star^2}{\mathbb{R}_\star}\approx
k\mathbb{T}_\star\mathbb{N}_\star\label{b30}.
\end{equation}
Where $\mathbb{N}_\star=\frac{4\pi}{3}\mathbb{R}_\star^3 n_\star$ is
full number of particle into the star core.

With using determinations derived above ({\ref{eta1}) and
({\ref{Ts1})  derived before, we obtain
\begin{equation}
\mathbb{M}_\star \approx \frac{\mathbb{M}_{Ch}}{(A/Z)^2}\label{Ms}
\end{equation}
where  $\mathbb{M}_{Ch}=\left(\frac{\hbar c}{G
m_p^2}\right)^{3/2}m_p$ is the Chandrasekhar mass.

The radius of the core is approximately equal
\begin{equation}
\mathbb{R}_\star\approx\biggl(\frac{\hbar c}{G m_p^2}\biggr)^{1/2}
\frac{a_B}{Z(A/Z)}.\label{RN}
\end{equation}
where  $A$ and  $Z$ are the mass and the charge number of atomic nuclei
the plasma consisting of.

\subsection[The equilibrium inside the star atmosphere]{The equilibrium state of the plasma inside the star atmosphere}
The star core is characterized by the constant mass density, the
charge density, the temperature and the pressure. At a temperature typical for a star core, the plasma can be considered as ideal gas,
as interactions between its particles are small in comparison with
$k\mathbb{T}_\star$. In atmosphere, near surface of a star, the temperature is
approximately by $3\div 4$ orders smaller. But the plasma density is
lower. Accordingly, interparticle interaction is lower too and we
can continue to consider this plasma as ideal gas.

In the absence of the gravitation, the equilibrium state of ideal gas
in some volume comes with the pressure equalization, i.e. with the
equalization of its temperature $T$ and its density $n$. This
equilibrium state is characterized by the equalization of the chemical
potential of the gas $\mu$ (Eq.({\ref{chem})).

\subsection[The radial dependence of density and temperature]{The radial dependence of density and temperature of substance inside a star atmosphere}

For the equilibrium system, where different parts have  different temperatures, the
following relation  of the
chemical potential of particles to its temperature holds (\cite {LL},§25):
\begin{equation}
\frac{\mu}{kT}=const
\end{equation}
As thermodynamic (statistical) part of chemical potential of
monoatomic ideal gas is \cite{LL},{§45}:
\begin{equation}
\mu_T= kT~ln \biggl[\frac{n}{2}\biggl(\frac{2\pi \hbar^2}{m
kT}\biggr)^{3/2}\biggr],\label{muB}
\end{equation}
we can conclude that at the equilibrium
\begin{equation}
n\sim T^{3/2}.\label{b32}
\end{equation}
In external fields the chemical potential of a gas \cite{LL}§25 is equal to
\begin{equation}
\mu=\mu_T + \mathcal{E}^{potential}
\end{equation}
where $\mathcal{E}^{potential}$ is the potential energy of particles in
the external field. Therefore in addition to fulfillment of condition
Eq. ({\ref{b32}), in a field with Coulomb potential, the equilibrium
needs a fulfillment of the condition
\begin{equation}
-\frac{GM_r \gamma}{r kT_r}+
\frac{\mathfrak{P}_r^2}{2kT_r}=const\label{gP}
\end{equation}
(where $m$ is the particle mass, $M_r$ is the mass of a star inside a
sphere with radius $r$, $\mathfrak{P}_r$ and $T_r$ are the polarization
and the temperature on its surface. As on the core surface, the left
part of Eq.(\ref{gP}) vanishes, in the atmosphere
\begin{equation}
M_r \sim r kT_r.\label{m-tr}
\end{equation}
Supposing  that a decreasing  of temperature inside the atmosphere is
a power function with the exponent $x$, its value on a  radius $r$ can be
written as
\begin{equation}
T_r=\mathbb{T}_\star\biggl(\frac{\mathbb{R_\star}}{r}\biggr)^x\label{Tr}
\end{equation}
and in accordance with  ({\ref{b32}}), the density
\begin{equation}
n_r=n_\star\biggl(\frac{\mathbb{R}_\star}{r}\biggr)^{3x/2}.\label{nr}
\end{equation}
Setting the  powers of $r$ in the right and the left parts of the
condition ({\ref{m-tr}) equal, one can obtain $x=4$.

Thus, at using power dependencies for the description of radial
dependencies of density and temperature, we obtain
\begin{equation}
n_r={n}_\star\biggl(\frac{\mathbb{R}_\star}{r}\biggr)^6\label{an-r}
\end{equation}
and
\begin{equation}
T_r=\mathbb{T}_\star\biggl(\frac{\mathbb{R}_\star}{r}\biggr)^4.\label{tr}
\end{equation}

\subsection[The mass of the star]{The mass of the star atmosphere and the full mass of a star}
After integration of ({\ref{an-r}}), we can obtain the mass of the star
atmosphere
\begin{equation}
\mathbb{M}_{A} = 4\pi \int_{\mathbb{R}_\star}^{\mathbb{R}_0} (A/Z) m_p
n_\star \biggl(\frac{\mathbb{R}_\star}{r}\biggr)^6 r^2 dr=
\frac{4\pi}{3}(A/Z)m_p n_\star
\mathbb{R_\star}^3\left[1-\left(\frac{\mathbb{R}_\star}{\mathbb{R}_0}\right)^3\right]\label{ma}
\end{equation}
It is equal to its core mass (to
$\frac{\mathbb{R}_\star^3}{\mathbb{R}_0^3}\approx 10^{-3}$), where
$\mathbb{R}_0$ is radius of a star.

Thus, the full mass of a star
\begin{equation}
\mathbb{M} = \mathbb{M}_A+ \mathbb{M}_\star\approx
2\mathbb{M}_\star\label{2mstar}
\end{equation}

\clearpage
\section [The main parameters of star]
{The virial theorem  and main parameters of a star} \label{Ch5}

\subsection{The energy of a star}

The virial theorem \cite{LL,VL} is applicable to a  system of
particles if they have a finite moving into a volume  $V$. If their
interaction obeys to the Coulomb's law, their potential energy
$\mathcal{E}^{potential}$, their kinetic energy
$\mathcal{E}^{kinetic}$ and pressure  $P$ are in the ratio:
\begin{equation}
2\mathcal{E}^{kinetic}+\mathcal{E}^{potential}= 3PV.
\end{equation}
On the star surface, the pressure is absent and for the particle
system as a whole:
\begin{equation}
2\mathcal{E}^{kinetic}=-\mathcal{E}^{potential}
\end{equation}
and the full energy of plasma particles into a star
\begin{equation}
\mathcal{E}(plasma)=\mathcal{E}^{kinetic}+\mathcal{E}^{potential}=-\mathcal{E}^{kinetic}.\label{ek}
\end{equation}
Let us calculate the separate items composing the full energy of a
star.
\subsubsection{The kinetic energy of plasma}

The kinetic energy of plasma into a core:
\begin{equation}
\mathcal{E}_\star^{kinitic}=\frac{3}{2}k\mathbb{T}_\star
\mathbb{N}_\star.
\end{equation}
The kinetic energy of atmosphere:
\begin{equation}
\mathcal{E}_a^{kinetic}=4\pi\int_{\mathbb{R}_\star}^{\mathbb{R}_0}
\frac{3}{2}k\mathbb{T}_\star
n_\star\left(\frac{\mathbb{R}_\star}{r}\right)^{10}
r^2dr\approx\frac{3}{7}\left(\frac{3}{2}k\mathbb{T}_\star
\mathbb{N}_\star\right)
\end{equation}
The total kinetic energy of plasma particles
\begin{equation}
\mathcal{E}^{kinetic}=\mathcal{E}_\star^{kinetic}+\mathcal{E}_a^{kinetic}=\frac{15}{7}k\mathbb{T}_\star
\mathbb{N}_\star\label{kes}
\end{equation}

\subsubsection{The potential energy of star plasma}
Inside a star core, the gravity force is balanced by the force of
electric nature. Correspondingly, the energy of electric
polarization can be considered as balanced by the gravitational
energy of plasma. As a result, the potential energy of a core can be
considered as equal to zero.

In a star atmosphere, this balance is absent.

The gravitational energy of an atmosphere
\begin{equation}
\mathcal{E}_a^{G}=- 4\pi G\mathbb{M}_\star \frac{A}{Z}m_p
n_\star\int_{\mathbb{R}_\star}^{\mathbb{R}_0}
\frac{1}{2}\left[2-\left(\frac{\mathbb{R}_\star}{r}\right)^{3}\right]\left(\frac{\mathbb{R}_\star}{r}\right)^{6}r
dr
\end{equation}
or
\begin{equation}
\mathcal{E}_a^{G}=\frac{3}{2}\left(\frac{1}{7}-\frac{1}{2}\right)\frac{G\mathbb{M}_\star^2}{\mathbb{R}_\star}
=-\frac{15}{28}\frac{G\mathbb{M}_\star^2}{\mathbb{R}_\star}
\end{equation}

The electric energy of atmosphere is
\begin{equation}
\mathcal{E}_a^{E}=-4\pi\int_{\mathbb{R}_\star}^{R_0}
\frac{1}{2}\varrho\varphi r^2 dr,
\end{equation}
where
\begin{equation}
\widetilde{\varrho}=\frac{1}{3r^2}\frac{d\mathfrak{P}r^2}{dr}
\end{equation}
and
\begin{equation}
\widetilde{\varphi}=\frac{4\pi}{3}\mathfrak{P}r.
\end{equation}
The electric energy:
\begin{equation}
\mathcal{E}_a^{E}=-
\frac{3}{28}\frac{G\mathbb{M}_\star^2}{\mathbb{R}_\star},
\end{equation}
and total potential energy of atmosphere:
\begin{equation}
\mathcal{E}_a^{potential}=\mathcal{E}_a^{G}+\mathcal{E}_a^{E}=
-\frac{9}{14}\frac{G\mathbb{M}_\star^2}{\mathbb{R}_\star}.\label{epa}
\end{equation}
The equilibrium in a star depends both on plasma energy and
energy of radiation.

\subsection{The temperature of a star core}
\subsubsection{The energy of the black radiation}
The energy of black radiation inside a star core is
\begin{equation}
\mathcal{E}_\star(br)= \frac{\pi^2}{15}k\mathbb{T}_\star
\left(\frac{k\mathbb{T}_\star}{\hbar c}\right)^{3}\mathbb{V}_\star.
\end{equation}
The energy of black radiation inside a star atmosphere  is
\begin{equation}
\mathcal{E}_a (br)=4\pi\int_{R_\star}^{R_0}
\frac{\pi^2}{15}k\mathbb{T}_\star
\left(\frac{k\mathbb{T}_\star}{\hbar
c}\right)^{3}\left(\frac{\mathbb{R}_\star}{r}\right)^{16}
r^2dr=\frac{3}{13}\frac{\pi^2}{15}k\mathbb{T}_\star
\left(\frac{k\mathbb{T}_\star}{\hbar c}\right)^{3}\mathbb{V}_\star.
\end{equation}
The total energy of black radiation inside a star is
\begin{equation}
\mathcal{E}(br)=\mathcal{E}_\star(br)+\mathcal{E}_a(br)=\frac{16}{13}\frac{\pi^2}{15}kT_\star
\left(\frac{kT_\star}{\hbar
c}\right)^{3}V_\star=1.23\frac{\pi^2}{15}kT_\star
\left(\frac{kT_\star}{\hbar c}\right)^{3}V_\star
\end{equation}

\subsubsection{The full energy of a star}
In accordance with  ({\ref{ek}}), the full
energy of a star
\begin{equation}
\mathcal{E}^{star}=-\mathcal{E}^{kinetic}+\mathcal{E}(br)
\end{equation}
i.e.
\begin{equation}
\mathcal{E}^{star}=-\frac{15}{7}k\mathbb{T}_\star
\mathbb{N}_\star+\frac{16}{13}\frac{\pi^2}{15}k\mathbb{T}_\star
\left(\frac{k\mathbb{T}_\star}{\hbar c}\right)^{3}\mathbb{V}_\star.
\end{equation}
The steady state of a star is determined by a minimum of its full
energy:
\begin{equation}
\left(\frac{d\mathcal{E}^{star}}{d\mathbb{T}_\star}\right)_{\mathbb{N}=const,\mathbb{V}=const}=0,
\end{equation}
it corresponds to the condition:
\begin{equation}
-\frac{15}{7} \mathbb{N}_\star+\frac{64\pi^2}{13\cdot 15}
\left(\frac{k\mathbb{T}_\star}{\hbar
c}\right)^{3}\mathbb{V}_\star=0.
\end{equation}
Together with Eq.({\ref{eta1}}) it defines the equilibrium
temperature of a star core:
\begin{equation}
\mathbb{T}_\star=\left(\frac{25\cdot
13}{28\pi^{4}}\right)^{1/3}\left(\frac{\hbar
c}{ka_B}\right)Z\approx Z\cdot 2.13\cdot 10^7 K\label{tcore}
\end{equation}

\subsection{Main star parameters}
\subsubsection{The star mass}
The virial theorem connect kinetic energy of a system with its
potential energy. In accordance with Eqs.({\ref{epa}}) and
({\ref{kes}})
\begin{equation}
\frac{9}{14}\frac{G\mathbb{M}_\star^2}{\mathbb{R}_\star}=\frac{30}{7}k\mathbb{T}_\star
\mathbb{N}_\star.
\end{equation}
Introducing the non-dimensional parameter
\begin{equation}
\eta=\frac{G\mathbb{M}_\star \frac{A}{Z}m_p}{\mathbb{R}_\star k\mathbb{T}_\star},\label{Bg}
\end{equation}
we obtain
\begin{equation}
\eta=\frac{20}{3}=6.67,\label{B}
\end{equation}
and at taking into account Eqs.({\ref{eta1}}) and ({\ref{tcore}}),
the core mass is
\begin{equation}
\mathbb{M}_\star=\left[\frac{20}{3}\left(\frac{25\cdot
13}{28}\right)^{1/3}\frac{3}{4\cdot
3.14}\right]^{3/2}\frac{\mathbb{M}_{Ch}}{{\left(\frac{A}{Z}\right)^2}}=6.84
\frac{\mathbb{M}_{Ch}}{{\left(\frac{A}{Z}\right)^2}}\label{Mcore}
\end{equation}
The obtained equation plays a very important role, because together
with Eq.(\ref{2mstar}), it gives a possibility to predict the total mass
of a star:
\begin{equation}
\mathbb{M}=2\mathbb{M}_\star=\frac{13.68
\mathbb{M}_{Ch}}{\left(\frac{A}{Z}\right)^2}\approx\frac{25.34
\mathbb{M}_\odot}{\left(\frac{A}{Z}\right)^2}.\label{M}
\end{equation}

The comparison of obtained prediction Eq.({\ref{M}}) with
measuring data gives a method to check our theory.
Although there is no way to determine  chemical
composition of cores of far stars, some predictions can be made in this way.
At first, there must be no stars which masses exceed the mass of the Sun by
more than one and a half orders, because it accords to limiting mass of
stars consisting from hydrogen with $A/Z=1$.
Secondly, the action of a specific mechanism (see. Sec.\ref{Ch13}) can make neutron-excess nuclei stable, but it don't give a base to suppose that stars with $A/Z>10$ (and with mass in
hundred times less than hydrogen stars) can exist. Thus, the theory
predicts that the whole mass spectrum must be placed in the interval
from 0.25 up to approximately 25 solar masses. These predications
are verified by measurements quite exactly. The mass distribution of
binary stars\footnote{The use of these data is caused by the fact
that only the measurement of parameters of binary star rotation
gives a possibility to determine their masses with satisfactory
accuracy.} is shown in Fig.{\ref{starM}} \cite{Heintz}.

\begin{figure}
\hspace{-3cm}
\includegraphics[scale=0.9]{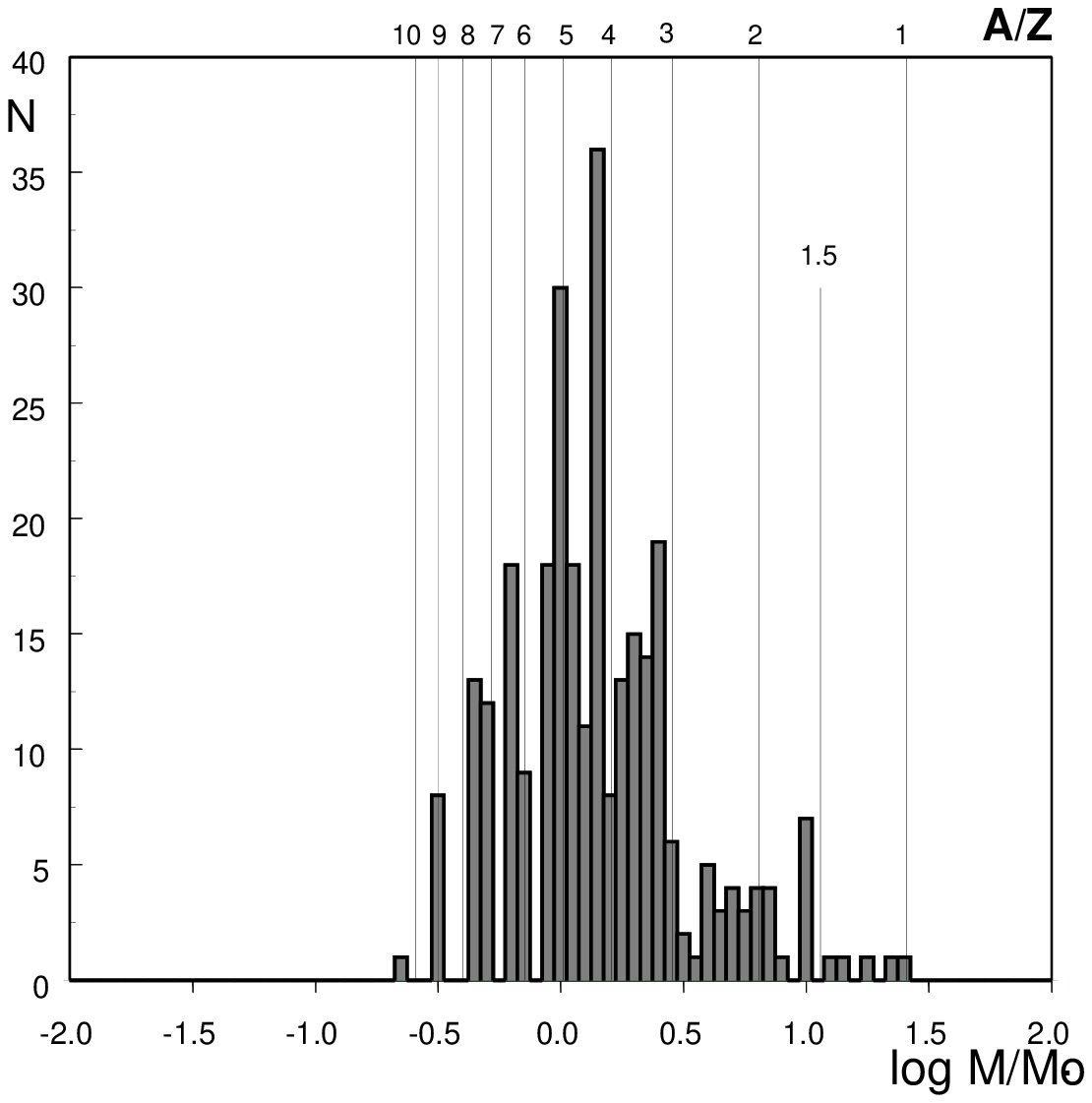}
\caption{The mass distribution of binary stars \cite{Heintz}. On
abscissa, the logarithm of the star mass over the Sun mass is shown.
Solid lines mark masses, which agree with selected values of A/Z from
Eq.(\ref{M}).}\label{starM}
\end{figure}

\begin{figure}
\hspace{-3cm}
\includegraphics[scale=0.9]{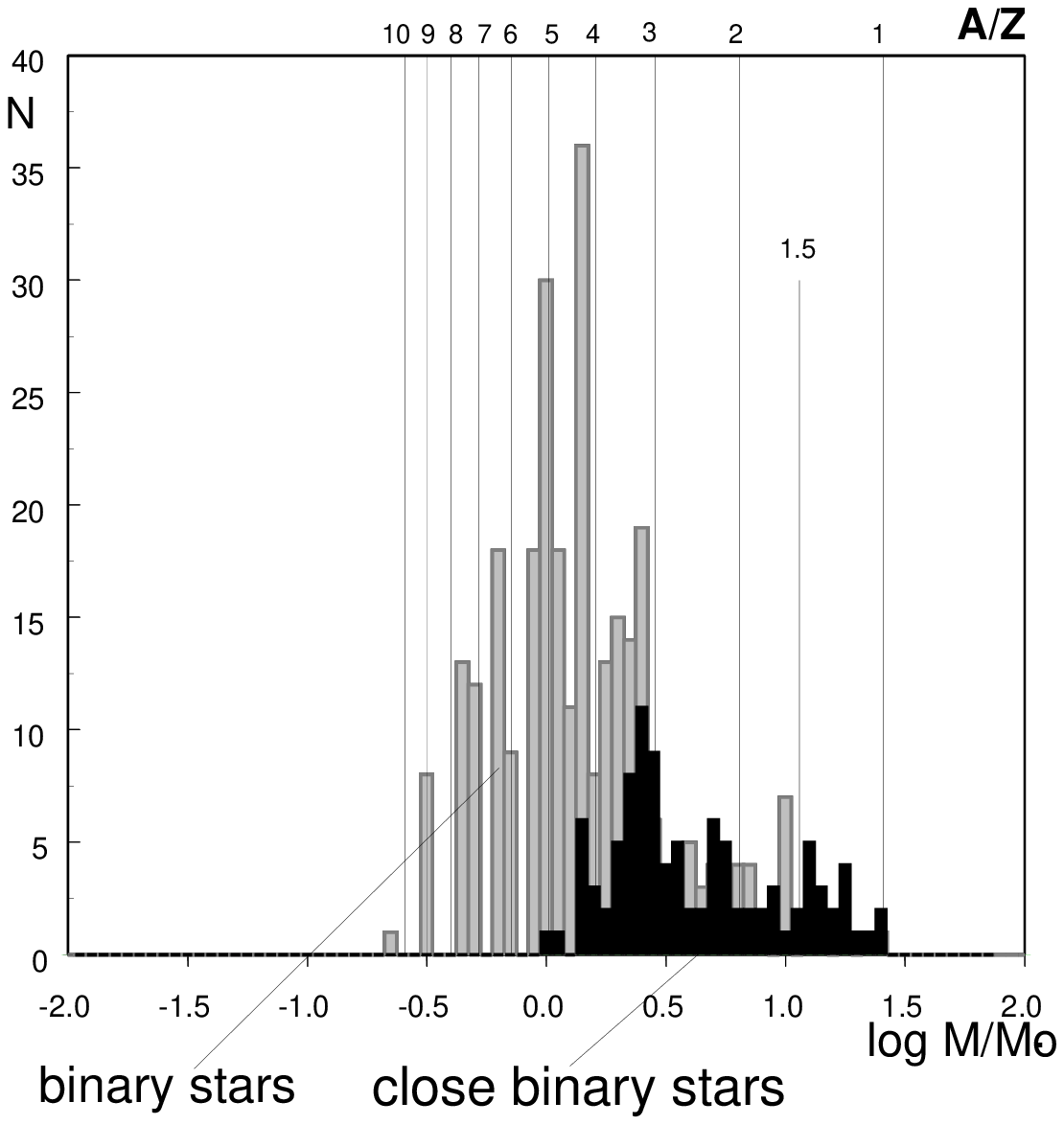}
\caption{The mass distribution of close binary stars \cite{Kh}. On
abscissa, the logarithm of the star mass over the Sun mass is shown.
Solid lines mark masses, which agree with selected values of A/Z from
Eq.(\ref{M}). The binary star spectrum is shown for
comparison.}\label{starM2}
\end{figure}
The mass spectrum of close binary stars\footnote{The data of these
measurements were obtained in different observatories of the world.
The last time the summary table with these data was gathered by
Khaliulilin Kh.F. (Sternberg Astronomical Institute) \cite{Kh} in
his dissertation (in Russian) which has unfortunately has a restricted
access. With his consent and for readers convenience, we place
that table in Appendix on http://astro07.narod.ru.}  is shown in Fig.{\ref {starM2}}.

 The very important element of the direct and clear confirmation of the theory is obviously  visible on these  figures  - both spectra is dropped near the value $A/Z=1$.

Beside it, one can easy see it, that  the mass spectrum of binary stars (Fig.({\ref{starM}})) consists of series of well-isolated lines which  are representing the stars with integer values of ratios
$A/Z=3,4,5 ...$, corresponding hydrogen-3,4,5 ... or
helium-6,8,10 ... (also line with the half-integer ratio $A/Z=3/2$, corresponding, probably, to helium-3,~Be-6,~C-9...).
The existence of stable stars with ratios $A/Z\geq 3$ raises questions.
It is generally assumed that stars are composed of hydrogen-1, deuterium, helium-4 and other heavier elements with $A/Z\approx 2$. Nuclei with $A/Z\geq 3$ are the neutron-excess and so short-lived, that they can not build a long-lived stars.
Neutron-excess nuclei can become stable under the action of mechanism of neutronization, which is acting  inside the dwarfs. It is accepted to think that this mechanism must not  work into the stars. The consideration of  the effecting of the electron gas of a dense plasma on the nucleus is described in chapter (\ref{Ch13}). The calculations of Ch.(\ref{Ch13}) show that the electron gas of dense plasma should also lead to the neutronization mechanism and to the stabilization of  the neutron-excess nuclei. This explains the existence of a stable of stars, where the plasma consists of nuclei with
$A/Z\geq 3$.

    Beside it, at considering of Fig.({\ref{starM}}), the question is arising:
why there are so few stars, which are composed by very stable nuclei of helium-4? At the same time, there are many stars with $A/Z=4$, i.e. consisting apparently of a hydrogen-4, as well as stars with $A/Z=3/2$, which hypothetically could be composed by another isotope of helium - helium-3.
This equation is discussed in Ch.(\ref{Ch13}).

Beside it, it is important to note, that according to Eq.(\ref{M}) the
Sun must consist from a substance with $A/Z=5$. This conclusion is in
a good agreement with results of consideration of solar oscillations
(Chapter \ref{Ch9}).

\subsubsection{Radii of stars}
Using Eq.({\ref{eta1}}) and Eq.({\ref{Mcore}}), we can determine the star core
radius:
\begin{equation}
\mathbb{R}_\star=1.42\frac{a_B}{Z(A/Z)}\left(\frac{\hbar c}{G
m_p^2}\right)^{1/2}  \approx \frac{9.79\cdot 10^{10}}{{Z(A/Z)}}
cm.\label{Rcore}
\end{equation}
The temperature near the star surface is relatively small. It is
approximately  by 3 orders smaller than it is inside the core.
Because of it at calculation of surface parameters, we must take into consideration effects of this
order, i.e. it is necessary to take into account the gravity action on the electron gas. At that it
is convenient to consider the plasma cell as some neutral quasi-atom
(like the Thomas-Fermi atom). Its electron shell is formed by a cloud of
free electrons.

Each such quasi-atom is retained on the star surface by its
negative potential energy
\begin{equation}
\left(\mathcal{E}_{gravitational}+\mathcal{E}_{electric}\right)<0.
\end{equation}
The electron cloud of the cell is placed in the volume $\delta
V=\frac{4\pi}{3}r_s^3$, (where  $r_s\approx\left(\frac{Z}{n_e}\right)^{1/3}$) under pressure $P_e$.
The evaporation of plasma cell releases energy $\mathcal{E}_{PV}=P_e V_s$, and the balance equation takes the form:
\begin{equation}
\mathcal{E}_{gravitational}+\mathcal{E}_{electric}+\mathcal{E}_{PV}=0\label{srf}.
\end{equation}
In cold plasma, the electron cloud of the cell has energy $\mathcal{E}_{PV}\approx{e^2}{n_e}^{1/3}$.
in very hot plasma at $kT\gg\frac{Z^2 e^2}{r_s}$, this energy is equal to $\mathcal{E}_{PV}=\frac{3}{2}ZkT$.
On the star surface these energies are approximately equal:
\begin{equation}
\frac{k\mathbb{T}_0}{e^2 n_e^{1/3}}\approx \frac{1}{\alpha}\left(\frac{\mathbb{R}_0}{\mathbb{R}_\star}\right)^2\approx 1.
\end{equation}
One can show it easily, that in this case
\begin{equation}
\mathcal{E}_{PV}\approx 2Z\sqrt{\frac{3}{2}kT\cdot{e^2}{n_e}^{1/3}}.
\end{equation}
And if to take into account Eqs.({\ref{an-r}})-({\ref{tr}}), we obtain
\begin{equation}
\mathcal{E}_{PV}\approx 1.5 Z k\mathbb{T}_\star\left(\frac{\mathbb{R}_\star}{\mathbb{R}_0}\right)^3\sqrt{\alpha\pi}
\end{equation}
The energy of interaction of a nucleus with its electron cloud does not change at evaporation of the cell
and it can be neglected. Thus, for the surface
\begin{equation}
\mathcal{E}_{electric}=\frac{2\pi\mathfrak{P}^2}{3n_s}=\frac{2G\mathbb{M}_\star}{\mathbb{R}_0}
\left({A}m_p-{Z}m_e\right).
\end{equation}
The gravitational energy of the cell on the surface
\begin{equation}
\mathcal{E}_{gravitational}=-\frac{2G\mathbb{M}_\star}{\mathbb{R}_0}\left({A}m_p+{Z}m_e\right).
\end{equation}
Thus, the balance condition Eq.({\ref{srf}}) on the star surface obtains the form
\begin{equation}
-\frac{4G\mathbb{M_\star}Zm_e}{\mathbb{R}_0}+
1.5Z k\mathbb{T}_\star\left(\frac{\mathbb{R}_\star}{\mathbb{R}_0}\right)^3\sqrt{\alpha\pi}=0.
\end{equation}

\subsubsection{The ${\mathbb{R}_\star}/{\mathbb{R}_0}$ ratio and ${\mathbb{R}_0}$}
With account of Eq.({\ref{tr}) and Eqs.({\ref{B})-({\ref{Bg}), we can
write
\begin{equation}
\frac{\mathbb{R}_0}{\mathbb{R}_\star}=\left(\frac{\sqrt{\alpha\pi}}{2\eta}\frac{\frac{A}{Z}m_p}{m_e}\right)^{1/2}
\approx {4.56}{\sqrt{\frac{A}{Z}}}\label{RRs}
\end{equation}
As the star core radius is known Eq.({\ref{Rcore}}), we can obtain the star surface radius:
\begin{equation}
\mathbb{R}_0\approx\frac{4.46\cdot 10^{11}}{Z(A/Z)^{1/2}}
cm.\label{R0}
\end{equation}

\subsubsection{The temperature of a star surface}
At known Eq.({\ref{tr}}) and Eq.({\ref{tcore}}), we can calculate
the temperature of external surface of a star
\begin{equation}
\mathbb{T}_0=\mathbb{T}_\star
\left(\frac{\mathbb{R}_\star}{\mathbb{R}_0}\right)^4 \approx
4.92\cdot 10^5 \frac{Z}{(A/Z)^2}\label{T0}
\end{equation}

\subsubsection{The comparison with measuring data}
The mass spectrum (Fig.{\ref{starM}}) shows that the Sun
consists basically from plasma with $A/Z=5$.
The radius of the Sun and its surface temperature are functions of Z too.
This values calculated at A/Z=5 and differen Z are shown in Table ({\ref{RT00}})

Table ({\ref{RT00}})}
{\hspace{3cm}\begin{tabular}
{||c|c|c||}\hline\hline
&$\mathbb{R_\odot},cm$&$\mathbb{T_\odot}$,K\\
Z&(calculated&(calculated\\
&ïî ({\ref{R0}})) & ïî ({\ref{T0}}))\\
\hline 1&$2.0\cdot 10^{11}$& 1961\\
\hline 2&$1.0\cdot 10^{11}$&3923\\
\hline 3&$6.65\cdot 10^{10}$&5885\\
\hline 4&$5.0\cdot 10^{10}$&7845\\
\hline\hline
\end{tabular}\label{RT00}}

\bigskip
One can see that these calculated data have a satisfactory agreement the measured radius of the Sun
\begin{equation}
\mathbb{R}_\odot=6.96 \cdot 10^{10} cm
\end{equation}
and the measured surface temperature
\begin{equation}
\mathbb{T}_\odot=5850~K
\end{equation}
at  $Z=3$.

The calculation shows that the mass of core of the Sun
\begin{equation}
\mathbb{M}_\star(Z=3,A/Z=5)\approx 9.68\cdot 10^{32}~g\label{MtS}
\end{equation}
i.t. almost exactly equals to one half of full mass of the Sun
\begin{equation}
\frac{\mathbb{M}_\star(Z=3,A/Z=5)}{\mathbb{M}_\odot}\approx 0.486\label{MrS}
\end{equation}
in full agreement with Eq.({\ref{2mstar}}).

\vspace{1cm}

In addition to obtained determinations of the mass of a star
Eq.({\ref{M}}), its temperature  Eq.({\ref{T0}}) and its radius
Eq.({\ref{R0}}) give possibility
 to check the calculation, if we compare these results with measuring
 data.  Really, dependencies measured by astronomers  can be
 described by functions:
\begin{equation}
\mathbb{M}=\frac{Const_1}{(A/Z)^2},
\end{equation}
\begin{equation}
{\mathbb{R}}_0=\frac{Const_2}{Z(A/Z)^{1/2}},
\end{equation}
\begin{equation}
{\mathbb{T}}_0=\frac{Const_3Z}{(A/Z)^{2}}.
\end{equation}
If to combine they in the way, to exclude unknown parameter $Z$,
one can obtain relation:
\begin{equation}
{\mathbb{T}}_0 \mathbb{R}_0=Const\, \mathbb{M}^{5/4},\label{5/4}
\end{equation}
Its accuracy can by checked. For this checking, let us use the
measuring data of parameters of masses, temperatures and radii of
close binary stars \cite{Kh}.  The results of these measurements are
shown in Fig.({\ref{RT-M}}), where the dependence according to
Eq.({\ref{5/4}}). It is not difficult to see that these data are well
described by the calculated dependence. It speaks about
successfulness of our consideration.
\begin{figure}
\begin{center}
\includegraphics[scale=0.7]{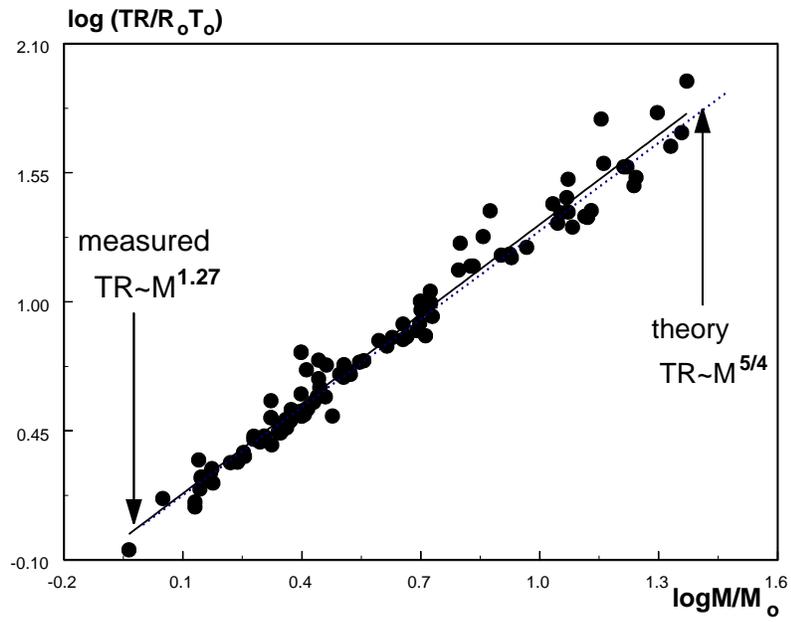}
\caption{The relation between main parameters of stars
(Eq.({\ref{5/4}})) and corresponding data of astronomical
measurements for close binary stars \cite{Kh} are shown.}
{\label{RT-M}}
\end{center}
\end{figure}

If main parameters of the star are  expressed through corresponding solar values
$\tau\equiv\frac{\mathbb{T}_0}{\mathbb{T}_\odot}$,$\rho\equiv\frac{\mathbb{R}_0}{\mathbb{R}_\odot}$
 and $\mu\equiv\frac{\mathbb{M}}{\mathbb{M}_\odot}$, that Eq.({\ref{5/4}}) can be rewritten as
\begin{equation}
\frac{\tau \rho}{\mu^{5/4}}=1\label{5/42}.
\end{equation}

  Numerical values of relations $\frac{\tau \rho}{\mu^{5/4}}$ for close binary stars \cite{Kh} are shown in the last column of the Table({\ref{mrt}})(at the end Chapter (\ref{Ch6})).

\clearpage
\section[The thermodynamic relations]{The thermodynamic relations of intra-stellar plasma}
\label{Ch6}

\subsection[The thermodynamic relations]{The thermodynamic relation of star atmosphere parameters}

Hot stars steadily generate energy and radiate it from their
surfaces. This  is non-equilibrium radiation in relation to a star.
But it may be a stationary radiation for a  star in steady state. Under this
condition, the star substance can be considered as an equilibrium.
This condition can be considered as quasi-adiabatic, because
the interchange of energy between both subsystems - radiation and
substance - is stationary and it does not result in a change of entropy
of substance. Therefore at consideration of state of a
star atmosphere, one can base it on equilibrium conditions of hot
plasma and the ideal gas law for adiabatic condition can be used for
it in the first approximation.

It is known, that thermodynamics can help to establish correlation
between  steady-state parameters of a system. Usually, the
thermodynamics considers systems at an equilibrium state with
constant temperature,  constant particle density and constant
pressure over all system. The characteristic feature of the considered system is
the existence of equilibrium at the absence of a constant temperature
and particle density over atmosphere of a star. To solve this
problem, one can introduce averaged pressure
\begin{equation}
\widehat{P}\approx\frac{G\mathbb{M}^2}{\mathbb{R}_0^4}\label{Pav},
\end{equation}
averaged temperature
\begin{equation}
\widehat{T}=\frac{\int_\mathbb{V} T dV}{V}\sim \mathbb{T}_0
\biggl(\frac{\mathbb{R}_0}{\mathbb{R}_\star}\biggr)\label{TR3}
\end{equation}
and averaged particle density
\begin{equation}
\widehat{n}\approx \frac{\mathbb{N}_A}{\mathbb{R}_0^3}
\end{equation}
After it by means of thermodynamical methods, one can find relation
between parameters of a star.

\subsubsection{The ${c_P}/{c_V}$ ratio}
At a movement of particles according to the theorem of the equidistribution,
the energy $kT/2$ falls at each degree of freedom. It gives
the heat capacity $c_v=3/2k$.

According to the virial theorem \cite{LL,VL}, the full energy of a
star should be equal to its kinetic energy (with opposite
sign)(Eq.({\ref{ek})), so as full energy related to one
particle
\begin{equation}
\mathcal{E}=-\frac{3}{2}kT
\end{equation}
In this case the heat capacity at constant volume (per particle over
Boltzman's constant $k$) by definition is
\begin{equation}
c_V=\biggl(\frac{dE}{dT}\biggr)_V=-\frac{3}{2}\label{cv}
\end{equation}
The negative heat capacity of stellar substance is not surprising.
It is a known fact and it is discussed in Landau-Lifshitz course \cite{LL}.
The own heat capacity of each particle of star substance is
positive. One obtains the negative heat capacity  at taking into
account the gravitational interaction between particles.

By definition the heat capacity of an ideal gas particle at
permanent pressure \cite{LL} is
\begin{equation}
c_P=\biggl(\frac{dW}{dT}\biggr)_P,\label{cp}
\end{equation}
where  $W$ is enthalpy of a gas.

As  for the ideal gas \cite{LL}
\begin{equation}
W-\mathcal{E}=NkT,
\end{equation}
and the difference between $c_P$ and $c_V$
\begin{equation}
{c_P-c_V}=1.
\end{equation}

Thus in the case considered, we have
\begin{equation}
c_P=-\frac{1}{2}.
\end{equation}
Supposing that conditions  are close to adiabatic ones, we can use
the equation of the Poisson's adiabat.

\subsubsection{The Poisson's adiabat}
The thermodynamical potential of a system consisting of $N$
molecules of ideal gas at temperature  $T$ and pressure $P$ can
be written as \cite{LL}:
\begin{equation}
\Phi=const\cdot N +NTlnP-Nc_P T lnT.
\end{equation}
The entropy of this system
\begin{equation}
S=const\cdot N -NlnP+Nc_P lnT.
\end{equation}
As at adiabatic process, the entropy remains constant
\begin{equation}
-NTlnP+Nc_P T lnT=const,
\end{equation}
we can write the equation for relation of averaged pressure in a
system with its volume (The Poisson's adiabat) \cite{LL}:
\begin{equation}
\widehat{P}V^{\widetilde{\gamma}}=const\label{Pua},
\end{equation}
where $\widetilde{\gamma}=\frac{c_P}{c_V}$ is the exponent of
 adiabatic constant. In considered case taking into account of
Eqs.({\ref{cp}}) and ({\ref{cv}}), we obtain
\begin{equation}
\widetilde{\gamma}=\frac{c_P}{c_V}=\frac{1}{3}.
\end{equation}
As $V^{1/3}\sim \mathbb{R}_0$, we have for equilibrium condition
\begin{equation}
\widehat{P}\mathbb{R}_0=const\label{Pub}.
\end{equation}

\subsection{The mass-radius ratio}
Using Eq.({\ref{Pav}}) from  Eq.({\ref{Pub}}), we use the
equation for dependence of masses of stars on their radii:
\begin{equation}
\frac{\mathbb{M}^2}{\mathbb{R}_0^3}=const\label{rm23}
\end{equation}
This equation shows the existence of internal constraint of
chemical parameters of equilibrium state of a star. Indeed, the
substitution of obtained determinations Eq.({\ref{R0}}) and
({\ref{T0})) into Eq.({\ref{rm23}}) gives:
\begin{equation}
Z\sim (A/Z)^{5/6}\label{Z-AZ}
\end{equation}
Simultaneously the observational data of  masses, of radii and their
temperatures was obtained by astronomers for close binary stars
\cite{Kh}. The dependence of radii of these stars over these masses
is shown in Fig.{\ref{RM}} on double logarithmic scale. The solid
line shows the result of fitting of measurement data
$\mathbb{R}_0\sim \mathbb{M}^{0.68}$. It is close to theoretical
dependence $\mathbb{R}_0\sim \mathbb{M}^{2/3}$~(Eq.{\ref{rm23}})
which is shown by dotted line.

\begin{figure}
\begin{center}
\includegraphics[scale=0.7]{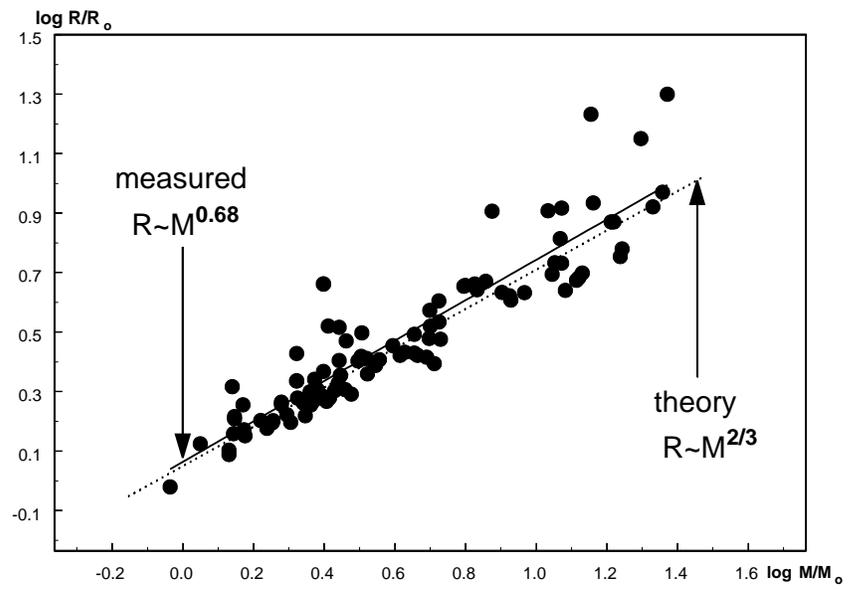}
\caption{The dependence of  radii of stars over the star mass
\cite{Kh}. Here the radius of stars is normalized to the sunny
radius, the stars masses are normalized to the mass of the Sum. The data
are shown on double logarithmic scale. The solid line shows the
result of fitting of measurement data $\mathbb{R}_0\sim
\mathbb{M}^{0.68}$. The theoretical dependence
$\mathbb{R}_0\sim\mathbb{M}^{2/3}$~({\ref{rm23}}) is shown by the
dotted line.}\label{RM}
\end{center}
\end{figure}

If parameters of the star are  expressed through corresponding solar values
$\rho\equiv\frac{\mathbb{R}_0}{\mathbb{R}_\odot}$
 and $\mu\equiv\frac{\mathbb{M}}{\mathbb{M}_\odot}$, that Eq.({\ref{rm23}}) can be rewritten as
\begin{equation}
\frac{\rho}{\mu^{2/3}}=1\label{2/3}.
\end{equation}
Numerical values of relations $\frac{\rho}{\mu^{2/3}}$ for close binary stars \cite{Kh} are shown in the Table({\ref{mrt}}).

\begin{table}

{The Table(\ref{mrt}).

The relations of main stellar parameters}

{

\tiny
\begin{tabular}
{||r|l|r|r|r|r|r|r|r||}\hline \hline
   & & & & & & & &  \\
N & {Star} & &$\mu\equiv\frac{\mathbb{M}}{\mathbb{M}_{\odot}}$
&$\rho\equiv\frac{\mathbb{R}_0}{\mathbb{R}_{\odot}}$
&$\tau\equiv\frac{\mathbb{T}_0}{\mathbb{T}_{\odot}}$
&$\frac{\rho}{\mu^{2/3}}$ & $\frac{\tau}{\mu^{7/12}}$
&$\frac{\rho\tau}{\mu^{5/4}}$
\\[0.4cm]\hline
& & & & & & & &  \\
&  & 1 & 1.48 & 1.803 & 1.043 & 1.38 & 0.83 & 1.15 \\[0.2cm]  \cline{3-9}
1&   BW Aqr  &&&&&&&\\
&  & 2 & 1.38 & 2.075 & 1.026 & 1.67 & 0.85  & 1.42 \\
& & & & & & & &  \\
\hline
& & & & & & & &  \\
&  & 1 & 2.4  & 2.028 & 1.692 & 1.13  & 1.01 & 1.15 \\[0.2cm]  \cline{3-9}
2 & V 889 Aql &       &  &&     &  &  &  \\
& &  2 & 2.2  & 1.826 & 1.607 &  1.08 & 1.01 & 1.09 \\
& & & & & & & &  \\
\hline
 & & & & & & & &  \\
& &1 & 6.24   & 4.512 & 3.043 & 1.33 & 1.04 & 1.39 \\[0.2cm]  \cline{3-9}
3 & V 539 Ara &       &   &&    &  &  &  \\
& & 2& 5.31   & 4.512 & 3.043 &  1.12 & 1.09  & 1.23 \\
 & & & & & & & &  \\
\hline
 & & & & & & & &  \\
& &1 & 3.31   & 2.58  & 1.966 & 1.16  & 0.98 & 1.13 \\[0.2cm]  \cline{3-9}
  4 & AS Cam  &       &     &&  &  &  &  \\
& & 2& 2.51   &1.912  & 1.709 & 1.03 & 1.0 & 1.03 \\
 & & & & & & & &  \\
\hline
 & & & & & & & &  \\
& & 1 &  22.8 & 9.35  & 5.658 & 1.16 & 0.91 & 1.06 \\[0.2cm]  \cline{3-9}
5 & EM Car  &       &       & && &  &  \\
& & 2 & 21.4  & 8.348 & 5.538 & 1.08 & 0.93 & 1.00\\
 & & & & & & & &  \\
\hline
 & & & & & & & &  \\
& &1 & 13.5  & 4.998 & 5.538 & 0.88  & 1.08 & 0.95 \\[0.2cm]  \cline{3-9}
  6 & GL Car  &&  &&&&&\\
& & 2& 13 & 4.726 & 4.923 & 0.85 & 1.1 & 0.94  \\
 & & & & & & & &  \\
\hline
 & & & & & & & &  \\
& &1 & 9.27 & 4.292 & 4 & 0.97 & 1.09 & 1.06  \\[0.2cm]  \cline{3-9}
  7 & QX Car    &&&&&&&\\
& & 2& 8.48 & 4.054 & 3.829 & 0.975 & 1.1 & 1.07 \\
 & & & & & & & &  \\
\hline
 & & & & & & & &  \\
& &1 & 6.7 & 4.591 & 3.111 & 1.29 & 1.02& 1.32\\[0.2cm]  \cline{3-9}
  8 & AR Cas &&&&&&&\\
& & 2&  1.9 & 1.808 & 1.487 & 1.18 & 1.02& 1.21 \\
 & & & & & & & &  \\
\hline
 & & & & & & & &  \\
& &1 &  1.4 & 1.616 & 1.102 & 1.29 & 0.91 & 1.17 \\[0.2cm]  \cline{3-9}
  9 & IT Cas    &&&&&&&\\
& & 2&  1.4 & 1.644 & 1.094 & 1.31 & 0.90& 1.18 \\
 & & & & & & & &  \\
\hline
 & & & & & & & &  \\
& &1 &  7.2 & 4.69 & 4.068 & 1.25 & 1.29 & 1.62 \\[0.2cm]  \cline{3-9}
  10 & OX Cas   &&&&&&&\\
& & 2&  6.3 & 4.54 & 3.93 & 1.33 & 1.34 & 1.79  \\
 & & & & & & & &  \\
\hline
 & & & & & & & &  \\
& &1 &  2.79 & 2.264 & 1.914 & 1.14 & 1.05 & 1.20\\[0.2cm]  \cline{3-9}
  11 & PV Cas   &&&&&&&\\
& & 2 &  2.79 & 2.264 & 2.769 & 1.14 & 1.05 & 1.20 \\
 & & & & & & & &  \\
\hline
 & & & & & & & &  \\
& &1 & 5.3 & 4.028 & 2.769  & 1.32 & 1.05 & 1.39 \\[0.2cm]  \cline{3-9}
  12 & KT Cen   &&&&&&&\\
& & 2 & 5 & 3.745 & 2.701 & 1.28 & 1.06 & 1.35 \\
 & & & & & & & &  \\
\hline
 & & & & & & & &  \\
& &1 &  11.8 & 8.26 & 4.05 & 1.59 & 0.96 & 1.53 \\[0.2cm]  \cline{3-9}
  13 & V 346 Cen &&&&&&&\\
& & 2 & 8.4 & 4.19 & 3.83 & 1.01 & 1.11 & 1.12 \\
 & & & & & & & &  \\
\hline
 & & & & & & & &  \\
& & 1 &  11.8 & 8.263 & 4.051 & 1.04 & 1.06 & 1.11 \\[0.2cm]  \cline{3-9}
  14 & CW Cep   &&&&&&&\\
& & 2 & 11.1 & 4.954 & 4.393 & 1.0 & 1.08 & 1.07 \\
 & & & & & & & &  \\
\hline
 & & & & & & & &  \\
& & 1 & 2.02 & 1.574 & 1.709  & 0.98 & 1.13 & 1.12 \\[0.2cm]  \cline{3-9}
  15 & EK Cep   &&&&&&&\\
& & 2 &  1.12 & 1.332 & 1.094   & 1.23 & 1.02 & 1.26 \\
 & & & & & & & &  \\
\hline
 & & & & & & & &  \\
& & 1 & 2.58 & 3.314 & 1.555  & 1.76 & 0.89 & 1.57 \\[0.2cm]  \cline{3-9}
  16 & $\alpha$ Cr B &&&&&&&\\
& & 2 &  0.92 & 0.955 & 0.923 & 1.01 & 0.97 & 0.98 \\
 & & & & & & & &  \\
\hline
 & & & & & & & &  \\
& & 1 & 17.5 & 6.022 & 5.66 & 0.89 & 1.06 &  0.95 \\[0.2cm]  \cline{3-9}
  17 & Y Cyg    &&&&&&&\\
& & 2 & 17.3 & 5.68 & 5.54 & 0.85 & 1.05 & 0.89 \\
 & & & & & & & &  \\
\hline\hline

\end{tabular}}\label{mrt}
\end{table}

\clearpage

\begin{table}

{The Table(\ref{mrt})(continuation).

The relations of main stellar parameters}

{
\tiny
{
\hspace{-2.0cm}
\begin{tabular}
 {||r|l|r|r|r|r|r|r|r||}\hline
\hline
   & & & & & & & &  \\
 N & \tiny{Star} & n
 &$\mu\equiv\frac{\mathbb{M}}{\mathbb{M}_{\odot}}$ &
$\rho\equiv\frac{\mathbb{R}_0}{\mathbb{R}_{\odot}}$ &
$\tau\equiv\frac{\mathbb{T}_0}{\mathbb{T}_{\odot}}$ &
 $\frac{\rho}{\mu^{2/3}}$& $\frac{\tau}{\mu^{7/12}}$ & $\frac{\rho\tau}{\mu^{5/4}}$
\\[0.4cm]\hline\hline
& & & & & & & &  \\
& &1 &  14.3 & 17.08 & 3.54  & 2.89 & 0.75 & 2.17 \\[0.2cm]  \cline{3-9}
  18 & Y 380 Cyg &&&&&&&\\
& & 2&  8 & 4.3 & 3.69 & 1.07& 1.1& 1.18 \\
 & & & & & & & &  \\
 \hline
 & & & & & & & &  \\
& & 1 &  14.5 & 8.607 & 4.55 & 1.45 & 0.95 &  1.38 \\[0.2cm]  \cline{3-9}
  19 & V 453 Cyg &&&&&&&\\
& & 2 &  11.3 & 5.41 & 4.44 & 1.07 & 1.08 & 1.16 \\
 & & & & & & & &  \\
\hline
 & & & & & & & &  \\
& & 1 & 1.79 & 1.567 & 1.46 & 1.06 & 1.04 & 1.11 \\[0.2cm]  \cline{3-9}
  20 & V 477 Cyg &&&&&&&\\
& & 2 & 1.35 & 1.27 & 1.11& 1.04 & 0.93 & 0.97 \\
 & & & & & & & &  \\
\hline
 & & & & & & & &  \\
& &1 & 16.3 & 7.42 & 5.09& 1.15 & 1.0 & 1.15 \\[0.2cm]  \cline{3-9}
  21 & V 478 Cyg &&&&&&&\\
& & 2 & 16.6 & 7.42 & 5.09 & 1.14 & 0.99 & 1.13 \\
 & & & & & & & &  \\
\hline
 & & & & & & & &  \\
& & 1 & 2.69 & 2.013 & 1.86 & 1.04 & 1.05 & 1.09 \\[0.2cm]  \cline{3-9}
  22 & V 541 Cyg &&&&&&&\\
& & 2 & 2.6 & 1.9 & 1.85 & 1.0 & 1.6 & 1.06 \\
 & & & & & & & &  \\
\hline
 & & & & & & & &  \\
& &1 & 1.39 & 1.44 & 1.11 & 1.16 & 0.92 & 0.92 \\[0.2cm]  \cline{3-9}
  23 & V 1143 Cyg &&&&&&&\\
& & 2 &  1.35 & 1.23 & 1.09 & 1.0 & 0.91 & 0.92 \\
 & & & & & & & &  \\
\hline
 & & & & & & & &  \\
& &1 & 23.5 & 19.96 & 4.39 & 2.43 & 0.67 & 1.69 \\[0.2cm]  \cline{3-9}
  24 & V 1765 Cyg &&&&&&&\\
& & 2& 11.7 & 6.52 & 4.29 & 1.26 & 1.02 & 1.29 \\
 & & & & & & & &  \\
\hline
 & & & & & & & &  \\
& &1 & 5.15 & 2.48 & 2.91 & 0.83 & 1.12 & 0.93 \\[0.2cm]  \cline{3-9}
  25 & DI Her    &&&&&&&\\
& & 2& 4.52 & 2.69 & 2.58 & 0.98 & 1.07 & 1.05 \\
 & & & & & & & &  \\
\hline
 & & & & & & & &  \\
 & & & & & & & &  \\
&  & 1 & 4.25 & 2.71 & 2.61     &    1.03 & 1.12 & 1.16\\[0.2cm]\cline{3-9}
26 &   HS Her  &       &     &&  &  & &  \\
&  & 2 & 1.49 & 1.48 & 1.32 & 1.14 & 1.04  & 1.19 \\
& & & & & & & &  \\
\hline
& & & & & & & &  \\
&  & 1 & 3.13  & 2.53 & 1.95 & 1.18  & 1.00 & 1.12 \\[0.2cm]  \cline{3-9}
27 & CO Lac &       &      && &  &  &  \\
& &  2 & 2.75  & 2.13 & 1.86 &  1.08 & 1.01 & 1.09 \\
& & & & & & & &  \\
\hline
& & & & & & & &  \\
& &1 & 6.24   & 4.12 & 2.64 & 1.03 & 1.08 & 1.11 \\[0.2cm]  \cline{3-9}
28 & GG Lup  &       &   &&    &  &  &  \\
& & 2& 2.51   & 1.92 & 1.79 &  1.04 & 1.05  & 1.09 \\
& & & & & & & &  \\
\hline
& & & & & & & &  \\
& &1 & 3.6   & 2.55 & 2.20 & 1.09  & 1.04 & 1.14 \\[0.2cm]  \cline{3-9}
29  & RU Mon  &       &&&       &  &  &  \\
& & 2& 3.33   & 2.29  & 2.15 & 1.03 & 1.07 & 1.10 \\
& & & & & & & &  \\
\hline
& & & & & & & &  \\
& & 1 & 2.5 & 4.59  & 1.33 & 2.49 & 0.78 & 1.95 \\[0.2cm]  \cline{3-9}
30 & GN Nor  &       &&&       &  &  &  \\
& & 2 & 2.5  & 4.59 & 1.33 & 2.49 & 0.78 & 1.95\\
& & & & & & & &  \\
\hline
& & & & & & & &  \\
& &1 & 5.02 & 3.31 & 2.80 & 1.13  & 1.09 & 1.23 \\[0.2cm]  \cline{3-9}
31 & U Oph    &&&&&&&\\
& & 2& 4.52 & 3.11 & 2.60 & 1.14 & 1.08 & 1.23  \\
& & & & & & & &  \\
\hline
& & & & & & & &  \\
& &1 & 2.77 & 2.54 & 1.86 & 1.29 & 1.03 & 1.32  \\[0.2cm]  \cline{3-9}
32 & V 451 Oph    &&&&&&&\\
& & 2& 2.35 & 1.86 & 1.67 & 1.05 & 1.02 & 1.07 \\
& & & & & & & &  \\
\hline
& & & & & & & &  \\
& &1 & 19.8 & 14.16 & 4.55 & 1.93 & 0.80 & 1.54 \\[0.2cm]  \cline{3-9}
33 & $\beta$ Ori &&&&&&&\\
& & 2&  7.5 & 8.07 & 3.04 & 2.11 & 0.94 & 1.98 \\
& & & & & & & &  \\
\hline
& & & & & & & &  \\
& &1 &  2.5 & 1.89 & 1.81 & 1.03 & 1.06 & 1.09 \\[0.2cm]  \cline{3-9}
34 & FT Ori    &&&&&&&\\
& & 2&  2.3 & 1.80 & 1.62 & 1.03 & 1.0 & 1.03 \\
& & & & & & & &  \\
\hline\hline

\end{tabular}}}\label{mrt}
\end{table}

\clearpage

\begin{table}

{The Table(\ref{mrt})(continuation).

The relations of main stellar parameters}

{
\tiny
{
\hspace{-2.0cm}
\begin{tabular}
 {||r|l|r|r|r|r|r|r|r||}\hline
\hline
   & & & & & & & &  \\
 N & \tiny{Star} & n
 &$\mu\equiv\frac{\mathbb{M}}{\mathbb{M}_{\odot}}$ &
$\rho\equiv\frac{\mathbb{R}_0}{\mathbb{R}_{\odot}}$ &
$\tau\equiv\frac{\mathbb{T}_0}{\mathbb{T}_{\odot}}$ &
 $\frac{\rho}{\mu^{2/3}}$& $\frac{\tau}{\mu^{7/12}}$ & $\frac{\rho\tau}{\mu^{5/4}}$
\\[0.4cm]\hline\hline
& & & & & & & &  \\
& &1 &  5.36 & 3.0 & 2.91 & 0.98 & 1.09 & 1.06 \\[0.2cm]  \cline{3-9}
35 & AG Per   &&&&&&&\\
& & 2&  4.9 & 2.61 & 2.91 & 0.90 & 1.15 & 1.04  \\
& & & & & & & &  \\
\hline
& & & & & & & &  \\
& &1 &  3.51 & 2.44 & 2.27 & 1.06 & 1.09 & 1.16\\[0.2cm]  \cline{3-9}
36 & IQ Per   &&&&&&&\\
& & 2 &  1.73 & 1.50 & 2.27 & 1.04 & 1.00 & 1.05 \\
& & & & & & & &  \\
\hline
& & & & & & & &  \\
& &1 & 3.93 & 2.85 & 2.41  & 1.14 & 1.08 & 1.24 \\[0.2cm]  \cline{3-9}
37 & $\varsigma$ Phe   &&&&&&&\\
& & 2 & 2.55 & 1.85 & 1.79 & 0.99 & 1.04 & 1.03 \\
& & & & & & & &  \\
\hline
& & & & & & & &  \\
& &1 &  2.5 & 2.33 & 1.74 & 1.27 & 1.02 & 1.29 \\[0.2cm]  \cline{3-9}
38 & KX Pup &&&&&&&\\
& & 2 & 1.8 & 1.59 & 1.38 & 1.08 & 0.98 & 1.06 \\
& & & & & & & &  \\
\hline
& & & & & & & &  \\
& & 1 &  2.88 & 2.03 & 1.95 & 1.00 & 1.05 & 1.05 \\[0.2cm]  \cline{3-9}
39 & NO Pup   &&&&&&&\\
& & 2 & 1.5 & 1.42 & 1.20 & 1.08 & 0.94 & 1.02 \\
& & & & & & & &  \\
\hline
& & & & & & & &  \\
& & 1 & 2.1 & 2.17 & 1.49  & 1.32 & 0.96 & 1.27 \\[0.2cm]  \cline{3-9}
40 & VV Pyx   &&&&&&&\\
& & 2 &  2.1 & 2.17 & 1.49   & 1.32 & 0.96 & 1.27 \\
& & & & & & & &  \\
\hline
& & & & & & & &  \\
& & 1 & 2.36 & 2.20 & 1.59  & 1.24 & 0.96 & 1.19 \\[0.2cm]  \cline{3-9}
41 & YY Sgr &&&&&&&\\
& & 2 &  2.29 & 1.99 & 1.59 & 1.15 & 0.98 & 1.12 \\
& & & & & & & &  \\
\hline
& & & & & & & &  \\
& & 1 & 2.1 & 2.67 & 1.42 & 1.63 & 0.92 &  1.50 \\[0.2cm]  \cline{3-9}
42 & V 523 Sgr   && &&&&&\\
& & 2 & 1.9 & 1.84 & 1.42 & 1.20 & 0.98 & 1.17 \\
& & & & & & & &  \\
\hline
& & & & & & & &  \\
& &1 & 2.11 & 1.9 & 1.30 & 1.15 & 0.84 & 0.97 \\[0.2cm]  \cline{3-9}
43 & V 526 Sgr &&&&&&&\\
& & 2&  1.66 & 1.60 & 1.30 & 1.14 & 0.97 & 1.10 \\
& & & & & & & &  \\
\hline
& & & & & & & &  \\
& & 1 &  2.19 & 1.83 & 1.52 & 1.09 & 0.96 &  1.05  \\[0.2cm]  \cline{3-9}
44 & V 1647 Sgr &&&&&&&\\
& & 2 &  1.97 & 1.67 & 4.44 & 1.06 & 1.02 & 1.09 \\
& & & & & & & &  \\
\hline
& & & & & & & &  \\
& & 1 & 3.0 & 1.96 & 1.67 & 0.94 & 0.88 & 0.83 \\[0.2cm]  \cline{3-9}
45 & V 2283 Sgr &&&&&&&\\
& & 2 & 2.22 & 1.66 & 1.67 & 0.97 & 1.05 & 1.02 \\
& & & & & & & &  \\
\hline
& & & & & & & &  \\
& &1 & 4.98 & 3.02 & 2.70 & 1.03 & 1.06 & 1.09 \\[0.2cm]  \cline{3-9}
46 & V 760 Sco &&&&&&&\\
& & 2 & 4.62 & 2.64 & 2.70 & 0.95 & 1.11 & 1.05 \\
& & & & & & & &  \\
\hline
& & & & & & & &  \\
& & 1 & 3.2 & 2.62 & 1.83 & 1.21 & 0.93 & 1.12 \\[0.2cm]  \cline{3-9}
47 & AO Vel &&&&&&&\\
& & 2 & 2.9 & 2.95 & 1.83 & 1.45 & 0.98 & 1.43 \\
& & & & & & & &  \\
\hline
& & & & & & & &  \\
& &1 & 3.21 & 3.14 & 1.73 & 1.44 & 0.87 & 1.26 \\[0.2cm]  \cline{3-9}
48 & EO Vel &&&&&&&\\
& & 2 &  2.77 & 3.28 & 1.73 & 1.66 & 0.95 & 1.58 \\
& & & & & & & &  \\
\hline
& & & & & & & &  \\
& &1 & 10.8 & 6.10 & 3.25 & 1.66 & 0.81 & 1.34 \\[0.2cm]  \cline{3-9}
49 & $\alpha$ Vir &&&&&&&\\
& & 2& 6.8 & 4.39 & 3.25 & 1.22 & 1.06 & 1.30 \\
& & & & & & & &  \\
\hline
& & & & & & & &  \\
& &1 & 13.2 & 4.81 & 4.79 & 0.83 & 1.06 & 0.91 \\[0.2cm]  \cline{3-9}
50 & DR Vul    &&&&&&&\\
& & 2& 12.1 & 4.37 & 4.79 & 0.83 & 1.12 & 0.93 \\
& & & & & & & &  \\
\hline\hline

\end{tabular}}}

\end{table}

\clearpage

\subsection[The mass-temperature-luminosity relations]{The mass-temperature and mass-luminosity relations}

Taking into account Eqs.({\ref{tr}}), ({\ref{Ts1}}) and
({\ref{RN}}) one can obtain the relation between surface temperature
and the radius of a star
\begin{equation}
\mathbb{T}_0\sim \mathbb{R}_0^{7/8},
\end{equation}
or accounting for  ({\ref{rm23}})
\begin{equation}
\mathbb{T}_0\sim \mathbb{M}^{7/12}\label{tm}
\end{equation}
The dependence of the temperature on the star surface over the star
mass of close binary stars \cite{Kh} is shown in Fig.(\ref{TM}).
Here the temperatures of stars are normalized to the sunny surface
temperature (5875~C), the stars masses are normalized to the mass of the
Sum. The data are shown on double logarithmic scale. The solid line
shows the result of fitting of measurement data ($\mathbb{T}_0\sim
\mathbb{M}^{0.59}$). The theoretical dependence $\mathbb{T}_0\sim
\mathbb{M}^{7/12}$~(Eq.{\ref{tm}}) is shown by dotted line.} 

\begin{figure}
\begin{center}
\includegraphics[scale=0.7]{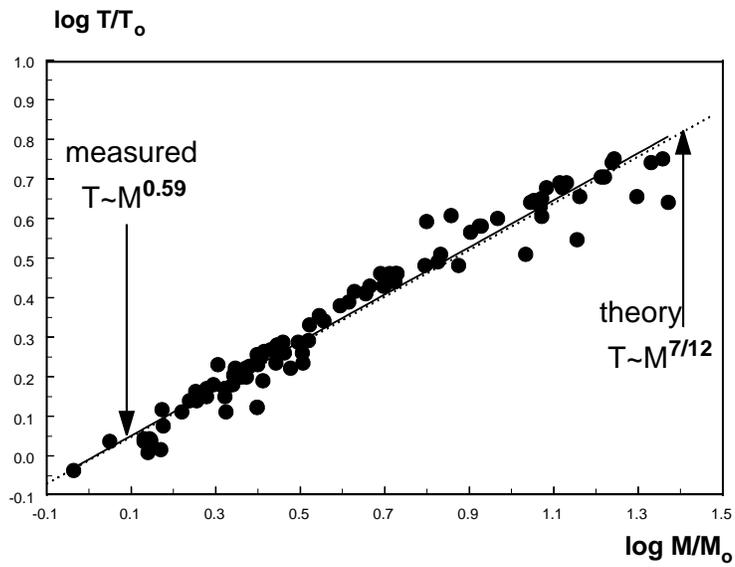}
\caption{The dependence of the temperature on the star surface over
the star mass of close binary stars \cite{Kh}. Here the temperatures
of stars are normalized to surface temperature of the Sun (5875~C),
the stars masses are normalized to the mass of Sum. The data are
shown on double logarithmic scale. The solid line shows the result of
fitting of measurement data ($\mathbb{T}_0\sim \mathbb{M}^{0.59}$).
The theoretical dependence $\mathbb{T}_0\sim
\mathbb{M}^{7/12}$~(Eq.{\ref{tm}}) is shown by dotted line.}
\label{TM}
\end{center}
\end{figure}

If parameters of the star are  expressed through corresponding solar values
$\tau\equiv\frac{\mathbb{T}_0}{\mathbb{T}_\odot}$
 and $\mu\equiv\frac{\mathbb{M}}{\mathbb{M}_\odot}$, that Eq.({\ref{tm}}) can be rewritten as
\begin{equation}
\frac{\tau}{\mu^{7/12}}=1\label{7/12}.
\end{equation}
Numerical values of relations $\frac{\tau}{\mu^{7/12}}$ for close binary stars \cite{Kh} are shown in the Table({\ref{mrt}}).

    The analysis of these data leads to few conclusions.
The averaging over all tabulated stars gives
\begin{equation}
<\frac{\tau}{\mu^{7/12}}>=1.007\pm 0.07 .
\end{equation}
and we can conclude that  the variability of measured data of surface temperatures and stellar masses has statistical character. Secondly, Eq.({\ref{7/12}}) is valid for all hot stars (exactly for all stars which are gathered in Tab.({\ref{mrt}})).

The problem with the averaging of $\frac{\rho}{\mu^{2/3}}$ looks different. There are a few of giants and super-giants in this Table.  The values of ratio  $\frac{\rho}{\mu^{2/3}}$  are more than 2 for them. It seems that, if to  exclude these stars  from consideration,  the averaging over stars of the main sequence gives value close to 1. Evidently, it needs in more detail consideration.

\vspace{0.1cm}

\underline{The luminosity} of a star
\begin{equation}
\mathbb{L}_0\sim \mathbb{R}_0^2 \mathbb{T}_0^4.
\end{equation}
at taking into account (Eq.{\ref{rm23}}) and (Eq.{\ref{tm}}) can be
expressed as
\begin{equation}
\mathbb{L}_0\sim \mathbb{M}^{11/3}\sim \mathbb{M}^{3.67}\label{lm}
\end{equation}
This dependence is shown in Fig.(\ref{LM})
\begin{figure}
\begin{center}
\includegraphics[scale=0.7]{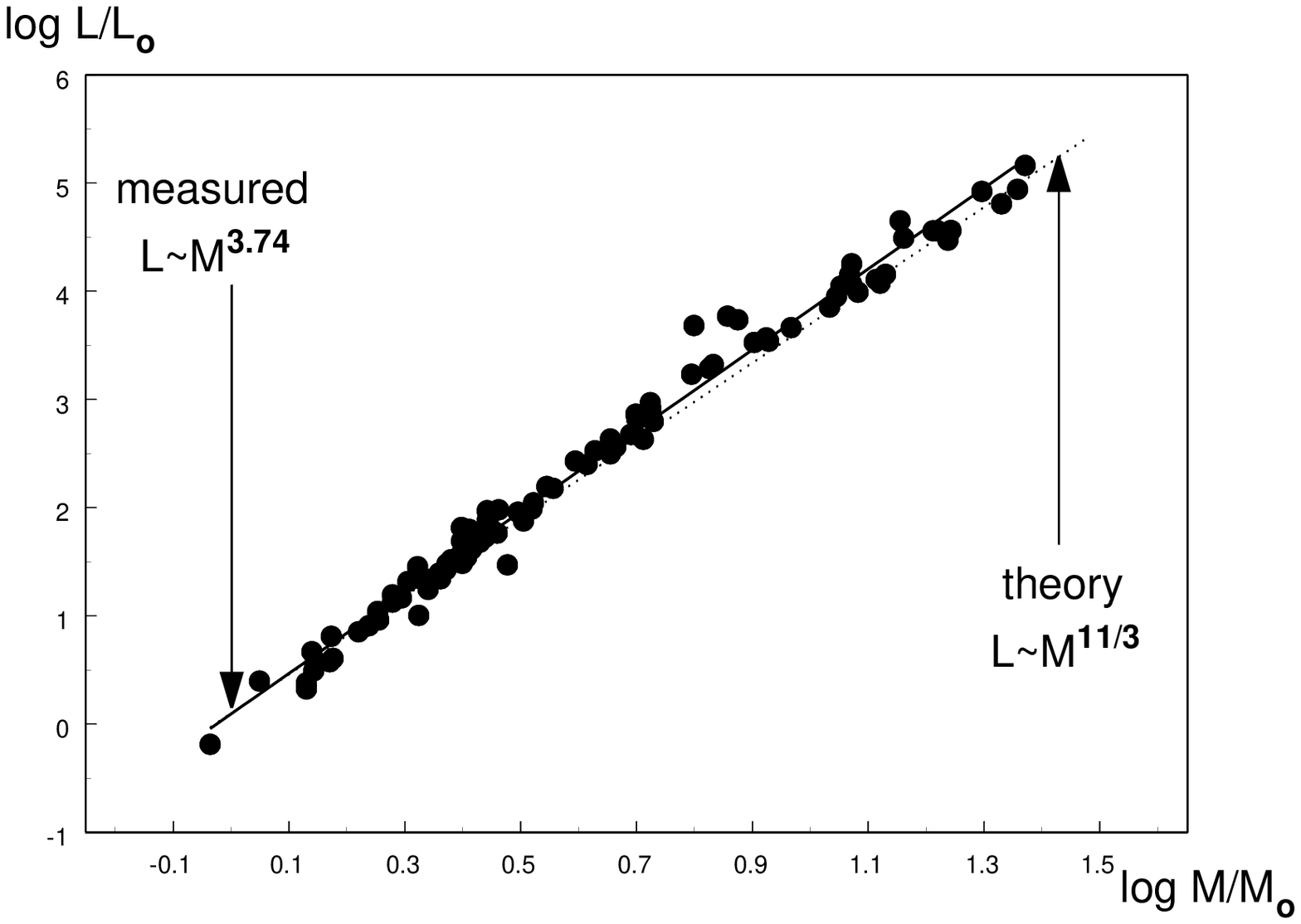}
\caption{The dependence of star luminosity on the star mass of
close binary stars \cite{Kh}. The luminosities are normalized to the
luminosity of the Sun, the stars masses are normalized to the mass
of the Sum. The data are shown on double logarithmic scale. The solid
line shows the result of fitting of measurement data $\mathbb{L}\sim
\mathbb{M}^{3.74} $. The theoretical dependence $\mathbb{L}\sim
\mathbb{M}^{11/3}$~(Eq.{\ref{lm}}) is shown by dotted line.}
\label{LM}
\end{center}
\end{figure}
It can be seen that all calculated interdependencies
$\mathbb{R(\mathbb{M})}$,$\mathbb{T(\mathbb{M})}$ and
$\mathrm{L(\mathbb{M})}$ show a good qualitative agreement with the
measuring data. At that it is important, that the quantitative explanation of
mass-luminosity dependence discovered at the beginning of 20th
century is obtained.

\subsubsection{The compilation of the results of calculations}
Let us put together the results of calculations.
it is energetically favorable for the star to be divided into two areas:
the core is located in the central area of the star
and the atmosphere is surrounding it from the outside.
(Fig.{\ref{star0}}).
\begin{figure}
\begin{center}
\includegraphics[scale=0.5]{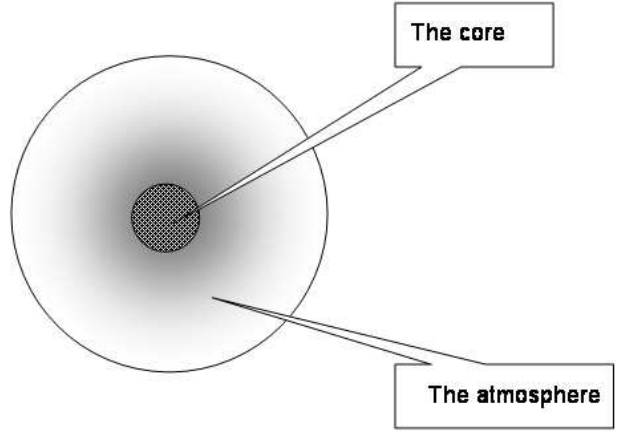}
\caption{The 	
schematic of the star interior\label{star0}}
\end{center}
\end{figure}
The core has the radius:
\begin{equation}
\mathbb{R}_\star=2.08\frac{a_B}{Z(A/Z)}\left(\frac{\hbar c}{G
m_p^2}\right)^{1/2}  \approx \frac{1.41\cdot 10^{11}}{{Z(A/Z)}}
cm.
\end{equation}
It is roughly equal to 1/10 of the stellar radius.

    At that the mass of the core is equal to
\begin{equation}
\mathbb{M}_\star=6.84
\frac{\mathbb{M}_{Ch}}{{\left(\frac{A}{Z}\right)^2}}.
\end{equation}
It is almost exactly equal to one half of the full mass of the star.

    The plasma inside the core has the constant density
\begin{equation}
{n_\star}=\frac{16}{9\pi}\frac{Z^3}{a_B^3}\approx
1.2 \cdot 10^{24} Z^3 cm^{-3}
\end{equation}
and constant temperature
\begin{equation}
\mathbb{T}_\star=\left(\frac{25\cdot
13}{28\pi^{4}}\right)^{1/3}\left(\frac{\hbar
c}{ka_B}\right)Z\approx Z\cdot 2.13\cdot 10^7
K.
\end{equation}

The plasma density and its temperature are  decreasing at an approaching to the stellar surface:
\begin{equation}
n_e(r)={n}_\star\biggl(\frac{\mathbb{R}_\star}{r}\biggr)^6
\end{equation}
and
\begin{equation}
T_r=\mathbb{T}_\star\biggl(\frac{\mathbb{R}_\star}{r}\biggr)^4.
\end{equation}

The external radius of the star is determined as
\begin{equation}
\mathbb{R}_0=
\left(\frac{\sqrt{\alpha\pi}}{2\eta}
\frac{\frac{A}{Z}m_p}{m_e}\right)^{1/2}{\mathbb{R}_\star}
\approx \frac{6.44\cdot 10^{11}}{Z(A/Z)^{1/2}}
cm
\end{equation}
and the temperature on the stellar surface is equal to
\begin{equation}
\mathbb{T}_0=\mathbb{T}_\star
\left(\frac{\mathbb{R}_\star}{\mathbb{R}_0}\right)^4 \approx
4.92\cdot 10^5 \frac{Z}{(A/Z)^2}
\end{equation}

\clearpage
\section[Magnetic fields of stars]
{Magnetic fields and magnetic moments of stars} \label{Ch7}

\subsection[Magnetic moments of stars]{Magnetic moments of celestial bodies}

A thin spherical surface with radius  $r$ carrying an electric
charge $q$ at the rotation around its axis with frequency  $\Omega$
obtains the magnetic moment
\begin{equation}
{{\boldsymbol{\mathfrak{m}}}}=\frac{ r^2}{3c}q\boldsymbol\Omega.
\end{equation}
The rotation of a ball charged at density $\varrho(r)$ will induce the
magnetic moment
\begin{equation}
\boldsymbol{\mathfrak{\mu}}=\frac{\boldsymbol\Omega}{3c}\int_0^R r^2\varrho(r) ~ 4\pi
r^2dr.
\end{equation}
Thus the positively charged core of a star induces the magnetic moment
\begin{equation}
\boldsymbol{\mathfrak{m}}_+=\frac{\sqrt{G}\mathbb{M}_\star
\mathbb{R}_\star^2}{5c}\boldsymbol\Omega.
\end{equation}
A negative charge will be concentrated in the star atmosphere. The
absolute value of atmospheric charge is equal to the positive
charge of a core. As the atmospheric charge is placed near the
surface of a star, its magnetic moment will be more than the core
magnetic moment. The calculation shows that as a result,
the total magnetic moment of the star will have the
same order of magnitude as the core but it will be negative:
\begin{equation}
\boldsymbol{\mathfrak{m}}_\Sigma\approx -\frac{\sqrt{G}}{c}\mathbb{M}_\star
\mathbb{R}_\star^2\boldsymbol\Omega.
\end{equation}
Simultaneously, the torque of a ball with mass $\mathbb{M}$ and
radius  $\mathbb{R}$ is
\begin{equation}
\boldsymbol{\mathcal{L}}\approx \mathbb{M}_\star \mathbb{R}_\star^2\boldsymbol\Omega.
\end{equation}
As a result, for celestial bodies where the force of their gravity
induces the electric polarization according to Eq.({\ref{roe}}), the
giromagnetic ratio will depend on world constants only:
\begin{equation}
\frac{{\boldsymbol{\mathfrak{m}}}_\Sigma}{\boldsymbol{\mathcal{L}}}\approx
-\frac{\sqrt{G}}{c}\label{ML}.
\end{equation}
This relation was obtained for the first time by P.M.S.Blackett
\cite{Blackett}. He shows that giromagnetic ratios of the Earth, the
Sun and the star 78 Vir are really near to $\sqrt{G}/c$.

By now the magnetic fields, masses, radii and velocities of rotation
are known for all planets of the Solar system and for a some stars
\cite{Sirag}. These measuring data are shown in Fig.({\ref{black}}),
which is taken from \cite{Sirag}. It is possible to see that these
data are in satisfactory agreement with Blackett's ratio.
\begin{figure}
\begin{center}
\includegraphics[scale=0.7]{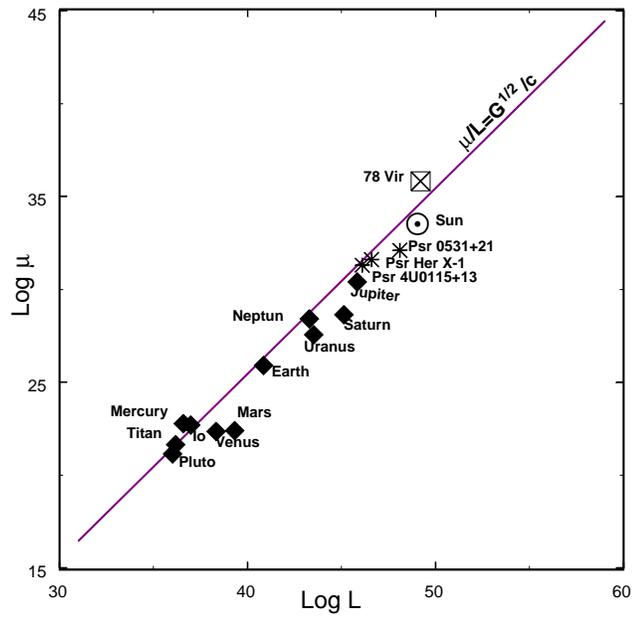}
\vspace{6cm} \caption {The observed values of magnetic moments
of celestial bodies vs. their angular momenta \cite{Sirag}. In
ordinate, the logarithm of the magnetic moment (in $Gs\cdot{cm^3}$)
is plotted; in abscissa the logarithm of the angular momentum (in
$erg\cdot{s}$) is shown. The solid line illustrates Eq.({\ref{ML}}).
The dash-dotted line fits of observed values.}
\label{black}
\end{center}
\end{figure}
At some assumption, the same parameters can be calculated for
pulsars. All measured masses of pulsars are equal by the order of
magnitude \cite{Thor}. It is in satisfactory agreement with the
condition of equilibrium of relativistic matter (see \label{sec11}).
It gives a possibility to consider that masses and radii of pulsars
are determined. According to generally accepted point of view,
pulsar radiation is related with its rotation, and it gives their
rotation velocity. These assumptions permit to calculate the
giromagnetic ratios for three pulsars with known magnetic fields on
their poles \cite{Beskin}. It is possible to see from
Fig.({\ref{black}}), the giromagnetic ratios of these pulsars are in
agreement with Blackett's ratio.

\subsection[Magnetic fields of stars]{Magnetic fields of hot stars}
To make  the above rough estimation for magnetic
fields induced by the star atmosphere more accurate, we can take into account the
distribution of electric charges inside the star atmosphere
In accordance with this distribution, the atmosphere
at its rotation induces the moment
\begin{equation}
{\boldsymbol{\mathfrak{m}}}_-=-\frac{\sqrt{G}\mathbb{M}_\star\boldsymbol\Omega{\mathbb{R}_\star}^2}{9c}\int_{\mathbb{R}_\star}^{R_0}
r^2\frac{d}{dr}\sqrt{\xi^3(2-\xi^3)}~dr,
\end{equation}
Where  $\xi=\frac{\mathbb{R_\star}}{r}$.

As $\mathbb{R}_\star\ll R_0$, the magnetic moment of a core can be
neglected. In this case, the total magnetic moment of a star
\begin{equation}
{\boldsymbol{\mathfrak{m}}}\approx
-2\frac{\sqrt{G}\mathbb{M_\star}}{3c}\left(\frac{R_0^{7/2}}{\mathbb{R}_\star^{3/2}}\right)\boldsymbol\Omega.
\end{equation}
and magnetic field near the pole of a star
\begin{equation}
\boldsymbol{\mathcal{H}}\approx
-\frac{2\sqrt{G}\mathbb{M_\star}}{9c}\frac{R_0^{1/2}}{\mathbb{R}_\star^{3/2}}\boldsymbol{\Omega}.
\end{equation}
Using of substitution of  relations obtained above, one can
conclude that the magnetic field on a star pole must not depend
on the parameter $Z$ and very slightly depends on $A/Z$, and this
dependence can be neglected. I.e. the calculations show that this field practically is not depending on the mass, on the radius and on the temperature of a hot star and must depend on the velocity of star rotation only:
\begin{equation}
{\boldsymbol{\mathcal{H}}}\approx
-0.47\left(\frac{m_e}{m_p}\right)^{3/4}\frac{\alpha
c}{{G}^{1/2}\eta^{1/4}}{\boldsymbol\Omega}\approx -8.8\cdot10^8 \boldsymbol{\Omega}
{\quad}Oe.\label{Ht}
\end{equation}
The magnetic fields are measured for stars of Ap-class \cite{rom}.
These stars are characterized by changing their brightness in
time. The periods of these changes are measured too. At present
there  is no full understanding of causes of these visible
changes of the luminosity. If these luminosity changes caused by some internal reasons
will occur not uniformly on a star surface, one can conclude that the measured period of the luminosity change can depend on star rotation.
It is possible to think that at relatively rapid rotation of a star, the period of a visible
change of the luminosity can be determined by this rotation in general.
To check this suggestion, we can compare the calculated
dependence (Eq.{\ref{Ht}}) with measuring data \cite{rom} (see Fig.
{\ref{H-W}}). Evidently one must not expect very good coincidence  of calculations  and measuring data, because calculations were made for the case of a spherically symmetric model and measuring data are obtained for stars where this symmetry is obviously violated. So getting consent on order of the value can be considered as wholly satisfied.
\begin{figure}
\begin{center}
\includegraphics[scale=0.7]{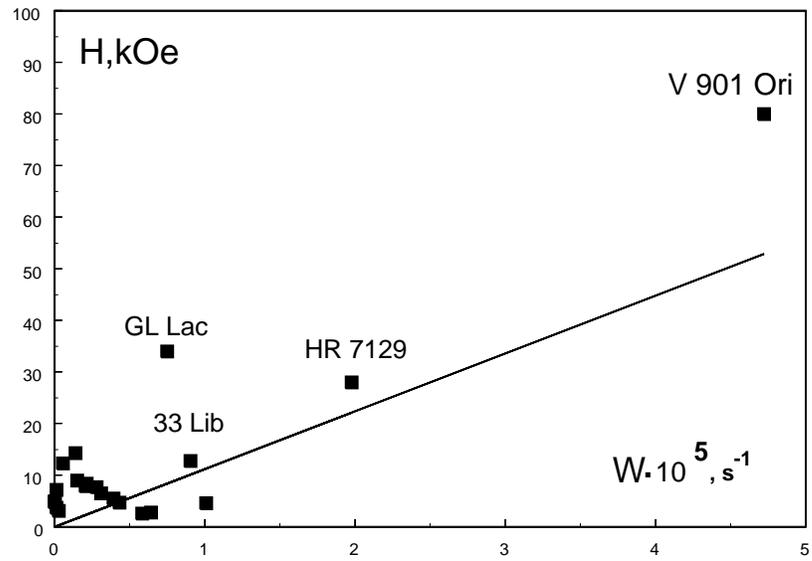}
\vspace{6cm} \caption {The dependence of magnetic fields on poles of
Ap-stars as a function of  their rotation velocity \cite{rom}. The
line shows Eq.(\ref{Ht})).}\label{H-W}
\end{center}
\end{figure}
It should be said that Eq.(\ref{Ht}) does not working well in case with the Sun.
The Sun surface rotates with period $T\approx 25\div 30$
days. At this velocity of rotation, the magnetic field on the Sun
pole calculated accordingly to Eq.(\ref{Ht}) must be about 1
kOe. The dipole field of Sun according to experts estimation is
approximately 20 times lower. There can be several reasons for that.
\clearpage

\section[The apsidal rotation]
{The angular velocity of the apsidal rotation in binary stars}
\label{Ch8}

\subsection[The apsidal rotation]{The apsidal rotation of close binary stars}
The apsidal rotation (or periastron rotation) of close binary stars
is a result of their non-Keplerian movement which originates from the
non-spherical form of stars. This non-sphericity has been produced
by rotation of stars around their axes or by their mutual tidal
effect. The second effect is usually smaller  and can be neglected.
The first and basic theory of this effect was developed by
A.Clairault at the beginning of the XVIII century. Now this effect was
measured for approximately 50 double stars. According to Clairault's
theory the velocity of periastron rotation must be approximately
100 times faster if matter is uniformly distributed inside a star.
Reversely, it would be absent if all star mass is concentrated in the
star center. To reach an agreement between the measurement data and
calculations, it is necessary to assume that the density of
substance grows in direction to the center of a star and  here it
runs up to a value which is hundreds times greater than  mean density
of a star. Just the same mass concentration of the stellar
substance is supposed by all standard theories of a star interior.
It has been usually considered as a proof of astrophysical models.
But it can be considered as a qualitative argument. To obtain a
quantitative agreement between theory and measurements, it is
necessary to fit parameters of the stellar substance distribution in
each case separately.

Let us consider this problem with taking into account the gravity
induced electric polarization of plasma in a star. As it was shown
above, one half of full mass of a star is concentrated in its plasma
core at a permanent density. Therefor, the effect of periastron
rotation of close binary stars must be reviewed with the account of
a change of forms of these star cores.

According to \cite{peri2},\cite{peri1} the ratio of the angular
velocity $\omega$ of rotation of periastron which is produced by the
rotation of a star around its axis with  the angular velocity
$\Omega$ is
\begin{equation}
\frac{\omega}{\Omega}=\frac{3}{2}\frac{(I_A-I_C)}{Ma^2}
\end{equation}
where $I_A$ and $I_C$ are the moments of inertia relatively to
principal axes of the ellipsoid. Their difference is
\begin{equation}
I_A-I_C=\frac{M}{5}(a^2-c^2),
\end{equation}
where $a$ and $c$ are the equatorial and polar radii of the star.

Thus we have
\begin{equation}
\frac{\omega}{\Omega}\approx \frac{3}{10}\frac{(a^2-c^2)}{a^2}.
\end{equation}

\subsection[The equilibrium form of the core]{The equilibrium form of the core of a rotating star}
In the absence of rotation the equilibrium equation  of plasma
inside star core (Eq.{\ref{Eu2}} is
\begin{equation}
\gamma {\bf g}_G+\rho_G {\bf E}_G=0\label{qm}
\end{equation}
where $\gamma$,${\bf g}_G$, $\rho_G$ and ${\bf E}_G$ are the
substance density the acceleration of gravitation, gravity-induced
density of charge and intensity of gravity-induced electric field
($div~{\bf g}_G=4\pi~ G~ \gamma$, $div~{\bf E}_G=4\pi \rho_G$ and
$\rho_G=\sqrt{G}\gamma$).

One can suppose, that at rotation,  under action of a rotational
acceleration  ${\bf g}_\Omega$, an additional electric charge with
density $\rho_\Omega$ and electric field ${\bf E}_\Omega$ can
exist, and the equilibrium equation obtains the form:

\begin{equation}
(\gamma_G+\gamma_\Omega)({\bf g}_G+{\bf
g}_\Omega)=(\rho_G+\rho_\Omega)({\bf E}_G+{\bf E}_\Omega),
\end{equation}

where

\begin{equation}
div~({\bf E}_G+{\bf E}_\Omega)=4\pi(\rho_G+\rho_\Omega)
\end{equation}

or

\begin{equation}
div~{\bf E}_\Omega=4\pi\rho_\Omega.
\end{equation}

We can look for a solution for electric potential in the form

\begin{equation}
\varphi=C_\Omega~r^2(3cos^2\theta-1)
\end{equation}

or in Cartesian coordinates

\begin{equation}
\varphi=C_\Omega(3z^2-x^2-y^2-z^2)
\end{equation}

where $C_\Omega$ is a constant.

 Thus

\begin{equation}
E_x=2~C_\Omega~x,~ E_y=2~C_\Omega~y,~ E_z=-4~C_\Omega~z
\end{equation}

and

\begin{equation}
div~{\bf E}_\Omega=0
\end{equation}

and we obtain important equations:

\begin{equation}
\rho_\Omega=0;
\end{equation}

\begin{equation}
\gamma g_\Omega=\rho {\bf E}_\Omega.
\end{equation}

Since centrifugal force must be contra-balanced by electric
force

\begin{equation}
\gamma~2\Omega^2~x=\rho~2C_\Omega~x
\end{equation}

and

\begin{equation}
C_\Omega=\frac{\gamma~\Omega^2}{\rho}=\frac{\Omega^2}{\sqrt{G}}
\end{equation}

The potential of a positive uniform charged ball is

\begin{equation}
\varphi(r)=\frac{Q}{R}\biggl(\frac{3}{2}-\frac{r^2}{2R^2}\biggr)
\end{equation}

The negative  charge on the surface of a sphere induces inside the
sphere the potential

\begin{equation}
\varphi(R)=-\frac{Q}{R}
\end{equation}

where according to Eq.({\ref{qm}}) $Q=\sqrt{G}M$, and $M$ is the
mass of the star.

Thus the total potential inside the considered star is

\begin{equation}
\varphi_\Sigma=\frac{\sqrt{G}M}{2R}\biggl(1-\frac{r^2}{R^2}\biggr)+\frac{\Omega^2}{\sqrt{G}}r^2(3cos^2\theta-1)
\end{equation}

Since the  electric potential must be equal to zero on the surface
of the star, at $r=a$ and $r=c$

\begin{equation}
\varphi_\Sigma=0
\end{equation}

and  we obtain the equation which describes the equilibrium form
of the core of a rotating star (at $\frac{a^2-c^2}{a^2}\ll 1$)

\begin{equation}
\frac{a^2-c^2}{a^2}\approx\frac{9}{2\pi}\frac{\Omega^2}{G\gamma}\label{ef}.
\end{equation}

\subsection[The angular velocity]{The angular velocity of the apsidal rotation}

Taking into account of Eq.({\ref{ef}}) we have

\begin{equation}
\frac{\omega}{\Omega}\approx
\frac{27}{20\pi}\frac{\Omega^2}{G\gamma}\label{oo}
\end{equation}
If both stars of a close pair induce a rotation of periastron,
this equation transforms to

\begin{equation}
\frac{\omega}{\Omega}\approx
\frac{27}{20\pi}\frac{\Omega^2}{G}\biggl(\frac{1}{\gamma_1}+\frac{1}{\gamma_2}\biggr),
\end{equation}
where $\gamma_1$ and $\gamma_2$ are densities of star cores.

The equilibrium density of star cores is known (Eq.({\ref{eta1}})):

\begin{equation}
\gamma=\frac{16}{9\pi^2}\frac{A}{Z}m_p\frac{Z^3}{a_B^3}\label{go}.
\end{equation}

If we introduce  the period of ellipsoidal rotation
$P=\frac{2\pi}{\Omega}$ and  the period of the rotation of
periastron $U=\frac{2\pi}{\omega}$,  we obtain from
Eq.({\ref{oo}})

\begin{equation}
\frac{\mathcal{P}}{\mathcal{U}}\biggl(\frac{\mathcal{P}}{\mathcal{T}}\biggr)^2\approx\sum_1^2\xi_i\label{2},
\end{equation}
where
\begin{equation}
\mathcal{T}=\sqrt{\frac{243~\pi^3}{80}}~\tau_0\approx 10 \tau_0,
\end{equation}
\begin{equation}
\tau_0=\sqrt{\frac{a_B^3}{G~m_p}}\approx 7.7\cdot 10^2 sec
\end{equation}
and
\begin{equation}
\xi_i=\frac{Z_i}{A_i(Z_i+1)^3}\label{2P}.
\end{equation}

\subsection[The comparison  with observations]{The comparison of the calculated angular velocity of the periastron rotation with observations}

Because the substance density (Eq.({\ref{go}})) is depending
approximately on the second power of the nuclear charge, the
periastron movement of stars consisting of heavy elements will fall
out from the observation as it is very slow. Practically the
obtained equation ({\ref{2}}) shows that it is possible to observe
the periastron rotation of a star consisting of light elements
only.

The value $\xi=Z/[AZ^3]$ is equal to $1/8$ for hydrogen,
$0.0625$ for deuterium, $1.85\cdot 10^{-2}$ for helium. The
resulting value of the periastron rotation of double stars will be
the sum of separate stars rotation. The possible combinations of a
couple and their value of $\sum_1^2\xi_i$ for stars consisting of
light elements is shown in Table {\ref{peri}}.

\bigskip

\begin{tabular}{||c|c|c||}\hline\hline
  star1&star2 &$\xi_1+\xi_2$\\
  composed of &composed of&\\\hline
  H & H & .25\\
  H & D & 0.1875\\
  H & He & 0.143\\
  H & hn & 0.125\\
  D & D & 0.125 \\
  D & He & 0.0815 \\
  D & hn & 0.0625 \\
  He & He & 0.037 \\
  He & hn & 0.0185 \\ \hline\hline
\end{tabular}\label{peri}
Table {\ref{peri}}
\bigskip

The "hn" notation in Table {\ref{peri}} indicates that the second
component of the couple consists of heavy elements or it is a dwarf.

The results of measuring of main parameters for close binary stars
are gathered in \cite{Kh}. For reader convenience, the data of these
measurement is applied in the Table in Appendix. One can compare our
calculations with  data of these measurements. The distribution of
close binary stars on value of $(\mathcal{P}/\mathcal{U})
(\mathcal{P}/\mathcal{T})^2$ is shown in Fig.{\ref{periastr}} on
logarithmic scale. The lines mark the values of parameters
$\sum_1^2\xi_i$ for different light atoms in accordance  with
{\ref{2P}}. It can be seen that calculated values the periastron
rotation for stars composed by light elements which is summarized in
Table{\ref{peri}} are in  good agreement with separate peaks of
measured data. It confirms that our approach to interpretation of
this effect is adequate to produce  a satisfactory accuracy of
estimations.
\begin{figure}
\hspace{-1.5cm}
\includegraphics[scale=0.7]{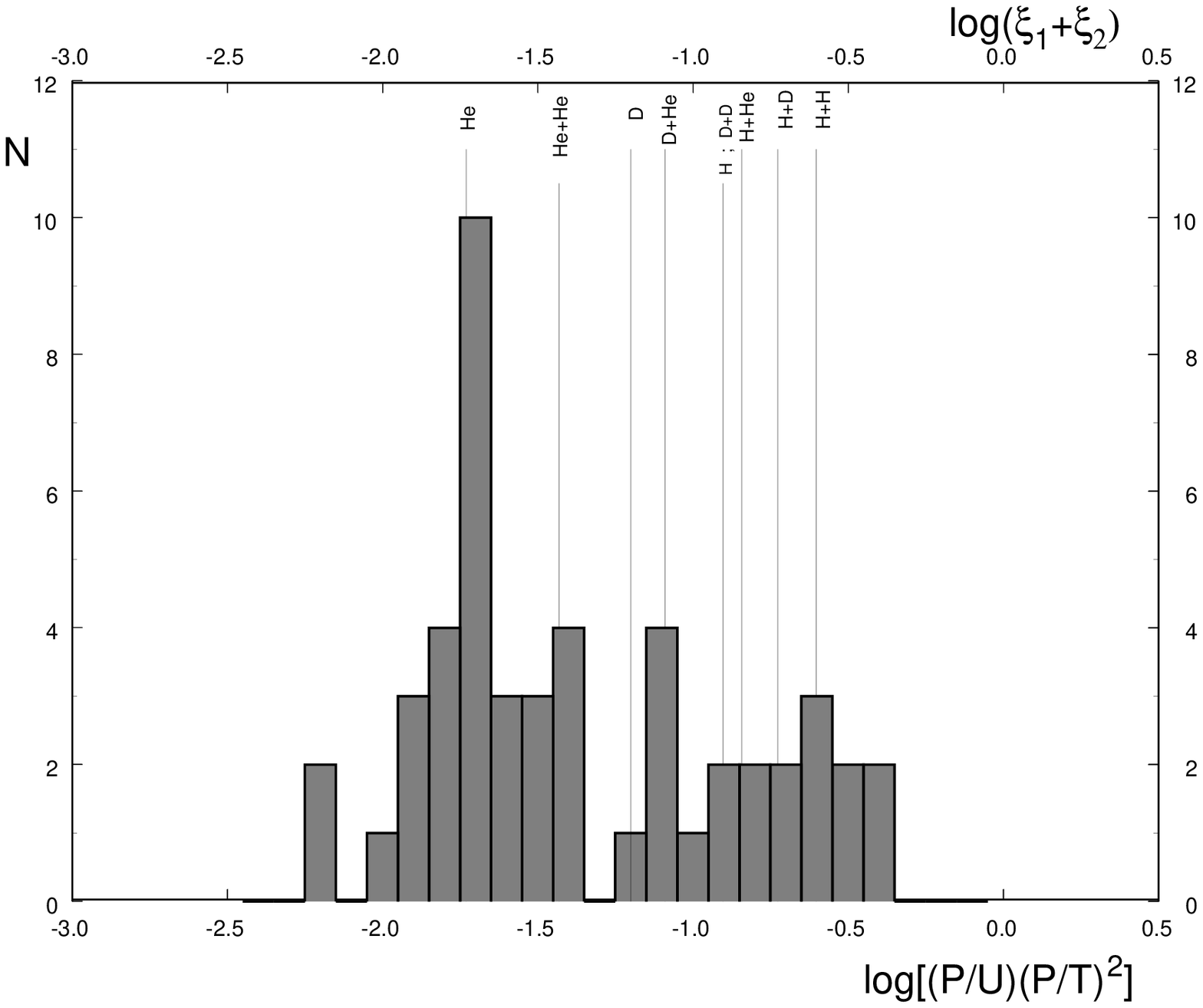}
\caption{The distribution of close binary stars \cite{Kh} on value
of $(\mathcal{P}/\mathcal{U})(\mathcal{P}/\mathcal{T})^2$. Lines
show  parameters $\sum_1^2\xi_i$ for different light atoms in
according with {\ref{2P}}.} \label{periastr}
\end{figure}
\clearpage

\section{The solar seismical oscillations}\label{Ch9}

\subsection [The spectrum  of solar oscillations]{The spectrum  of solar seismic oscillations}

The measurements \cite{bison} show that the Sun surface is subjected
to a seismic vibration. The most intensive oscillations have the
period about five minutes and the wave length about $10^4$km or
about hundredth part of the Sun radius. Their spectrum obtained by
BISON collaboration is shown in Fig.{\ref{bison}}.

\begin{figure}
\begin{center}
\includegraphics[scale=0.7]{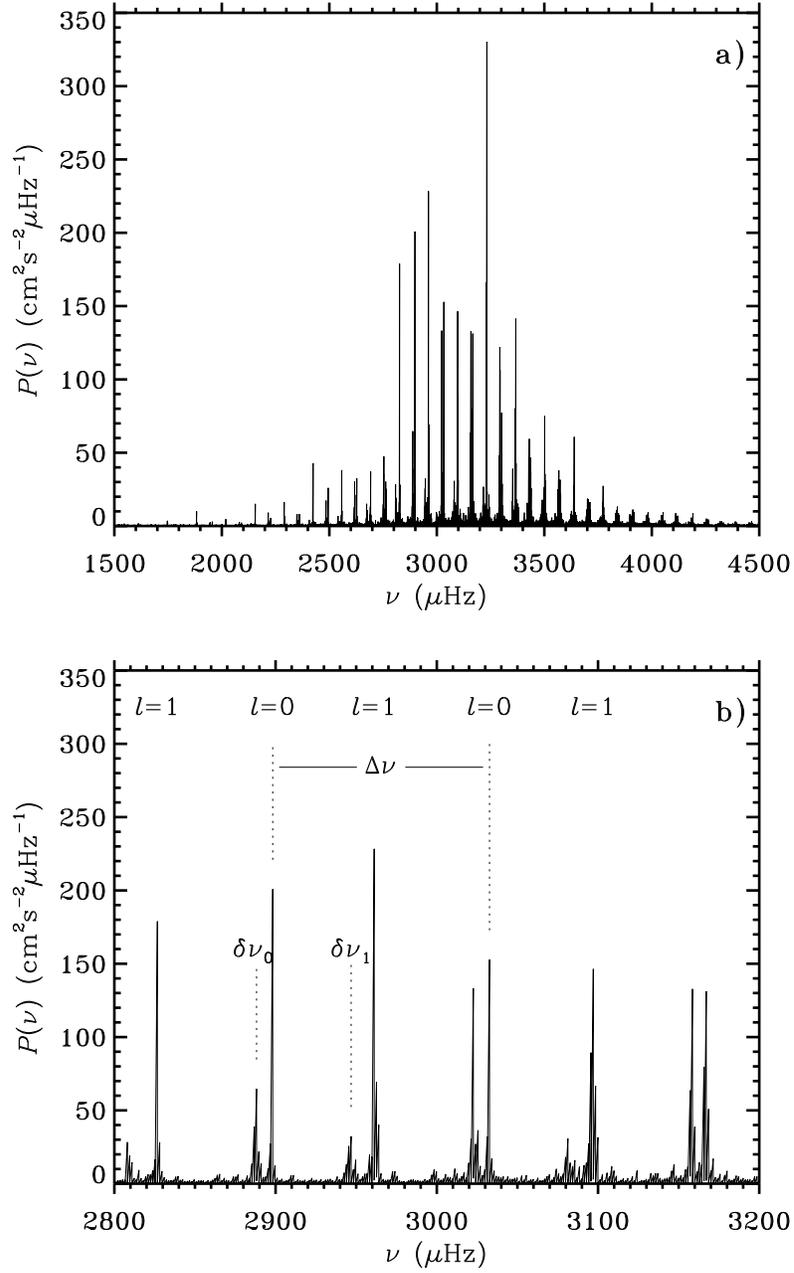}
\caption{$(a)$ The power spectrum of solar
oscillation obtained by means of Doppler velocity measurement in
light integrated over the solar disk. The data were obtained from
the BISON network \cite{bison}.
$(b)$ An expanded view of a part of frequency
range.}\label{bison}
\end{center}
\end{figure}

It is supposed, that these oscillations are a superposition of a big
number of different modes of resonant acoustic vibrations, and that
acoustic waves propagate in different trajectories in the interior
of the Sun and they have multiple reflection from surface. With these
reflections trajectories of same waves can be closed and as a result
standing waves are forming.

Specific features of spherical body oscillations are described by
the expansion in series on spherical functions. These oscillations
can have a different number of wave lengths on the radius of a
sphere ($n$) and  a different number of wave lengths on its surface
which is determined by the $l$-th spherical harmonic. It is accepted
to describe the sunny surface oscillation spectrum as the expansion
in series \cite{CD}:
\begin{equation}
\nu_{nlm} \simeq \Delta \nu_0(n+\frac{l}{2}+\epsilon_0)-l(l+1)D_0 +
m\Delta \nu_{rot}.\label{nu}
\end{equation}
Where the last item is describing the effect of the Sun rotation and
is small. The main contribution is given by the first item which
creates a large splitting in the spectrum (Fig.{\ref{bison}})
\begin{equation}
\triangle\nu=\nu_{n+1,l}-\nu_{n,l}.
\end{equation}
The small splitting of spectrum (Fig.{\ref{bison}}) depends on the
difference
\begin{equation}
\delta\nu_l=\nu_{n,l}-\nu_{n-1,l+2}\approx (4l+6)D_0.
\end{equation}
A satisfactory agreement of these estimations and measurement data
can be obtained at \cite{CD}

\begin{equation}
\Delta \nu_0=120~\mu Hz,~ \epsilon_0=1.2,~ D_0=1.5~\mu Hz,~ \Delta
\nu_{rot}=1\mu Hz.\label{del}
\end{equation}

 To obtain these values of parameters $\Delta \nu_0,~ \epsilon_0
è ~D_0$ from theoretical models is not possible. There are a lot of
qualitative and quantitative assumptions used at a model
construction and a direct calculation of spectral frequencies
transforms into a unresolved complicated problem.

Thus, the current interpretation of the measuring spectrum by the
spherical harmonic  analysis does not make it clear. It gives no
hint to an answer to the question: why oscillations close to hundredth
harmonics are really excited and there are no waves near
fundamental harmonic?

The measured spectra have a very  high resolution (see
Fig.({\ref{bison}})). It means that an oscillating system has high
quality. At this condition, the system must have oscillation on a
fundamental frequency. Some peculiar mechanism must exist to force
a system to oscillate on a high harmonic. The current explanation
does not clarify it.

It is important, that now the solar oscillations are measured by
means of two different methods. The solar oscillation spectra which
was obtained on program "BISON", is shown on Fig.({\ref{bison}})).
It has a very high resolution, but (accordingly to the Liouville's theorem)
it was obtained with some loss of luminosity, and as a result not
all lines are well statistically worked.
\begin{figure}
\hspace{.5cm}
\includegraphics[scale=1]{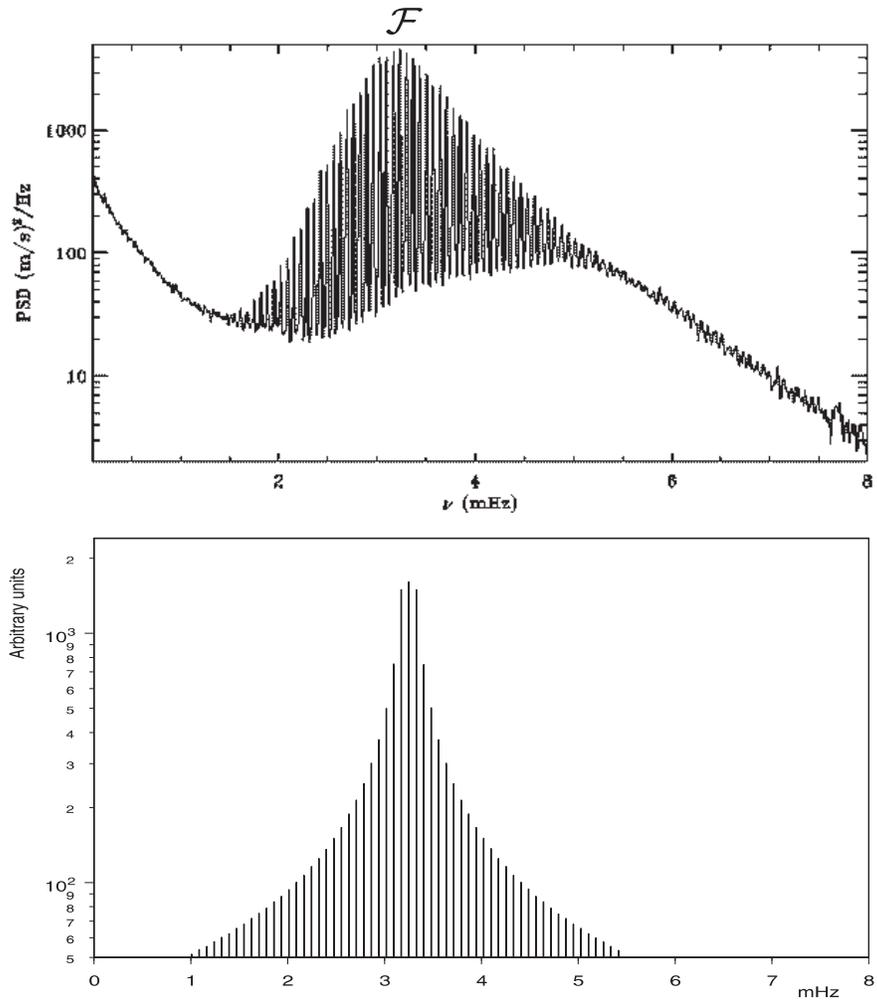}
\caption{$(a)$ The measured power spectrum of solar oscillation. The
data were obtained from the SOHO/GOLF measurement \cite{soho}. $(b)$
The calculated spectrum described by Eq.({\ref{ur}}) at $<Z>=3.4$ and
$A/Z=5$.} \label{soho}
\end{figure}

Another spectrum was obtained in the program  "SOHO/GOLF". Conversely,
it is not characterized by high resolution, instead it gives
information about general character of the solar oscillation
spectrum (Fig.{\ref{soho}})).

The existence of this spectrum requires  to change the view at all
problems of solar oscillations. The theoretical explanation of this
spectrum must give answers at least to four questions :

1.Why does the whole spectrum consist from a large number of
equidistant spectral lines?

2.Why does the central frequency of this spectrum ${\cal F}$ is approximately
equal  to $\approx 3.23~ mHz$?

3. Why does  this spectrum splitting $f$ is approximately equal  to
$67.5~ \mu Hz$?

4. Why does the intensity of spectral lines decrease  from the
central line to the periphery?

The answers to these questions can be obtained if we take into
account  electric polarization of a solar core.

The description  of measured spectra by means of spherical
analysis does not make clear of the physical meaning of this
procedure. The reason of difficulties lies in attempt to consider
the oscillations of a Sun as a whole. At existing dividing of a star
into core and atmosphere, it is easy to understand that the core
oscillation must form a measured spectrum. The fundamental mode of
this oscillation must be determined by its spherical mode when the
Sun radius oscillates without changing of the spherical form of the
core. It gives a most low-lying mode with frequency:
\begin{equation}
\Omega_s\approx \frac{c_s}{\mathbb{R}_\star},
\end{equation}
where $c_s$ is sound velocity in the core.

It is not difficult to obtain the numerical estimation of this
frequency by order of magnitude. Supposing that the sound velocity
in dense matter is $10^7 cm/c$ and radius is close to  $\frac{1}{10}$ of
external radius of a star, i.e. about $10^{10}cm$, one can obtain as
a result
\begin{equation}
F=\frac{\Omega_s}{2\pi}\approx 10^{-3}Hz
\end{equation}
It gives possibility to conclude that this estimation is in
agreement with measured frequencies. Let us consider this mechanism
in more detail.

\subsection[The sound speed in plasma]{The sound speed in hot plasma}

The pressure of high temperature plasma is a sum of the plasma
pressure (ideal gas pressure) and the pressure of black radiation:
\begin{equation}
P=n_e kT +\frac{\pi^2}{45\hbar^3 c^3}(kT)^4.\label{pr2}
\end{equation}
and its entropy is
\begin{equation}
S=\frac{1}{\frac{A}{Z}m_p}~ln~\frac{(kT)^{3/2}}{n_e}
+\frac{4\pi^2}{45\hbar^3 c^3n_e}(k T)^3,\label{s}
\end{equation}
The sound speed $c_s$ can be expressed by Jacobian \cite{LL}:
\begin{equation}
c_s^2=\frac{D(P,S)}{D(\rho,S)}=\frac{\biggl(\frac{D(P,S)}{D(n_e,T)}\biggr)}{\biggl(\frac{D(\rho,S)}{D(n_e,T)}\biggr)}
\end{equation}
or
\begin{equation}
c_s=\biggl\{\frac{5}{9}\frac{k{T}}{A/Z m_p}
\biggl[1+\frac{2\left(\frac{4\pi^2}{45\hbar^3
c^3}\right)^2(kT)^6}{5n_e[n_e+\frac{8\pi^2}{45\hbar^3
c^3}(kT)^3]}\biggr]\biggr\}^{1/2}
\end{equation}
For  $T={\mathbb{T}_\star}$ and $n_e=n_\star$ we have:
\begin{equation}
{\frac{4\pi^2(k{\mathbb{T}_\star})^3}{45\hbar^3
c^3n_\star}}=\approx0.18~.
\end{equation}
 Finally we obtain:
\begin{equation}
c_s=\biggl\{\frac{5}{9}\frac{\mathbb{T}_\star}{(A/Z)m_p}[1.01]\biggr)^{1/2}
\approx 3.14~10^7~\biggl(\frac{Z}{A/Z}\biggr)^{1/2}~
cm/s~.\label{cs}
\end{equation}
\subsection[The basic elastic oscillation]{The basic elastic oscillation of a spherical core}
Star cores consist of dense high temperature plasma which is a
compressible matter. The basic mode of elastic vibrations of a
spherical core is related with its radius oscillation. For the
description of this type of oscillation, the potential $\phi$ of
displacement velocities $v_r=\frac{\partial \psi}{\partial r}$ can
be introduced  and the motion equation can be reduced to the wave
equation expressed through $\phi$ \cite{LL}:
\begin{equation}
c_s^2\Delta\phi=\ddot\phi,
\end{equation}
and a spherical derivative for periodical in time oscillations
$(\sim e^{-i\Omega_s t})$
 is:
\begin{equation}
\Delta\phi=\frac{1}{r^2}\frac{\partial}{\partial
r}\biggl(r^2\frac{\partial \phi}{\partial
r}\biggr)=-\frac{\Omega_s^2}{c_s^2}\phi~.
\end{equation}
It has the finite solution for the full core volume including its
center
\begin{equation}
\phi=\frac{A}{r}sin \frac{\Omega_s r}{c_s},
\end{equation}
where $A$ is a constant. For  small oscillations, when displacements
on the surface $u_R$ are small $(u_R/R=v_R/ \Omega_s R\rightarrow
0)$ we obtain the equation:
\begin{equation}
tg \frac{\Omega_s {\mathbb{R}}}{c_s}=\frac{\Omega_s
{\mathbb{R}}}{c_s}
\end{equation}
which has the solution:
\begin{equation}
\frac{\Omega_s {\mathbb{R}}}{c_s}\approx4.49.
\end{equation}
Taking into account  Eq.({\ref{cs}})), the main frequency of the
core radial elastic oscillation is
\begin{equation}
\Omega_s =
4.49\biggl\{1.4\biggl[\frac{Gm_p}{r_B^3}\biggr]\frac{A}{Z}
\biggl(Z +1\biggr)^3\biggr\}^{1/2}.\label{qb}
\end{equation}
It can be seen that this frequency depends on ${Z}$ and ${A}/{Z}$
only.
Some values of  frequencies of radial sound oscillations ${\cal
F}=\Omega_s/2\pi$  calculated from this equation for selected $A/Z$
and $Z$ are shown in third column of Table
({\ref{st-osc}}).

\newpage
{\hspace{8cm} Table ({\ref{st-osc}})}
\begin{tabular}{||c|c|c||c|c||}\hline\hline
&&${\cal F},mHz$&&${\cal F},mHz$\\
Z&A/Z&(calculated&star&\\
& & ïî ({\ref{qb}}))
&&measured\\
\hline 1&1&0.23&$\xi~Hydrae$&$\sim 0.1$\\ \hline
1&2&0.32& $\nu~Indus$&0.3\\
\hline 2&2&0.9&$\eta~Bootis$&0.85\\ \hline
&&& The Procion$(A\alpha~CMi)$&1.04\\
2&3&1.12& & \\
&&&$\beta~Hydrae$&1.08\\ \hline
3&4&2.38&$\alpha~Cen~A$&2.37\\ \hline\hline
3&5&2.66&&\\
\hline 3.4&5&\bf3.24&The Sun&\bf3.23\\ \hline
4&5&4.1&&\\\hline\hline
\end{tabular}\label{st-osc}

\bigskip

\bigskip
The star mass spectrum {(ðèñ.\ref{starM}) shows that the ratio  $A/Z$ must be $\approx 5$ for the Sum. It is in accordance with the calculated frequency of solar core oscillations if the averaged charge of nuclei  $Z\approx 3.4$. It is not a confusing conclusion, because  the plasma electron gas  prevents the decay of $\beta$-active nuclei (see Sec.\ref{Ch13}). This mechanism can probably to stabilize neutron-excess nuclei.

\subsection[The low frequency oscillation]{The low frequency oscillation of the density of a neutral
plasma}

Hot plasma has the density ${n_\star}$ at its equilibrium state.
The local deviations from this state induce processes of density
oscillation since plasma tends to return to its steady-state
density. If we consider small periodic oscillations of core radius
\begin{equation}
R={\mathbb{R}}+ u_R \cdot sin~ \omega_{n_\star}t,
\end{equation}
where a radial displacement of plasma particles is small
($u_R\ll{\mathbb{R}}$), the oscillation process of plasma density
can be described by the equation

\begin{equation}
\frac{d{\mathcal{E}}}{dR}=\mathbb{M}\ddot R~.
\end{equation}
Taking into account
\begin{equation}
\frac{d{\mathcal{E}}}{dR}=\frac{d\mathcal{E}_{plasma}}{dn_e}\frac{dn_e}{dR}
\end{equation}
and
\begin{equation}
\frac{3}{8}\pi^{3/2} \mathbb{N}_e \frac{e^3 a_0^{3/2}}
{(k\mathbb{T})^{1/2}}\frac{n_\star}{\mathbb{R}^2}=\mathbb{M}\omega_{n_\star}^2
\end{equation}
From this we obtain
\begin{equation}
\omega_{n_\star}^2=\frac{3}{\pi^{1/2}}
k{\mathbb{T}}\biggl(\frac{e^2}{a_B k{\mathbb{T}}}\biggr)^{3/2}
\frac{Z^3}{{\mathbb{R}}^2 A/Z m_p}
\end{equation}\label{qm2}
and finally
\begin{equation}
\omega_{n_\star}=\biggl\{\frac{2^8}{3^5}\frac{{\pi}^{1/2}}{{10}^{1/2}}
\alpha^{3/2}\biggl[\frac{Gm_p}{a_B^3}\biggr]\frac{A}{Z}Z^{4.5}\bigg\}^{1/2},\label{qm3}
\end{equation}
where $\alpha=\frac{e^2}{\hbar c}$ is the fine structure constant.
These low frequency oscillations of neutral plasma density are
similar to phonons in solid bodies. At that oscillations with
multiple frequencies $k\omega_{n_\star}$ can exist. Their power is
proportional to $1/{\kappa}$, as the occupancy these levels in
energy spectrum must be reversely proportional to their energy
$k\hbar\omega_{n_\star}$. As result, low frequency oscillations of
plasma density constitute set of vibrations
\begin{equation}
\sum_{\kappa=1} \frac{1}{\kappa}~sin(\kappa\omega_{n_\star}t)~.
\end{equation}

\subsection[The spectrum of solar oscillations]{The spectrum of solar core oscillations}
The set of the low frequency oscillations with $\omega_\eta$ can be
induced by sound oscillations with $\Omega_s$. At that, displacements
obtain the spectrum:
\begin{equation}
u_R\sim \sin~\Omega_s t\cdot\sum_{\kappa=0}
\frac{1}{\kappa}~\sin~\kappa\omega_{n_\star}t\cdot \sim
\xi\sin~\Omega_s t + \sum_{\kappa=1}
\frac{1}{\kappa}~\sin~(\Omega_s \pm \kappa \omega_{n_\star}
)t,\label{ur}
\end{equation}
where $\xi$ is a coefficient $\approx 1$.

This spectrum is shown in Fig.({\ref{soho}}).

The central frequency of experimentally measured distribution of solar oscillations  is approximately equal to (Fig.({\ref{bison}}))
\begin{equation}
{\cal F}_\odot\approx 3.23~ mHz\label{ffs}
\end{equation}
and the experimentally measured frequency splitting in this spectrum
is approximately equal to
\begin{equation}
{f}_\odot\approx 68~ \mu Hz.\label{ffg}
\end{equation}
 A good agreement of the calculated frequencies of basic modes of
oscillations (from Eq.({\ref{qb}}) and Eq.({\ref{qm}})) with
measurement can be obtained at $Z=3.4$ and $A/Z=5$:
\begin{equation}
{\cal F}_{_{_{Z=3.4;\frac{A}{Z}=5}}} =
\frac{\Omega_s}{2\pi}=3.24~mHz;~f_{_{_{Z=3.4;\frac{A}{Z}=5}}}=\frac{\omega_{n_\star}}{2\pi}=68.1~\mu
Hz.
\end{equation}

\clearpage

\section[Addendum I: A mechanism of stabilization nuclei in plasma]
{Addendum I:  A mechanism of stabilization for neutron-excess nuclei in plasma}
\label{Ch13}

\subsection{ Neutron-excess nuclei and the neutronization}
The form of the star mass spectrum
(Fig.\ref{starM}) indicates  that plasma of many stars consists of neutron-excess nuclei with the ratio $A/Z=3,4,5$ and so on. These nuclei are subjects of a decay under the "terrestrial" conditions.
Hydrogen isotopes  $_{ 1}^4H,~ _1^5H,~ _1^6H$ have short time of life and emit electrons with energy more than 20 Mev. The decay of helium isotopes $_2^6He,~ _2^8He, ~_2^{10}He$ have the times of life, which can reach tenth part of the seconds.

Stars have the time of life about billions years and the lines of their spectrum of masses are not smoothed. Thus we should suppose that there is  some mechanism of stabilization of  neutron-excess nuclei inside  stars. This mechanism is well known - it is neutronization \cite{LL}\S106. It is accepted to think that this mechanism is characteristic for dwarfs with density of particles about $10^{30}$ per cm$^3$ and
pressure of relativistic electron gas
\begin{equation}
{P}\approx\hbar c\cdot n_e^{4/3}\approx 10^{23}dyne/cm^2.\label{Pn}
\end{equation}

The possibility of realization of neutronization in dense plasma is considered below in detail.
At thus, we must try to find an explanation to characteristic features of the star mass spectrum.
At first, we can see that, there is actually quite a small number of stars with $A/Z=2$ exactly.
The question is arising: why there are so few stars, which are composed by very stable nuclei of helium-4? At the same time, there are many stars with $A/Z=4$, i.e. consisting apparently of a hydrogen-4, as well as stars with $A/Z=3/2$, which hypothetically could be composed by another isotope of helium - helium-3.

\subsubsection{The electron cloud of plasma cell}
Let us consider a possible mechanism of the action of the electron gas effect on the plasma nuclear subsystem.
It is accepted to consider  dense plasma to be  divided in plasma cells. These cells are filled by electron gas and they have positively charged nuclei in their centers \cite{Le}.

This construction is non stable from the point of view of the classical mechanics because  the opposite charges collapse is "thermodynamic favorable". To avoid a divergence in the theoretical description of this problem, one can  artificially  cut off the integrating at some small distance characterizing the particles interaction. For example, nuclei can be considered as  hard cores with the finite radii.

It is more correctly, to consider this structure as a quantum-mechanical object and to suppose that the electron can not approach the nucleus closer than its own de Broglie's radius $\lambda_e$.

Let us consider the behavior of the electron gas inside the plasma cell. If to express the number of electrons in the volume $V$ through the density of electron $n_e$, then the maximum value of electron momentum \cite{LL}:
\begin{equation}
p_F=\left(3\pi^2~n_e\right)^{1/3}\hbar.\label{pFn}
\end{equation}

The kinetic energy of the electron gas can be founded from the general expression
for the energy of the Fermi-particles, which fills the volume $V$ \cite{LL}:
\begin{equation}
\mathcal{E}=\frac{Vc}{\pi^2 \hbar^3}\int_0^{p_F}p^2\sqrt{m_e^2
c^2+p^2}dp.
\end{equation}
After the integrating of this expression and the subtracting of the energy at rest, we can calculate the kinetic energy of the electron:
\begin{equation}
\mathcal{E}_{kin}=\frac{3}{8}m_e c^2
\left[\frac{\xi(2\xi^2+1)\sqrt{\xi^2+1}-Arcsinh(\xi)-\frac{8}{3}\xi^3}{\xi^3}\right]\label{ekk}
\end{equation}
(where $\xi=\frac{p_F}{m_ec}$).

The potential energy of an electron is determined by the value of the attached electric field.
The electrostatic potential of this field $\varphi(r)$ must be equal to zero at infinity.\footnote{In general, if there is an uncompensated electric charge inside the cell, then we would have to include it to the potential
$\varphi (r)$. However, we can do not it, because will consider only electro-neutral cell, in which the charge of the nucleus exactly offset by the electronic charge, so  the electric field on the cell border is equal to zero.}
With this in mind, we can write the energy balance equation of electron
\begin{equation}
\mathcal{E}_{kin}=e\varphi(r).\label{ek-p}
\end{equation}
	
The potential energy of an electron at its moving in an electric field of the nucleus  can be evaluated on the basis of the Lorentz transformation \cite{LL2}\S24.		
If in the laboratory frame of reference, where an electric charge placed, it creates an electric potential $\varphi_0$, the potential in the frame of reference moving with velocity $v$ is
\begin{equation}
\varphi=\frac{\varphi_0}{\sqrt{1-\frac{v^2}{c^2}}}.
\end{equation}
Therefore, the potential energy of the electron in the field of the nucleus can be written as:
\begin{equation}
\mathcal{E}_{pot}= -\frac{Ze^2}{r}\frac{\xi}{\beta}\label{epp}.
\end{equation}
Where
\begin{equation}
\beta=\frac{v}{c}.
\end{equation}
and
\begin{equation}
\xi\equiv\frac{p}{m_e c},
\end{equation}
$m_e$ is the mass of electron in the rest.

And one can  rewrite the energy balance Eq.(\ref{ek-p}) as follows:
\begin{equation}
\frac{3}{8}m_ec^2\xi\mathbb{Y}=e\varphi(r)\frac{\xi}{\beta}
\label{ek-p2}.
\end{equation}
where
\begin{equation}
\mathbb{Y}=\left[\frac{\xi(2\xi^2+1)\sqrt{\xi^2+1}-Arcsinh(\xi)-\frac{8}{3}\xi^3}{\xi^4}\right].
\end{equation}
Hence
\begin{equation}
\varphi(r)=\frac{3}{8}\frac{m_ec^2}{e}\beta\mathbb{Y}.\label{fy}
\end{equation}

In according with Poisson's electrostatic equation
\begin{equation}
\Delta\varphi(r)=4\pi e n_e
\end{equation}
or at taking into account that the
electron density is depending on momentum (Eq.(\ref{pFn})),
we obtain
\begin{equation}
\Delta\varphi(r)=\frac{4e}{3\pi}\left(\frac{\xi}
{\lambda_C\hspace{-0.35cm}\widetilde{}}{~~}\right)^3,
\end{equation}
where
$\lambda_C{\hspace{-0.35cm}\widetilde{}}{~~}=\frac{\hbar}{m_e c}$
is the Compton radius.
\vspace{1cm}

At introducing of the new variable
\begin{equation}
\varphi(r)=\frac{\chi(r)}{r},
\end{equation}
we can 	transform the Laplacian:
\begin{equation}
\Delta\varphi(r)=\frac{1}{r}\frac{d^2\chi(r)}{dr^2}.
\end{equation}
As (Eq.(\ref{fy}))
\begin{equation}
\chi(r)=\frac{3}{8}\frac{m_ec^2}{e}\mathbb{Y}\beta r~,
\end{equation}
the differential equation can be rewritten:
\begin{equation}
\frac{d^2\chi(r)}{dr^2}=\frac{\chi(r)}{\mathbb{L}^2}~,\label{dL}
\end{equation}
where
\begin{equation}
\mathbb{L}=\left(\frac{9\pi}{32}\frac{\mathbb{Y}\beta}
{\alpha\xi^3}\right)^{1/2}
\lambda_C\hspace{-0.35cm}\widetilde{}~~~,
\end{equation}
$\alpha=\frac{1}{137}$ is the fine structure constant.
		
This differential equation has the solution:
\begin{equation}
\chi(r)= C\cdot exp\left(-\frac{r}{\mathbb{L}}\right).
\end{equation}
Thus, the equation of equilibrium of the electron gas inside a cell (Eq.(\ref{ek-p2})) obtains the form:
\begin{equation}
\frac{Ze}{ r}\cdot e^{-r/\mathbb{L}}=\frac{3}{8}m_ec^2\beta\mathbb{Y}~~.\label{epp3}
\end{equation}

\subsection{The Thomas-Fermi screening}
Let us consider the case when an ion is placed at the center of a cell, the external shells don't permit  the plasma electron to approach to the nucleus on the distances much smaller than the Bohr radius.
The electron moving is non-relativistic in this case.
At that $\xi\rightarrow 0$,
the kinetic energy of the electron
\begin{equation}
\mathcal{E}_{kin}=\frac{3}{8}m_ec^2\xi\mathbb{Y}\rightarrow \frac{3}{5}E_F~~,
\end{equation}
and the screening length
\begin{equation}
\mathbb{L}\rightarrow \sqrt{\frac{\mathcal{E}_F}{6\pi e^2 n_e}}.
\end{equation}
	Thus, we get the Thomas-Fermi screening in the case of the non-relativistic motion of an electron.
\subsection{The screening with relativistic electrons}
In the case the $\ll $bare$ \gg $ nucleus, there is nothing to prevent the electron to approach it at an extremely small distance $\lambda_{min}$, which is limited by its own than its de Broglie's wavelength. Its movement in this case becomes relativistic at $\beta\rightarrow 1$ è $\xi\gg 1$.
In this case, at not too small $\xi$, we obtain
\begin{equation}
\mathbb{Y}\approx 2\left(1-\frac{4}{3\xi}\right),
\end{equation}
and at $\xi\gg 1$
\begin{equation}
\mathbb{Y}\rightarrow 2 ~.
\end{equation}

In connection with it, at the distance $r\rightarrow \lambda_{min}$ from a nucleus, the equilibrium equation  (\ref{epp3})
reforms to
\begin{equation}
\lambda_{min}\simeq Z\alpha \lambda_C~~.
\end{equation}
and the density of electron gas in a layer of thickness $\lambda_{min}$ can be determined from the condition of normalization.
As there are Z electrons into each cell, so
\begin{equation}
Z\simeq n_e^{\lambda}\cdot{\lambda_{min}}^3
\end{equation}
From this condition it follows that
\begin{equation}
\xi_{\lambda}\simeq \frac{1}{2\alpha Z^{2/3}}\label{xil}
\end{equation}
Where $n_e^\lambda$ and $\xi_{\lambda}$ are the density of electron gas and the relative momentum of electrons at the distance $\lambda_{min}$ from the nucleus.
In accordance with Eq.({\ref{ekk}}), the energy of all the Z electrons in the plasma cell is
\begin{equation}
\mathcal{E}\simeq Zm_ec^2 \xi_\lambda
\end{equation}
At substituting of Eq.({\ref{xil}}), finally we obtain the energy of the electron gas in a plasma cell:
\begin{equation}
\mathcal{E}\simeq \frac{m_e c^2}{2\alpha} Z^{1/3}\label{z13}
\end{equation}
This layer provides the pressure on the nucleus:
\begin{equation}
{P}\simeq \mathcal{E}\left(\frac{\xi}{\lambda_C\hspace{-0.35cm}
\widetilde{}}{~~}\right)^3\approx 10^{23}dyne/cm^2
\end{equation}
This pressure is in order of value with pressure of neutronization (\ref{Pn}).
\subsection{The neutronization}.
The considered above $\ll$ attachment$\gg$ of the electron to the nucleus in a dense plasma should lead to a phenomenon neutronization of the nucleus, if it is energetically favorable. The $\ll$ attached $\gg$ electron  layer should have a stabilizing effect on the neutron-excess nuclei, i.e. the neutron-excess nucleus, which is instable into substance with the atomic structure,  will became stable inside the dense plasma. It explains the stable existence of stars with a large ratio of A/Z.

These formulas allow to answer questions related to the characteristics of the star mass distribution.
The numerical evaluation of energy the electron gas in a plasma cell gives:
\begin{equation}
\mathcal{E}\simeq\frac{m_e c^2}{2\alpha} Z^{1/3}\approx 5 \cdot 10^{-5}Z^{1/3} erg
\end{equation}
The mass of nucleus of helium-4 $M(_2^4He)=4.0026~ a.e.m.$, and the mass hydrogen-4 $M(_1^4H)=4.0278~ a.e.m.$. The mass defect
$\approx 3.8\cdot 10^{-5} egr$.
Therefore, the  reaction
\begin{equation}
_2^4He+e\rightarrow _1^4 H,
\end{equation}
is energetically favorable. At this reaction the nucleus captures the electron from gas  and proton  becomes a neutron.

There is the visible line of stars with the ratio $A/Z=3/2$ at the star mass spectrum. It can be attributed to the stars, consisting of $_2^3He$,~$_4^6Be$,~$_6^9C$, etc.

It is not difficult to verify at direct calculation that the reactions of neutronization  and transforming of $_2^3He$ into $_1^3H$ and $_4^6Be$ into $_3^6Li$ are energetically allowed. So the nuclei $_2^3He$ and $_4^6Be$ should be  conversed by neutronization into  $_1^3H$ and $_3^6Li$. The line $A/Z=3/2$ of the star mass spectrum  can not be formed by these nuclei.
However, the reaction
\begin{equation}
_6^9 C+e\rightarrow _5^9B~,
\end{equation}
is not energetically allowed
and therefore it is possible to believe that the stars of the above line mass spectrum may consist of carbon-9.

\vspace{1.6cm}

It is interesting, is it possible to study this effect under laboratory conditions?
One can use gaseous tritium for it  and can measure the rate of its  decay at two states.
First the rate of tritium decay  at atomic (molecular) state. Second, the same rate  under the effect of ionizing electric discharge. Essentially, one must aim to use the dense gas and the power discharge
in order to ionize more atoms of tritium. Taking into account the preceding consideration, such a test does not
look as a hopeless project.

\clearpage
\section[Addendum II: Other stars]
{Addendum II: Other stars, their classification and some cosmology}
\label{Ch11}

The Schwarzsprung-Rassel diagram is now a generally accepted base for
star classification. It seems that a classification based on the EOS of
substance may be more justified from physical point of view. It can
be  emphasized by possibility to determine the number of classes of
celestial bodies.

The matter can totally have eight states (Fig.({\ref{substA}})).
\begin{figure}
\begin{center}
\hspace{-0.1cm}\includegraphics{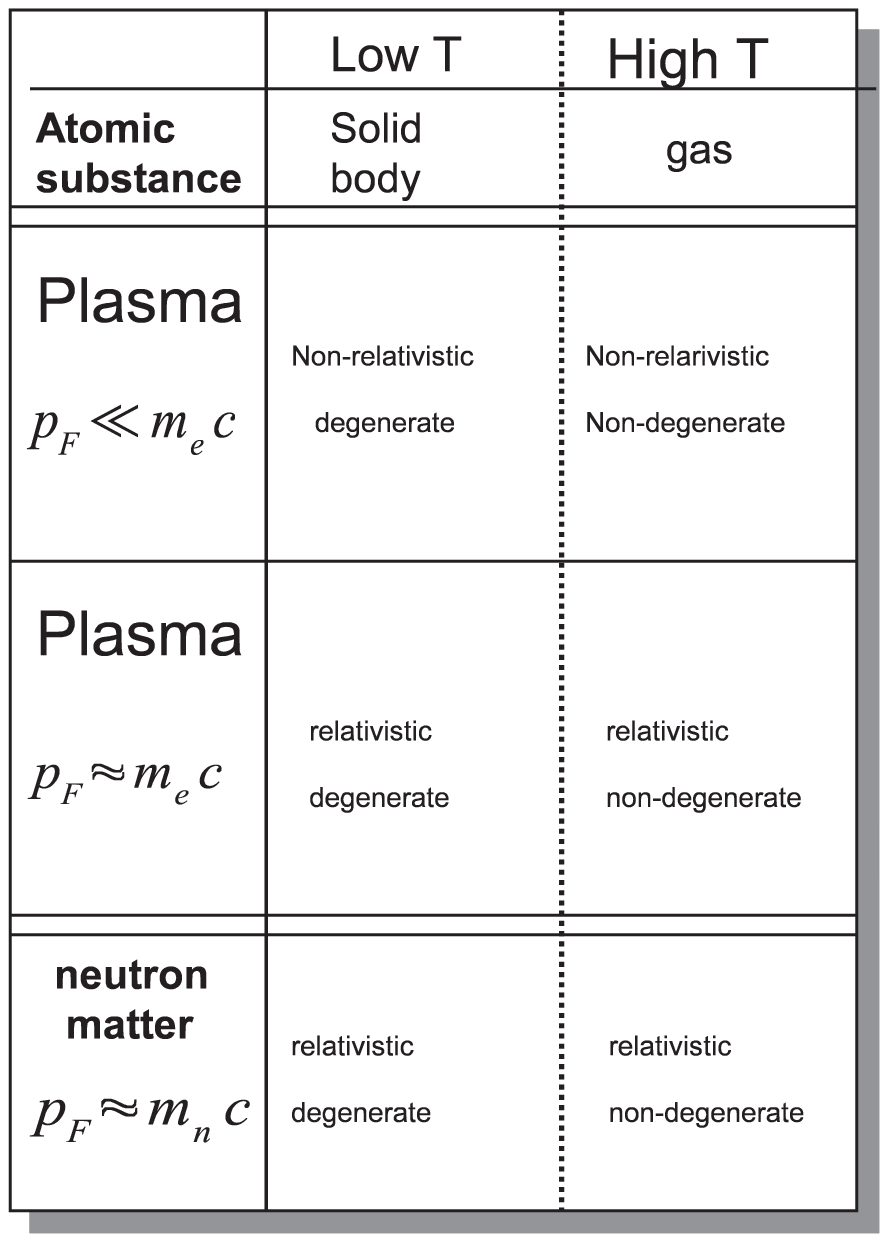}
\caption{The steady states of substance}\label{substA}
\end{center}
\end{figure}

The atomic substance at low temperature exists as  condensed matter
(solid or liquid). At high temperature it transforms into gas phase.

The electron -nuclear plasma can exist in four states. It can be
relativistic or non-relativistic. The electron gas of
non-relativistic plasma can be degenerate (cold) or non-degenerate
(hot). The relativistic electron gas is degenerate at temperature below $T_F$. Very high temperature can remove the degeneration even for relativistic  electrons (if, certainly, electronic gas was not ultra-relativistic originally).

In addition to that, a substance can exist as  neutron matter with the nuclear density approximately.

At present,  assumptions about existence of matter at different
states, other than the above-named,  seem unfounded. Thus, these above-named
states of matter show a possibility of classification  of celestial
bodies in accordance with this dividing.

\subsection{The atomic substance}
\subsubsection{Small bodies} Small celestial bodies - asteroids and
satellites of planets - are usually considered as bodies consisting from
atomic matter.

\subsubsection{Giants} The transformation of atomic matter into plasma can
be induced by action of high pressure, high temperature or both
these factors. If these factors inside a body are not high enough,
atomic substance can transform into gas state. The characteristic
property of this celestial body is absence of electric polarization
inside it. If temperature of a body is below ionization temperature
of atomic substance but higher than its evaporation temperature, the
equilibrium equation comes to
\begin{equation}
-\frac{dP}{dr}=\frac{G\gamma}{r^2}M_r\approx\frac{P}{R}\approx
\frac{\gamma}{m_p} \frac{kT}{R}.
\end{equation}
Thus, the radius of the body
\begin{equation}
R \approx \frac{GM m_p}{kT}.
\end{equation}
If its mass $M\approx 10^{33}~g$ and temperature at its center
$T\approx 10^5~K$, its radius is  $R\approx 10^2~R_{\odot}$. These
proporties are characteristic for giants, where pressure at center
is about $P\approx 10^{10}~din/cm^2$ and it is not enough for substance
ionization.

\subsection{Plasmas}
\subsubsection[Stars]{The non-relativistic non-degenerate plasma. Stars.}
Characteristic properties of hot stars consisting of
non-relativistic non-degenerate plasma was considered above in
detail. Its EOS is ideal gas equation.

\subsubsection[Planet]{Non-relativistic degenerate plasma. Planets.}
At cores of large planets, pressures are large enough to transform
their substance into plasma. As temperatures are not very high here,
it can be supposed, that  this plasma can  degenerate:
\begin{equation}
T<<T_F
\end{equation}
The pressure which is induced by gravitation must be balanced by
pressure of non-relativistic degenerate electron gas
\begin{equation}
\frac{G\mathbb{M}^2}{6\mathbb{R}\mathbb{V}}\approx\frac{(3\pi^2)^{2/3}}{5}
\frac{\hbar^2}{m_e} \biggl(\frac{\gamma}{m_p A/Z}\biggr)^{5/3}
\end{equation}
It opens  a way to estimate the mass of this body:
\begin{equation}
\mathbb{M}\approx\mathbb{M}_{Ch}
\biggl(\frac{\hbar}{mc}\biggr)^{3/2}
\biggl(\frac{\gamma}{m_p}\biggr)^{1/2}
\frac{6^{3/2}9\pi}{4(A/Z)^{5/2}}
\end{equation}

At density about  $\gamma\approx 1~g/cm^3 $, which is characteristic
for large planets, we obtain their masses
\begin{equation}
\mathbb{M}\approx 10^{-3} \frac{\mathbb{M}_{Ch}}{(A/Z)^{5/2}}
\approx\frac{4\cdot 10^{30}}{(A/Z)^{5/2}} ~ g
\end{equation}
Thus, if we suppose that large planets consist of hydrogen
(A/Z=1), their masses must not be over  $4\cdot 10^{30}g$. It is in
agreement with the Jupiter's mass, the biggest planet of the Sun
system.

\subsubsection{The cold relativistic substance}\label{sec11}
The kinetic energy of $N$ relativistic particles ({\ref{ekk}}):
\begin{equation}
\mathcal{E}^{kinetic}=\frac{3}{8}Nmc^2
\left[\frac{\xi(2\xi^2+1)\sqrt{\xi^2+1}-Arcsinh(\xi)-\frac{8}{3}\xi^3}{\xi^3}\right]
\label{e-rs}
\end{equation}
where $\xi=\frac{p_F}{mc}$ and density of substance:
\begin{equation}
n=\frac{p_F^3}{3\pi^2\hbar^3}=\frac{\xi^3}{3\pi^2}
\left(\frac{mc}{\hbar}\right)^3.\label{nn-rs}
\end{equation}
The pressure of this system is
\begin{equation}
P=-\left(\frac{d\mathcal{E}^{kinetic}}{dV}\right)_{S=0}=\frac{mc^2}{8\pi^2}
\left(\frac{mc}{\hbar}\right)^3
\left[\xi\left(\frac{2}{3}\xi^2-1\right)\sqrt{\xi^2+1}+Arcsinh(\xi)\right]
\end{equation}
For simplicity we can suppose that main part of mass of relativistic
star is concentrated into its core and it is uniformly distributed
here. In this case we can write the pressure balance equation:
\begin{equation}
\frac{GM^2}{2R}=3PV.
\end{equation}
It permits to estimate the full number of particles into the
considered system:
\begin{equation}
N=\left(\frac{\hbar
c}{Gm_p^2}\right)^{3/2}\biggl(\frac{3}{4\pi}\biggr)^{2}\frac{1}{\pi^{3/2}}
\left[\frac{\xi\left(\frac{2}{3}\xi^2-1\right)\sqrt{\xi^2+1}+Arcsinh(\xi)}{\xi^4}\right]^{3/2}.
\label{m-puls}
\end{equation}
Accordingly to the virial theorem:
\begin{equation}
\mathcal{E}_{total}=-\mathcal{E}^{kinetic}
\end{equation}
and ({\ref{e-rs}}), the full energy of a relativistic star
\begin{eqnarray}
\mathcal{E}_{total}\sim
-\frac{1}{\xi^9}
\left[\xi(2\xi^2+1)\sqrt{\xi^2+1}-Arcsinh(\xi)-\frac{8}{3}\xi^3\right]\cdot
\nonumber \\
\cdot\left[\xi\left(\frac{2}{3}\xi^2-1\right)\sqrt{\xi^2+1}+Arcsinh(\xi)\right]^{3/2}.\label{e-rt}
\end{eqnarray}
Because this expression has a minimum at $\xi\approx 1.35$, a star
consisting of relativistic substance must have the equilibrium
particle density:
\begin{equation}
n_{\star}\approx 8.3\cdot10^{-2}\left(\frac{mc}{\hbar}\right)^3.
\label{n-rs}
\end{equation}

{\bf{The relativistic degenerate plasma. Dwarfs.}}

The degenerate relativistic plasma has a cold relativistic electron
subsystem. Its nuclear subsystem can be non-relativistic. The star
consisting of this plasma accordingly to Eq.({\ref{n-rs}}) at
$m=m_e$, must have an energy minimum at electron density
\begin {equation}
n_r\approx 1. 4\cdot 10^{30} cm^{-3}\label{n-dw}
\end {equation}
at a star radius
\begin {equation}
R\approx {10^{-2}R_{\odot}}
\end {equation}
It is not difficult to see that these densities and radii are
characteristic for dwarfs.

{\bf{The neutron matter. Pulsars.}} Dwarfs may be
considered as stars where a process of neutronization is just
beginning. At a nuclear density, plasma turns into neutron matter.
\footnote{At nuclear density neutrons and protons are
indistinguishable inside pulsars as inside a huge nucleus. It
permits to suppose a possibility of gravity induced electric
polarization in this matter. }

In accordance with Eq.({\ref{n-rs}}) at $m=m_n$, a star consisting
of neutron matter must have the minimal energy at neutron density
$n_\star\approx 9\cdot 10^{39}$ $particle/cm^3$. As result the
masses of neutron stars must be  approximately equal to $M_{Ch}$.
The measured mass distribution of pulsar composing binary stars
\cite{Thor} is shown on Fig.\ref{pulsar}. It can be considered as a
confirmation of the last conclusion.

\begin{figure}
\begin{center}
\centering\includegraphics[height=10cm]{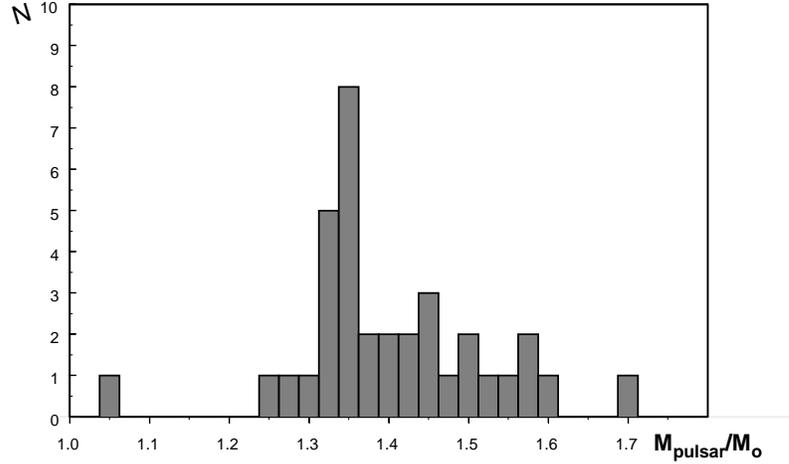} \caption{The
mass distribution of pulsars from binary systems \cite{Thor}.On
abscissa the logarithm of pulsar mass in solar mass is shown.}
\label{pulsar}
\end{center}
\end{figure}

\subsubsection[Quasars]{The hot relativistic plasma.Quasars}
Plasma is hot if its temperature  is higher than degeneration
temperature of its electron gas. The ratio of plasma temperature in
the core of a star to the temperature of degradation of its electron
gas for case of non-relativistic hot star plasma is
(Eq.({\ref{alf}}))
\begin{equation}
\frac{\mathbb{T}_\star}{T_F(n_\star)} \approx 40\label{r40}
\end{equation}
it can be supposed that the same ratio must be characteristic for
the case of a relativistic hot star. At this temperature, the
radiation pressure plays a main role and accordingly the equation of
the pressure balance takes the form:

\begin{equation}
\frac{GM^2}{6RV}\approx\frac{\pi^2}{45}\frac{(k\mathbb{T}_\star)^4}{(\hbar
c)^3}\approx \biggl(\frac{\mathbb{T}_\star}{T_F}\biggr)^3 kTn
\end{equation}

\begin{figure}
\begin{center}
\centering\includegraphics[scale=0.5]{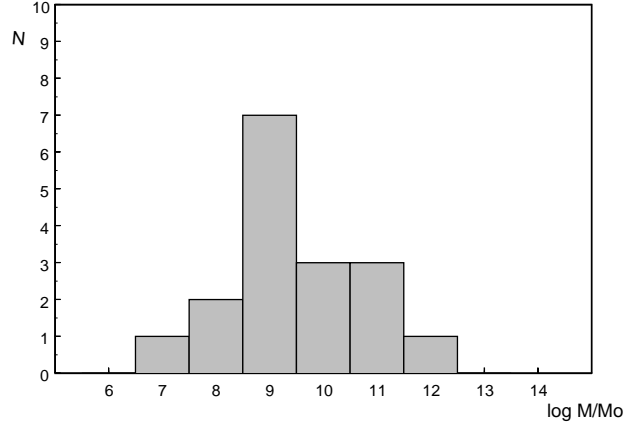}
\caption{The mass distribution of galaxies. \cite{Al}. On the
abscissa, the logarithm of the galaxy mass over the Sun mass is
shown.} \label{galaxy}
\end{center}
\end{figure}

This makes it possible to estimate the mass of a hot relativistic
star
\begin{equation}
M\approx \biggl(\frac{\mathbb{}T_\star}{T_F}\biggr)^6
\biggl(\frac{\hbar c}{G m_p^2} \biggr)^{3/2} m_p \approx 10^9
M_{\odot}
\end{equation}
According to the existing knowledge, among compact celestial objects
only quasars have masses of this level. Apparently it is an
agreed-upon opinion that quasars represent some relatively short
stage of evolution of galaxies. If we adhere to this hypothesis, the
lack of information about quasar mass distribution can be replaced
by the distribution of masses of galaxies
\cite{Al}(Fig.{\ref{galaxy}}). It can be seen, that this
distribution is in a qualitative agreement with supposition that
quasars are composed from the relativistic hot plasma.

Certainly the used estimation Eq.({\ref{r40}}) is very arbitrary. It is possible to expect the existing of  quasars which can have substantially lesser masses.

As the steady-state particle density of relativistic matter $n_r$ is known (Eq.({\ref{n-dw}})), one can estimate the quasar radius:
\begin{equation}
R_{qu}\approx \sqrt[3]{\frac{M_{qu}}{n_r m_p}}\approx 10^{12}cm.
\end{equation}
It is in agreement with the astronomer measuring data obtained from periods of the their luminosity changing.

\subsubsection{About the cosmic object categorizations}
Thus,  it seems possible under some assumptions to find
characteristic parameters of different classes of stars, if to
proceed from EOS of atomic, plasma and neutron substances. The EOS types
can be compared with classes of celestial bodies. As any other
EOS are unknown, it gives a reason to think that all classes of
celestial bodies are discovered (Fig.{\ref{substarA}}).
\begin{figure}
\begin{center}
\hspace{-0.1cm}
\includegraphics{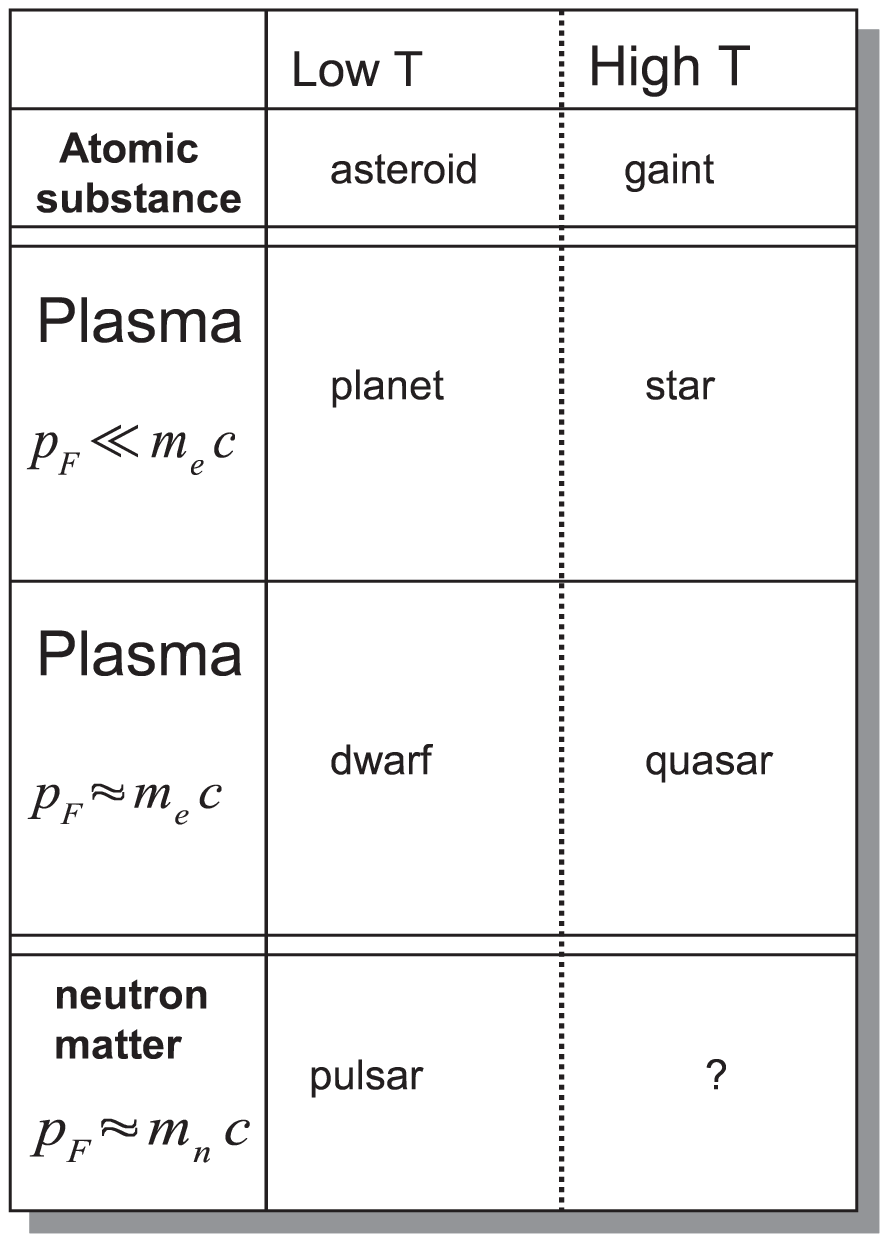}
\end{center}
\caption{The categorization of celestial bodies} \label{substarA}
\end{figure}

\subsection{Several words about star evolution}
There are not formulas in this section, which could serve as a handhold for suggestions. The formulas of the previous sections  can not help too in understanding that, as the evolution of the stars can proceed and as their transitions them from the one class to another are realized, since these formulas were received for the description of the star steady-state. There the comparison of the schemes to the substance categorization (Fig.({\ref{substA}})) and the categorization of star classes (Fig.({\ref{substarA}})) can serve as the base of next suggestions.

At analyzing of these schemes, one can suppose that the evolution of stellar objects is going at the reduction of their temperature. In light of it, it is possible to suppose  the existing of one more body, from which the development began. Really, neutron matter at nuclear density (Eq.({\ref{n-rs}})) is not ultra-relativistic. At this density, corresponding to $p_F\approx mc$, its energy and pressure can depend on the temperature if this temperature it is high enough.\footnote{The
ultra-relativistic matter with $p_F\gg mc$ is possessed  by limiting
pressure which is not depending on temperature.} It seems, there is
no thermodynamical prohibition to imagine this matter so hot when
the neutron gas is non-degenerate. An estimation
shows that it can be possible if mass of this body is near to
$10^{50}g$ or even to $10^{55}g$. As it is accepted to think that
full mass of the Universe is about $10^{53}g$, it can be assumed
that on an early stage of development of Universe, there was some
celestial body with the mass about  $10^{53}g$ composed by the neutron matter at the approximately nuclear density with at the temperature above $10^{12}K$.
After some time, with temperature decreased it has lost its
stability and decayed into quasars with mass up to $10^{12}M_{Ch}$,
consisting of the non-degenerate relativistic plasma  at
$T>10^{10}K$. After next cooling at loosing of stability they was
decaying on galaxies of hot stars with mass about $M\approx M_{Ch}$
and core temperature about $T\approx 10^{7}K$, composed by
non-relativistic hot plasma. A next cooling must leads hot stars to
decaying on dwarfs, pulsars, planets or may be on small bodies. The
substances of these bodies (in their cores) consists of degenerate
plasma (degenerate electron subsystem and cold nuclear subsystem) or
cold neutron matter, it makes them stable in expanding and cooling
Universe.\footnote{The temperature of plasma inside these bodies can
be really quite high as electron gas into dwarfs, for example, will
be degenerate even at temperature $T\approx 10^{9}K$.}
It is important to emphasize, that the  gravity induced electric polarization excludes the possibility of the gravitational collapse as the last stage of the star evolutions.

\subsection{About $\ll$black holes$\gg$ }
It seems that the idea about the  $\ll$black holes$\gg$ existence is organic related to  the suggestion about an inevitable collapse of  large cosmic bodies on the last stage of their evolutions.
However, the models of collapsing masses  were appearing as a consequence of the rejection from attention of the gravity induced electric polarization of the intra-stellar plasma, which was considered in previous chapters. If to take into account this  mechanism,  the possibility of collapse must be  excludes. It  allows newly to take a look on the $\ll$black hole$\gg$ problem.

In accordance with the standard approach, the Schwarzshield radius of  $\ll$black hole$\gg$ with M is
\begin{equation}
r_{bh}=\frac{2GM}{c^2}
\end{equation}
and accordingly the average density of $\ll$black holes$\gg$:
\begin{equation}
\gamma_{bh}=\frac{3c^6}{32\pi M^2 G^3}\label{gamma}.
\end{equation}

The estimations which was made in previous chapters are showing that all large inwardly-galactic objects of all classes - a stars, dwarves, pulsars, giants - possess the mass of the order $M_{Ch}$ (or 10$M_{Ch}$). As the density of these  objects are small relatively to the limit (\ref{gamma}). As result, a searching of $\ll$black hole$\gg$
inside  stellar objects of our Galaxy seems as hopeless.

On the other hand, stellar objects, consisting of hot relativistic plasma - a quasars, in accordance with their  mass and density, may stay $\ll$black holes$\gg$. The process of collapse is not needed for their creation.
As the quasar mass $M_{qu}\gg M_{Ch}$, all other stellar objects must organize
their moving around it and one can suppose that a $\ll$black hole$\gg$ can exist at the center of our Galaxy.

\clearpage

\section
{The conclusion} \label{Ch12}

Evidently, the  main conclusion from the above consideration
consists in statement of the fact that now there are quite enough measuring
data to place the theoretical astrophysics on a reliable foundation.
All above measuring data are known  for a relatively long time. The
traditional system of view based on the Euler equation in the form
({\ref{Eu}}) could not give a possibility to explain  and even to consider. with due
proper attention, to  these data. Taking into account  the gravity
induced electric polarization of plasma and  a change starting
postulate gives a possibility to obtain results for  explanation of
 measuring data considered above.

\hspace{1cm}

Basically these results are the following.

\hspace{1cm}

Using the standard method of plasma description leads to the conclusion
that at conditions characteristic for the central stellar region, the plasma
has the minimum energy  at constant density
$n_\star$ (Eq.(\ref{eta1})) and at the constant temperature  $\mathbb{T}_\star$
(Eq.(\ref{tcore})).

\hspace{1cm}

This plasma forms the core of a star, where the pressure is constant and
gravity action is balanced by the force of the gravity induced by the electric polarization.
The virial theorem gives a possibility to calculate the stellar core mass
$\mathbb{M}_\star$ (Eq.(\ref{Mcore})) and its radius $\mathbb{R}_\star$
({\ref{Rcore}}). At that the stellar core volume is approximately equal to  1/1000 part of full volume of a star.

\hspace{1cm}

The remaining mass of a star located over the core has a density
approximately thousand times smaller and it is convenient to name it
a star atmosphere. At using thermodynamical arguments, it is
possible to obtain the radial dependence of plasma density inside
the atmosphere $n_a\approx r^{-6}$ (Eq.({\ref{an-r}})) and the
radial dependence of its temperature $\mathbb{T}_a\approx r^{-4}$
(Eq.({\ref{tr}})).

\hspace{1cm}

It gives a possibility to conclude that  the mass of the stellar
atmosphere  $\mathbb{M}_a$ (Eq.({\ref{ma}})) is almost exactly equal
to the stellar core mass.Thus, the full stellar mass can be
calculated. It depends on the ratio of the mass and the charge of
nuclei composing the plasma. This claim is in a good agreement with
the measuring data of the mass distribution of both -  binary stars
and close binary stars (Fig.({\ref{starM}})-({\ref{starM2}}))
\footnote{The measurement of parameters of these stars has a
satisfactory  accuracy only.}. At that it is important that the
upper limit of masses of both - binary stars and close binary stars
- is in accordance   with the calculated value of the mass of the
hydrogen star (Eq.({\ref{M}})). The obtained formula explains the
origin of sharp peaks of stellar mass distribution - they evidence
that the substance of these stars have a certain value of the ratio
$A/Z$. In particular the solar plasma according to  (Eq.({\ref{M}}))
consists of nuclei with $A/Z=5$.

\hspace{1cm}

Knowing temperature and substance density on the core and knowing
their radial dependencies, it is possible to estimate the surface
temperature  $\mathbb{T}_0$ (\ref{T0}) and the radius of a star
$\mathbb{R}_0$ ({\ref{R0}}). It turns out that these measured
parameters must be related to the star mass with the  ratio
$\mathbb{T}_0 \mathbb{R}_0\sim \mathbb{M}^{5/4}$ ({\ref{5/4}}). It
is in a good agreement with measuring data (Fig.({\ref{RT-M}})).

\hspace{1cm}

Using another thermodynamical relation - the Poisson's adiabat -
gives a way to determine the relation between radii of stars and
their masses $\mathbb{R}_0^3\sim \mathbb{M}^2$ (Eq.({\ref{rm23}})),
and between their surface temperatures and masses  $\mathbb{T}_0\sim
\mathbb{M}^{5/7}$ (Eq.({\ref{tm}})). It gives the quantitative
explanation of the mass-luminosity dependence (Fig.({\ref{LM}})).

\hspace{1cm}

According to another familiar Blackett's dependence, the
giromagnetic ratios of celestial bodies are approximately equal to
$\sqrt{G}/c$. It has a simple explanation too. When there is the
gravity induced electric polarization of a substance of a celestial
body, its rotation must induce a magnetic field
(Fig.({\ref{black}})). It is important that all (composed by
eN-plasma) celestial bodies - planets, stars, pulsars - obey the
Blackett's dependence. It confirms a consideration that the gravity
induced electric polarization must be characterizing for all kind of
plasma. The calculation of magnetic fields of hot stars shows that
they must be proportional to rotation velocity of stars
({\ref{Ht}}). Magnetic fields of Ap-stars are measured, and they can
be compared with periods of changing of luminosity of these stars.
It is possible that this mechanism is characteristic for stars with
rapid rotation (Fig.({\ref{H-W}})), but obviously there are other
unaccounted factors.

\hspace{1cm}

Taking into account the gravity induced electric polarization and
coming from the Clairault's theory, we can describe the periastron
rotation of binary stars as effect descended from non-spherical
forms of star cores. It gives the quantitative explanation of this
effect, which is in a good agreement with measuring data
(Fig.({\ref{periastr}})).

\hspace{1cm}

The solar oscillations can be considered as elastic vibrations of
the solar core. It permits to obtain two basic frequencies of this
oscillation: the basic frequency of sound radial oscillation of the
core and the frequency of splitting depending on oscillations of
substance density near its equilibrium value (Fig.({\ref{soho}})).

\hspace{1cm}

The plasma can exists in four possible states. The  non-relativistic
electron gas of plasma can be degenerate and non-degenerate. Plasma
with relativistic electron gas can have a cold and a hot nuclear
subsystem.  Together with the atomic substance and neutron
substance, it gives seven possible states. It suggests a  way of a
possible classification of celestial bodies. The advantage of this
method of classification is in the possibility to estimate
theoretically main parameters characterizing the celestial bodies of
each class. And these predicted parameters are in agreement with
astronomical observations. It can be supposed hypothetically  that
cosmologic transitions between these classes go in direction of
their temperature being lowered. But these suppositions have no
formal base at all.

\hspace{1cm}

Discussing  formulas obtained, which  describe star properties, one
can note a important moment: these considerations  permit to look at
this problem from different points of view. On the one hand, the
series of conclusions follows from existence of spectrum of star
mass (Fig.({\ref{starM}})) and from known chemical composition
dependence. On the another hand, the calculation of natural
frequencies of the solar core gives a different approach to a
problem of chemical composition determination. It is important that
for the Sun, both these approaches lead to the same conclusion
independently and unambiguously. It gives a confidence in
reliability of obtained results.

\hspace{1cm}

In fact, the calculation of magnetic fields of hot stars
agrees with measuring data  on order of the value only. But one must not expect the best agreement in this case because calculations were made for the case of a spherically symmetric model and measuring data are obtained for stars where this symmetry is obviously violated.  But it is important, that the all remaining measuring data  (all known of today) confirm both - the new postulate and
formulas based on it. At that, the main stellar parameters - masses,
radii and temperatures - are expressed through combinations of world
constants and at that they show a good accordance with observation
data. It is important  that quite a satisfactory quantitative
agreement of obtained results and measuring data can be achieved by
simple and clear physical methods without use of any fitting
parameter. It gives a special charm and attractiveness to  star
physics.

\clearpage

\appendix
\begin{center}
\bf{Appendix}
\end{center}
\begin{center}{\bf{The Table of main parameters of close binary stars}}\\
\scriptsize{(cited on Kh.F.Khaliullin's dissertation,
Sternberg Astronomical Institute.
In Russian)}\end{center}

\hspace{-5.0cm}
{\tiny
{
\begin{tabular}
{||r|l|r|r|r|r|r|r|r|r|c||}\hline \hline
   & & & & & & & & & & \\
  N & \tiny{Name of star} & {U} & {P} & $\mathbb{M}_1/\mathbb{M}_{\odot}$ &
  $\mathbb{M}_2/\mathbb{M}_{\odot}$ &  $\mathbb{R}_1/\mathbb{R}_{\odot}$ &
  $\mathbb{R}_2/\mathbb{R}_{\odot}$& $\mathbb{T}_1$ & $\mathbb{T}_2$ &\tiny{References}\\
& &\tiny{period of}&\tiny{period of}&\tiny{mass of}&\tiny{mass of}&\tiny{radius of}&\tiny{radius of}&\tiny{temperature} &\tiny{temperature}& \\
&
&\tiny{apsidal}&\tiny{ellipsoidal} &\tiny{component 1},&\tiny{component 2},&\tiny{1 component}& \tiny{2 component}&\tiny{of}&\tiny{of}& \\
& &\tiny{rotation},&\tiny{rotation},&\tiny{in}&\tiny{in}&\tiny{in}&\tiny{in}&\tiny{1 component,}&\tiny{2 component}, & \\
& &\tiny{years} & \tiny{days}&\tiny{the Sun mass}&\tiny{the Sun mass} &\tiny{the Sun radius} &\tiny{the Sun
radius} &\tiny{K} &\tiny{K}&
\\\hline
  1 & BW Aqr & 5140 & 6.720 & 1.48 & 1.38 &  1.803 & 2.075 & 6100 & 6000 & 1,2\\
  2 & V 889 Aql & 23200 & 11.121 & 2.40 & 2.20& 2.028 & 1.826 & 9900 & 9400 & 3,4 \\
  3 & V 539 Ara & 150 & 3.169 & 6.24 & 5.31 &  4.512 & 3.425 & 17800 & 17000  &5,12,24,67 \\
  4 & AS Cam & 2250 & 3.431 & 3.31 & 2.51 &   2.580 & 1.912 & 11500 & 10000 & 7,13\\
  5 & EM Car & 42 & 3.414 & 22.80 & 21.40 &   9.350 & 8.348 & 33100 & 32400 & 8\\
  6 & GL Car & 25 & 2.422 & 13.50 & 13.00 &   4.998 & 4.726 & 28800 & 28800 & 9\\
  7 & QX Car & 361 & 4.478 & 9.27 & 8.48 &   4.292 & 4.054 & 23400 & 22400  & 10,11,12\\
  8 & AR Cas & 922 & 6.066 & 6.70 & 1.90 &   4.591 & 1.808 & 18200 & 8700 & 14,15\\
  9 & IT Cas & 404 & 3.897 & 1.40 & 1.40 &   1.616 & 1.644 & 6450 & 6400 & 84,85\\
  10 & OX Cas & 40 & 2.489 & 7.20 & 6.30 &   4.690 & 4.543 & 23800 & 23000 & 16,17\\
  11 & PV Cas & 91 & 1.750 & 2.79 & 2.79 &   2.264 & 2.264 & 11200 & 11200  & 18,19\\
  12 & KT Cen & 260 & 4.130 & 5.30 & 5.00 &  4.028 & 3.745 & 16200 & 15800 & 20,21\\
  13 & V 346 Cen & 321 & 6.322 & 11.80 & 8.40  & 8.263 & 4.190 & 23700 & 22400 & 20,22\\
  14 & CW Cep & 45 & 2.729  & 11.60 & 11.10 &  5.392 & 4.954 & 26300 & 25700  & 23,24\\
  15 & EK Cep & 4300 & 4.428 & 2.02 & 1.12 &  1.574 & 1.332 & 10000  & 6400  &25,26,27,6\\
  16 & $\alpha$ Cr B & 46000 & 17.360 & 2.58 & 0.92  & 3.314 & 0.955 & 9100 & 5400 & 28,29\\
  17 & Y Cyg & 48 & 2.997 & 17.50 & 17.30 &   6.022 & 5.680 & 33100 & 32400 & 23,30\\
  18 & Y 380 Cyg & 1550 & 12.426 & 14.30 & 8.00   & 17.080 & 4.300 & 20700 & 21600 &31 \\
  19 & V 453 Cyg & 71 & 3.890 & 14.50 & 11.30 &  8.607 & 5.410 & 26600 & 26000 &17,32,33\\
  20 & V 477 Cyg & 351 & 2.347 & 1.79 & 1.35 &   1.567 & 1.269 & 8550 & 6500 & 34,35\\
  21 & V 478 Cyg & 26 & 2.881 & 16.30 & 16.60 &  7.422 & 7.422 & 29800 & 29800 & 36,37\\
  22 & V 541 Cyg & 40000 & 15.338 & 2.69 & 2.60  & 2.013 & 1.900 & 10900 & 10800 & 38,39\\
  23 & V 1143 Cyg & 10300 & 7.641 & 1.39 & 1.35  & 1.440 & 1.226 & 6500 & 6400 & 40,41,42\\
  24 & V 1765 Cyg & 1932 & 13.374 & 23.50 & 11.70   & 19.960 & 6.522 & 25700 & 25100 & 28\\
  25 & DI Her & 29000 & 10.550  & 5.15 & 4.52 &   2.478 & 2.689 & 17000 & 15100 & 44,45,46,47\\
  26 & HS Her & 92 & 1.637  & 4.25 & 1.49 &   2.709 & 1.485 & 15300 & 7700 & 48,49\\
  27 & CO Lac & 44 & 1.542 & 3.13 & 2.75 &   2.533 & 2.128 & 11400 & 10900 & 50,51,52\\
  28 & GG Lup & 101 & 1.850 & 4.12 & 2.51 &  2.644 & 1.917 & 14400 & 10500 & 17\\
  29 & RU Mon & 348 & 3.585 & 3.60 & 3.33 &  2.554 & 2.291 & 12900 & 12600  & 54,55\\
  30 & GN Nor & 500 & 5.703 & 2.50 & 2.50 &  4.591 & 4.591 & 7800 & 7800  & 56,57\\
  31 & U Oph & 21 & 1.677 & 5.02 & 4.52 &   3.311 & 3.110 & 16400 & 15200 & 53,58,37\\
  32 & V 451 Oph & 170 & 2.197 & 2.77 & 2.35  & 2.538 & 1.862 & 10900 & 9800 & 59,60\\
  33 & $\beta$ Ori & 228 & 5.732 & 19.80 & 7.50   & 14.160 & 8.072 & 26600 & 17800 & 61,62,63\\
  34 & FT Ori & 481 & 3.150 & 2.50 & 2.30 &   1.890 & 1.799 & 10600 & 9500 & 64\\
  35 & AG Per & 76 & 2.029 & 5.36 & 4.90 &  2.995 & 2.606 & 17000 & 17000 &23,24\\
  36 & IQ Per & 119 & 1.744 & 3.51 & 1.73 &  2.445 & 1.503 & 13300 & 8100 &65,66\\
  37 & $\zeta$ Phe & 44 & 1.670 & 3.93 & 2.55  & 2.851 & 1.852 & 14100 & 10500 &11,67 \\
  38 & KX Pup & 170 & 2.147 & 2.50 & 1.80 &   2.333 & 1.593 & 10200 & 8100 & 21\\
  39 & NO Pup & 37& 1.257 & 2.88 & 1.50 &   2.028 & 1.419 & 11400 & 7000 &11,69\\
  40 & VV Pyx & 3200 & 4.596 & 2.10 & 2.10  & 2.167 & 2.167 & 8700 & 8700 &70,71\\
  41 & YY Sgr & 297 & 2.628 & 2.36 & 2.29 & 2.196 & 1.992 & 9300 & 9300 & 72\\
  42 & V 523 Sgr & 203 & 2.324 & 2.10 & 1.90   & 2.682 & 1.839 & 8300 & 8300 &73 \\
  43 & V 526 Sgr & 156 & 1.919 & 2.11 & 1.66 &   1.900 & 1.597 & 7600 & 7600 &74\\
  44 & V 1647 Sgr & 592 & 3.283 & 2.19 & 1.97 &  1.832 & 1.669 & 8900 & 8900 & 75\\
  45 & V 2283 Sgr & 570 & 3.471 & 3.00 & 2.22 &  1.957 & 1.656 & 9800 & 9800 & 76,77\\
  46 & V 760 Sco & 40 & 1.731 & 4.98 & 4.62 &   3.015 & 2.642 & 15800 & 15800 & 78\\
  47 & AO Vel & 50 & 1.585 & 3.20 & 2.90 &   2.623 & 2.954 & 10700 & 10700 & 79\\
  48 & EO Vel & 1600 & 5.330 & 3.21 & 2.77 &  3.145 & 3.284 & 10100 & 10100 & 21,63\\
  49 & $\alpha$ Vir & 140 & 4.015 & 10.80 & 6.80  & 8.097 & 4.394 & 19000 & 19000 &80,81,68 \\
  50 & DR Vul & 36 & 2.251 & 13.20 & 12.10 &   4.814 & 4.369 & 28000 & 28000 & 82,83\\ \hline\hline
\end{tabular}}}


\clearpage


\markboth{Bibliography}{Bibliography}
\addcontentsline{toc}{chapter}{Bibliography}

\begin{thebibliography}{18}


\bibitem {Al}   Allen C.W. - Astrophysical quantities, 1955,University of London, The Athlone Press.

\bibitem{Beskin} Beskin V.S., Gurevich A.V., Istomin Ya.N.:
$\newline{\textit{Physics of the Pulsar Magnetosphere }}$ (Cambridge
University Press) (1993)

\bibitem{Blackett} Blackett P.M.S.: $\textit{Nature}, \bf{159}$, 658, (1947)


\bibitem{peri2} Chandrasekhar S.: Monthly Notices of the RAS {\bf{93}}, 449, (1933)

\bibitem{CD} Christensen-Dalsgaard, J.: Stellar oscillation, Institut for
Fysik og Astronomi, Aarhus Universitet, Denmark, (2003)

\bibitem{bison}    {Elsworth, Y. at al.} - In Proc. {\cal GONG'94} Helio- and Astero-seismology from Earth and Space,
eds. Ulrich,R.K., Rhodes Jr,E.J. and D\"{a}ppen,W., Asrtonomical
Society of the Pasific  Conference Series, vol.76, San
Fransisco,{\bf 76}, 51-54.


\bibitem{Heintz} Heintz W.D.: $\textit{Double stars}$ In Geoph. and Astroph. monographs, {\bf{15}},
D.Reidel Publ. Corp., (1978)

\bibitem{Kh}  Khaliullin K.F.: $\textit {Dissertation}$, Sternberg
Astronomical Institute, Moscow, (Russian)(2004) (see Table in
Appendix on http://astro07.narod.ru)

\bibitem{LL} Landau L.D. and Lifshits E.M.: Statistical Physics,
{\bf{1}}, 3rd edition, Oxford:Pergamon, (1980)

\bibitem{LL8} Landau L.D. and Lifshits E.M.: Electrodynamics of condensed matter,
{\bf{1}}, 3rd edition, Oxford:Pergamon, (1980)

\bibitem{LL2} Landau L.D. and Lifshits E.M.: The Classical Theory of Fields.
{\bf{1}}, Pergamon Press, N.Y. (1971)

\bibitem{Le} Leung Y.C.:  $\textit{Physics of Dense Matter}$  In Science Press/World Scientific,
Beijing and Singapore, (1984)

\bibitem{rom}   {I.I.Romanyuk  at al.} Magnetic Fields of Chemically Peculiar and Related Stars,
Proceedings of the International Conference (Nizhnij Arkhyz, Special
Astrophysical Observatory of Russian Academy of Sciences, September
24-27, 1999), eds: Yu. V. Glagolevskij and I.I. Romanyuk, Moscow,
2000, pp. 18-50.

\bibitem{peri1} Russel H.N.: Monthly Notices of the RAS {\bf{88}}, 642, (1928)

\bibitem{Sirag} Sirag S.-P.: $\textit{Nature}, \bf{275}$, 535, (1979)


\bibitem{soho} Solar Physics, {\bf{175/2}}, (http://sohowww.nascom.nasa.gov/gallery/Helioseismology)


\bibitem{Thor} Thorsett S.E. and Chakrabarty D.: $\textit{E-preprint: astro-ph/9803260}$, (1998)



\bibitem{VL} Vasiliev B.V. and Luboshits V.L.: $\textit{Physics-Uspekhi}, \bf{37}$, 345, (1994)
\bibitem{V-01} Vasiliev B.V.: \textit{Nuovo Cimento B}, {\bf{116}}, pp.617-634,
(2001)



\end{thebibliography}

\clearpage

\begin{thebibliography}{85}


\bibitem {KhKo}   Khaliulilin Kh.F. and Kozyreva V.S.

\textit{Apsidal motion in the eclipsing binary system of BW Aqr}

Astrophys. and Space Sci., {\bf{120}}  (1986) 9-16.



\bibitem {Imbert}   Imbert M.

\textit{Photoelectric radial velosities of eclipsing binaries. IV.
Orbital elements of BW Aqr},

Astron.Astrophys.Suppl., {\bf {69}} (1987) 397-401.





\bibitem {KhKh}   Khaliulilin Kh.F. and Khaliulilina A.I.

\textit{Fotometricheskoe issledovanie zatmenno-dvoinoi sistemy s
relativistskim vrasheniem orbity V889 Aql,}

Astronom.zh.,{\bf{66}}(1989)76-83 (in Russian).




\bibitem {KhKh87}
Khaliulilin Kh.F. and Khaliulilina A.I.

\textit{K probleme vrashenia linii apsid v zatmennoi sisteme V889
Aql},

Astron.cirk., {\bf N{1486}}  (1987) 5-7 (in Russian).




\bibitem {Clausen}
Clausen J.V.

\textit{V 539 Arae: first accurate dimensions of eclipsing
binaries},

Astron.Astrophys., {\bf {308}} (1996) 151-169.




\bibitem {LavLav}
Lavrov M.I. and Lavrova N.V.

\textit{Revisia elementov fotometricheskoi orbity EK Cep},

Astron.cirk. {\bf {971}} (1977) 3-4 (in Russian).




\bibitem {KhKo94}   Khaliulilin Kh.F. and Kozyreva V.S.
\textit{Apsidal motion in the eclipsing binary AS Cam},

Astrophys. and Space Sci., {\bf{120}}  (1994) 115-122.




\bibitem {Anders89}   Andersen J. and Clausen J.V.,
\textit{Absolute dimensions of eclipsing binaries.XV. EM Cainae},

Astron.Astrophys. {\bf{213}} (1989) 183-194.




\bibitem {Gemenez}   Gemenez A. and Clausen J.V.,

\textit{Four-color photometry of eclipsing binaries. XXIIA.
Photometric elements and apsidal motion of GL Cainae},

Astron.Astrophys. {\bf{161}} (1986) 275-286.




\bibitem {Anders83}   Andersen J., Clausen J.V., Nordstrom B. and
Reipurth B.,

\textit{Absolute dimensions of eclipsing binaries.I. The early-type
detached system QX Cainae},

Astron.Astrophys. {\bf{121}} (1983) 271-280.




\bibitem {Gemenez86b}   Gemenez A., Clausen J.V. and Jensen K.S.

\textit{Four-color photometry of eclipsing binaries. XXIV. Aspidal
motion of QX Cainae, $\xi$ Phoenicis and NO Puppis},

Astron.Astrophys. {\bf{159}} (1986) 157-165.




\bibitem {DeGreve}   De Greve J.P.

\textit{Evolutionary models for detached close binaries: the system
Arae and QX Cainae},

Astron.Astrophys. {\bf{213}} (1989) 195-203.




\bibitem {Malony}   Malony F.P., Guinan E.F. and Mukherjec J.

\textit{Eclipsing binary star as test of gravity theories}

Astron.J. {\bf{102}} (1991) 256-261.



\bibitem {Mossak92}   Mossakovskaya L.V.

\textit{New photometric elements of AR Cas, an eclipsing binary
system with apsidal motion}

Astron. and Astroph. Trans. {\bf{2}} (1992) 163-167.



\bibitem {Haffer62}   Haffer C.M. and Collins G.M.

\textit{Computation of elements of eclipsing binary stars by
high-speed computing machines}

Astroph.J.Suppl., {\bf{7}} (1962) 351-410.





\bibitem {Crink89}   Crinklaw G. and Etzel P.

\textit{A photometric analisis of the eclipsing binary OX
Cassiopeiae}

Astron.J. {\bf{98}} (1989) 1418-1426.




\bibitem {Claret93}   Claret A. and Gimenez A.

\textit{The aspidal motion test of the internal stellar structure:
comparision between theory and observations}

Astron.Astroph. {\bf{277}} (1993) 487-502.




\bibitem {Wolf95a}   Wolf M.

\textit{Aspidal motion in the eclipsing binary PV Cassiopeiae}

Monthly Not.Roy.Soc. {\bf{286}} (1995) 875-878.




\bibitem {Popper87}   Popper D.M.

\textit{Rediscussion of eclipsing binaries.XVII.The detached early A
type binaries PV Cassiopeae and WX Cephei}

Astron.J. {\bf{93}} (1987) 672-677.



\bibitem {LavLav85}
Lavrov M.I. and Lavrova N.V.

\textit{Revisia fotometrichestih elementov u zatmennyh dvoinyh
sistem s ekscentricheskimi orbitami.2.KT Cen}

Trudy Kaz.Gor.AO {\bf {49}} (1985) 18-24 (in Russian).



\bibitem {Soder75}   Soderhjelm S.

\textit{Observations of six southern eclipsing  binaries for apsidal
motion}

Astron.Astroph.Suppl.Ser {\bf{22}} (1975) 263-283.



\bibitem {Gemenez86a}   Gemenez A., Clausen J.V. and Anderson J.

\textit{Four-color photometry of eclipsing binaries. XXIA.
Photometric  analysis and aspidal motion study of V346 Centauri},

Astron.Astrophys. {\bf{160}} (1986) 310-320.



\bibitem {Gemenez87}   Gemenez A., Chun-Hwey Kim and Il-Seong Nha

\textit{Aspidal motion in the early-type eclipsing binaries CW
Cephei, Y Cyg and AG Per}

Montly .Not.Roy.Astron.Soc. {\bf{224}} (1987) 543-555.



\bibitem {Boz}   Bocula R.A.

\textit{Peresmotr elementov fotometricheskoi orbity zatmennyh sistem
CW Cep, V 539 Ara, AG Per, AR Aur, RS Cha, ZZ Boo.}

Peremennye zvezdy{\bf{21}} (1983) 851-859 (in Russian).



\bibitem {Kh83}   Khaliulilin Kh.F.

\textit{Relativistskoe vrashenie orbity zatmennoi dvoinoi sistemy EK
Cep}

Astron.zh.{\bf{60}} (1983) 72-82 (in Russian).



\bibitem {Tomkin}   Tomkin J.

\textit{Secondaries of eclipsing binary. V. EK Cephei}

Astroph.J. {\bf{271}} (1983) 717-724.



\bibitem {Claret}   Claret A., Gemenez A. and Martin E.L.

\textit{A test case of stellar evolution the  eclipsing binary EK
Cephei}

Astron.Astroph. {\bf{302}} (1995) 741-744.



\bibitem {Volk}   Volkov I.M.

\textit{The discovery of apsidal motion in the  eclipsing binary
system $\alpha$ Cr B}

Inf.Bull.Var.Stars N3876,(1993) 1-2.



\bibitem {Quiroga}   Quiroga R.J., van L.P.R.

\textit{Angular momenta in binary systems}

Astroph.Space Sci. {\bf{146}} (1988) 99-137.



\bibitem {Hill95}   Hill G. and Holmgern D.E.

\textit{Studies of early-type varieble stars}

Asrton.Astroph.{\bf{297}} (1995) 127-134.



\bibitem {Hill84}   Hill G. and Batten A.H.

\textit{Studies of early-type varieble stars.III. The orbit and
physical dimensions for V 380 Cygni}

Asrton.Astroph.{\bf{141}} (1984) 39-48.



\bibitem {Zak}   Zakirov M.M.

\textit{Ob apsidalnom dvizhenii v dvoinoi sisteme  V 453 Cyg}

Astron.cirk.N1537,21 (in Russian).



\bibitem {Kar}   Karetnikov V.G.

\textit{Spectral investigation of eclipsing binary stars at the
stage of mass exchange}

Publ.Astron.Inst.Czech.{\bf{70}} (1987) 273-276.



\bibitem {Mossak87}   Mossakovskaya L.V. and Khaliulilin Kh.F.

\textit{Prichina anomalnogo apsidalnogo dvizhenia v sisteme V 477
Cyg}

Astron.cirk.N1536, 23-24 (in Russian).



\bibitem {Gemenez92}   Gemenez A. and Quintana J.M.

\textit{Apsidal motion and revised photometry elements of the
eccentric eclipsing binary V 477 Cyg}

Astron.Astrophys. {\bf{260}} (1992) 227-236.



\bibitem {Mossak96}   Mossakovskaya L.V. and Khaliulilin Kh.F.

\textit{Vrashenie linii apsid v sisteme  V 478 Cyg}

Pisma v Astron.zh.{\bf{22}} (1996) 149-152.



\bibitem {Popper91}   Popper D.M. and Hill G.

\textit{Rediscussion of eclipsing binaries.XVII.Spectroscopic orbit
of OB system with a cross-correlation procedure}

Astron.J. {\bf{101}} (1991) 600-615.



\bibitem {Kh85}   Khaliulilin Kh.F.

\textit{The unique eclipsing binary system V 541 Cygni with
relativistic apsidal motion}

Astrophys.J. {\bf{229}} (1985) 668-673.



\bibitem {Lin}   Lines R.D.,Lines H., Guinan E.F. and Carroll

\textit{Time  of minimum determination  of eclipsing binary V 541
Cygni}

Inf.Bull.Var.Stars N3286,1-3.



\bibitem {Kh83b}   Khaliulilin Kh.F.

\textit{Vrashenie linii apsid v zatmennoi sisteme V 1143 Cyg}

Asrton. cirk.N1262,1-3 (in Russian).



\bibitem {Anders87}   Andersen J., Garcia J.M., Gimenes A. and
Nordstom B.

\textit{Absolute dimension of eclipsing binaries.X. V1143 Cyg}

Astron.Astrophys. {\bf{174}} (1987) 107-115.



\bibitem {Burns}   Burns J.F., Guinan E.F. and Marshall J.J.

\textit{New apsidal motion determination of eccentric  eclipsing
binary V 1143 Cyg}

Inf.Bull.Var.Stars N4363,1-4.



\bibitem {Hill85}   Hill G. and Fisher W.A.

\textit{Studies of early-type varieble stars.II. The orbit and
masses ofHR 7551}

Astron.Astrophys. {\bf{139}} (1985) 123-131.



\bibitem {Mart80}   Martynov D.Ya. and  Khaliulilin Kh.F.

\textit{On the relativistic motion of periastron in the eclipsing
binary system DI Her}

Astrophys.and Space Sci. {\bf{71}} (1980) 147-170.



\bibitem {Popper82}   Popper D.M.

\textit{Rediscussion of eclipsing binaries.XVII. DI Herculis, a
B-tipe system with an accentric orbit}

Astron.J. {\bf{254}} (1982) 203-213.



\bibitem {Mart97}   Martynov D.Ia. è Lavrov M.I.

\textit{Revizia elementov fotometricheskoi orbity i skorosti
vrashenia linii apsid u zatmennoi dvoinoi sistemy DI Her}

Pisma v Astron.Zh. {\bf{13}} (1987) 218-222 (in Russian).


\bibitem {Kh91}   Khaliulilin Kh.F., Khodykin S.A. and Zakharov A.I.

\textit{On the nature of the anomalously slow apsidal motion  of DI
Herculis}

Astrophys.J. {\bf{375}} (1991) 314-320.



\bibitem {KhKh92}   Khaliulilina A.I. and Khaliulilin Kh.F.

\textit{Vrashenie linii apsid v zatmennoi dvoinoi sisteme  HS Her}

Astron.cirk.N 1552 (1992) 15-16(in Russian).


\bibitem {Mart88}   Martynov D.Ia., Voloshina I.E. and Khaliulilina A.I.

\textit{Fotometricheskie elementy zatmennoi sistemy  HS Her}

Asrton. zh. {\bf{65}} (1988) 1225-1229 (in Russian).



\bibitem {Mezz}   Mezzetti M., Predolin F., Giuricin G. and Mardirossian F.

\textit{Revised photometric elements of eight eclipsing binaries}

Astron.Astroph.Suppl. {\bf{42}} (1980) 15-22.


\bibitem {Mossak87a}   Mossakovskaya L.V. and Khaliulilin Kh.F.

\textit{Tret'e telo v zatmennoi sisteme s apsidalnym dvizheniem CO
Lac?}

Astron. cirk.N1495, 5-6 (in Russian).


\bibitem {Sem}   Semeniuk I.

\textit{Apsidal motion in binary systems. I. CO Lacertae, an
eclipsing variable with apsidal motion}

Acta Astron. {\bf{17}} (1967) 223-224.



\bibitem {Anders91}   Andersen J.

\textit{Accurate masses and radii of normal stars}

Astron.Astroph.Rev. {\bf{3}} (1991) 91-126.



\bibitem {Kh85}   Khaliulilina A.I., Khaliulilin Kh.F. and Martynov D.Ya.

\textit{Apsidal motion and the third body in the system RU
Monocerotis}

Montly .Not.Roy.Astron.Soc. {\bf{216}} (1985) 909-922.




\bibitem {Mart86}   Martynov D.Ya. and Khaliulilina A.I.

\textit{RU Monocerotis: poslednie resultaty}

Astron.zh.{\bf{63}} (1986) 288-297 (in Russian).






\bibitem {Sh62}   Shneller H.

\textit{Uber die periodenanderrungen bei bedeckungsveranderlichen}

Budd.Mitt. N53 (1962) 3-41.


\bibitem {Kort}   Kort S.J., J. de,
\textit{The orbit and motion of priastron of GN Normae}

Ricerche Astron. {\bf{3}} (1954) 119-128.



\bibitem {Kamp}   Kamper B.C.

\textit{Light-time orbit and apsidal motion of eclipsing binary U
Ophiuchi}

Astrophys. Space Sci. {\bf{120}} (1986) 167-189.


\bibitem {Claus86}   Clausen J.V., Gemenez A. and Scarfe C.

\textit{Absolute dimentions of eclipsing binaries.XI. V 451
Ophiuchi}

Astron.Astroph. {\bf{167}} (1986) 287-296.





\bibitem {KhKo89}   Khaliulilin Kh.F. and Kozyreva V.S.

\textit{Photometric light  curves and physical parameters of
eclipsing binary systems IT Cas, CO Cep, AI Hya with possible
apsidal motions}

Astrophys. and Space Sci., {\bf{155}}  (1989) 53-69.




\bibitem {Monet}   Monet D.G.

\textit{A discussion of apsidal motion detected in selected
spectroscopic  binary systems}

Astrophys. J., {\bf{237}}  (1980) 513-528.



\bibitem {Sv}   Svechnicov M.A.

\textit{Katalog orbitalnyh elementov, mass i svetimostei tesnyh
dvoinyh zvezd}

Irkutsk, Izd-vo Irkutsk. Univer.(In Russian).




\bibitem {Br}   Brancewicz H.K. and Dworak T.Z.

\textit{A Catalogue of parameters for eclipsing binaries}

Acta Astron., {\bf{30}}  (1980) 501-524.




\bibitem {Wolf95}   Wolf M. and Saronova L.,

\textit{Aspidal motion in the eclipsing binary FT Ori}

Astron.Astroph. Suppl.{\bf{114}} (1995) 143-146.



\bibitem {Dr}   Drozdz M., Krzesinski J. and Paydosz G.,

\textit{Aspidal motion of IQ Persei}

Inf. Bull.Var.Stars, N3494, 1-4.


\bibitem {Lacy}   Lacy C.H.S. and Fruch M.L.

\textit{Absolute dimentions and masses of eclipsing binaries. V. IQ
Persei}

Astroph.J.{\bf{295}} (1985) 569-579.


\bibitem {Anders83}   Andersen J.

\textit{Spectroscopic observations of eclipsing binaries.V. Accurate
mass determination for the B-type systems V 539 Arae and $\xi$
Phaenicis}

Astron.Astroph.{\bf{118}} (1983) 255-261.


\bibitem {Odell}   Odell A.P.

\textit{The structure of Alpha Virginis.II. The apsidal constant}

Astroph.J.{\bf{192}} (1974) 417-424.




\bibitem {Gron}   Gronbech B.

\textit{Four-color photometry of eclipsing binaries.V. photometric
elements of NO Puppis }

Astron.Astroph.{\bf{50}} (1980) 79-84.




\bibitem {Harm}   Harmanec P.

\textit{Stellar masses and radii based on motion binary data}

Bull.Astron.Inst.Czech.{\bf{39}} (1988) 329-345.



\bibitem {Anders84}   Andersen J., Clausen L.V. and Nordstrom B.

\textit{Absolute dinemtions of eclipsing binaries.V. VV Pyxidis a
detached early A-tipe system with equal components}

Astron.Astroph.{\bf{134}} (1984) 147-157.


\bibitem {lacy93b}   Lacy C.H.S.

\textit{The photometric orbit and apsidal motion of YY Sagittarii}

Astroph.J.{\bf{105}} (1993) 637-645.



\bibitem {lacy93a}   Lacy C.H.S.

\textit{The photometric orbit and apsidal motion of V 523
Sagittarii}

Astroph.J.{\bf{105}} (1993) 630-636.



\bibitem {lacy93c}   Lacy C.H.S.

\textit{The photometric orbit and apsidal motion of V 526
Sagittarii}

Astroph.J.{\bf{105}} (1993) 1096-1102.



\bibitem {Anders85}   Andersen J. and Gimenes A.

\textit{Absolute dinemtions of eclipsing binaries.VII. V 1647
Sagittarii}

Astron.Astroph.{\bf{145}} (1985) 206-214.



\bibitem {Sw}   Swope H.H.

\textit{V 2283 Sgr, an  eclipsing star with rotating apse}

Ric.Astron.{\bf{8}} (1974) 481-490.




\bibitem {O'K}   O'Konnell D.J.K.

\textit{The photometric orbit and apsidal motion of V2283
Sagittarii}

Ric.Astron.{\bf{8}} (1974) 491-497.




\bibitem {Anders85b}   Andersen J., Clausen L.V., Nordstrom B. and
Popper D.M.

\textit{Absolute dinemtions of eclipsing binaries.VIII. V 760
Scorpii}

Astron.Astroph.{\bf{151}} (1985) 329-339.




\bibitem {Claus95}   Clausen L.V., Gimenez A.  and
Houten C.J.

\textit{Four-color photometry of eclipsing binaries.XXVII. A
photometric anallysis of the (possible ) Ap system AO Velorum}

Astron.Astroph.{\bf{302}} (1995) 79-84.


\bibitem {Popp80}   Popper D.M.

\textit{Stellar masses}

Ann. Rev. Astron. and Astroph.{\bf{18}} (1980) 115-164.



\bibitem {Dukes}   Dukesr R.J.

\textit{The beta Cephei nature of Spica}

Astroph.J.{\bf{192}} (1974) 81-91.




\bibitem {Kh87}   Khaliulilina A.I.

\textit{DR Vulpeculae: the quadruple system}

Montly .Not.Roy.Astron.Soc. {\bf{225}} (1987) 425-436.




\bibitem {KhKh88}
Khaliulilin Kh.F. and Khaliulilina A.I.

\textit{Fotometricheskoe issledovanie zatmennoi zvezdy DR Vul.
Parametry sistemy i vrashenie linii apsid},

Astron.zh., {\bf N{65}}  (1988) 108-116 (in Russian).


\bibitem {KhKo89}   Khaliulilin Kh.F. and Kozyreva V.S.

\textit{Photometric light curves and physical parameters of
eclipsing binary systems IT Cas, CO Cep, AI Hya with possible
apsidal motions}

Astrophys. and Space Sci., {\bf{155}}  (1989) 53-69.


\bibitem {Holm}   Holmgren D. anf Wolf M.
\textit{Apsidal motion of the eclipsing binary  IT Cassiopeiae}

Observatory {\bf{116}}  (1996) 307-312.


\end{thebibliography}
\end{document}